\pdfoutput=1
\documentclass[journal,comsoc]{IEEEtran}
\usepackage[T1]{fontenc}% optional T1 font encoding
\usepackage{color}
\usepackage{url}
\usepackage{cite}
\usepackage{graphicx} 
\usepackage{subfigure}
\usepackage[table]{xcolor}
\usepackage{multirow}
\usepackage{cite}
\usepackage{makecell}
\usepackage{longtable}
\usepackage{tabularx}
\usepackage{ulem}
\usepackage{titlesec}
\usepackage[backref]{hyperref} 
\hypersetup{hidelinks}

\ifCLASSINFOpdf
\else
\fi
\usepackage{amsmath}
\usepackage[cmintegrals]{newtxmath}
\begin{document}
%
% paper title
% Titles are generally capitalized except for words such as a, an, and, as,
% at, but, by, for, in, nor, of, on, or, the, to and up, which are usually
% not capitalized unless they are the first or last word of the title.
% Linebreaks \\ can be used within to get better formatting as desired.
% Do not put math or special symbols in the title.
\title{The Security and Privacy of Mobile Edge Computing: An Artificial Intelligence Perspective}

\author{Cheng Wang,
        Zenghui Yuan,
        Pan Zhou, \IEEEmembership{Senior Member, ~IEEE},
        Zichuan Xu,
        Ruixuan Li, \IEEEmembership{Member,~IEEE},
        and Dapeng Oliver Wu, \IEEEmembership{Fellow,~IEEE}% <-this % stops a space

\thanks{Cheng Wang, Zenghui Yuan and Pan Zhou are with Hubei Engineering Research Center on Big Data Security, Key Laboratory of Distributed System Security of Hubei Province, School of Cyber Science and Engineering, Huazhong University of Science and Technology, Wuhan, 430074, China. (e-mail: wangcheng20, zenghuiyuan, panzhou@hust.edu.cn).}
\thanks{Zichuan Xu is with the School of Software, and the Key Laboratory for Ubiquitous Network and Service Software of Liaoning Province, Dalian University of Technology, Dalian 116024, China. (e-mail: z.xu@dlut.edu.cn).}
\thanks{Ruixuan Li is with the Intelligent and Distributed Computing Laboratory, School of Computer Science and Technology,
Huazhong University of Science and Technology, Wuhan 430074, China.
(e-mail: rxli@hust.edu.cn).}
\thanks{Dapeng Oliver Wu is with the Department of Computer Science, City University of Hong Kong, Kowloon 999077, Hong Kong. (e-mail: dpwu@ieee.org).}
% <-this % stops a space
\thanks{Cheng Wang and Zenghui Yuan contributed equally to this work.}
\thanks{\textit{(Corresponding authors: Pan Zhou and Ruixuan Li.)}}
}

\maketitle

% As a general rule, do not put math, special symbols or citations
% in the abstract or keywords.
\begin{abstract}
Mobile Edge Computing (MEC) is a new computing paradigm that enables cloud computing and information technology (IT) services to be delivered at the network's edge. By shifting the load of cloud computing to individual local servers, MEC helps meet the requirements of ultralow latency, localized data processing, and extends the potential of Internet of Things (IoT) for end-users. However, the crosscutting nature of MEC and the multidisciplinary components necessary for its deployment have presented additional security and privacy concerns. Fortunately, Artificial Intelligence (AI) algorithms can cope with excessively unpredictable and complex data, which offers a distinct advantage in dealing with sophisticated and developing adversaries in the security industry. Hence, in this paper we comprehensively provide a survey of security and privacy in MEC from the perspective of AI.
% based on two interesting viewpoints. 
On the one hand, we use European Telecommunications Standards Institute (ETSI) MEC reference architecture as our based framework while merging the Software Defined Network (SDN) and Network Function Virtualization (NFV) to better illustrate a serviceable platform of MEC. On the other hand, we focus on new security and privacy issues, as well as potential solutions from the viewpoints of AI. Finally, we 
comprehensively discuss the opportunities and challenges associated with applying AI to MEC security and privacy
% as well as the security and privacy implications of machine learning in the MEC environment,
as possible future research directions.
\end{abstract}

% Note that keywords are not normally used for peerreview papers.

% For peer review papers, you can put extra information on the cover
% page as needed:
% \ifCLASSOPTIONpeerreview
% \begin{center} \bfseries EDICS Category: 3-BBND \end{center}
% \fi
%
% For peerreview papers, this IEEEtran command inserts a page break and
% creates the second title. It will be ignored for other modes.
\IEEEpeerreviewmaketitle

\section{Introduction}
The number of end devices, which include smartphones, wearable gadgets, tablets, and Internet-of-Things (IoT) devices, etc., has exploded in recent years.
% According to Ericsson's report \cite{Peter2019ericsson}, the global IoT market will reach 24.9 billion devices by the end of 2025.
Simultaneously, a growing number of mobile and IoT applications, such as online gaming, virtual reality, and self-driving vehicles, have become more resource-intensive and latency-sensitive. These applications render the conventional cloud computing paradigm obsolete for its long propagation delays \cite{mao2017survey}. To meet the exponentially growing resource and latency requirements of these applications, the Mobile Edge Computing (MEC) paradigm was proposed \cite{hu2015mobile}. The purpose of MEC is to push the powerful cloud computing capacities onto the edge servers that are in close proximity with end-users. MEC enables the provision of IT services and cloud computing capabilities at the mobile network's edge, within the Radio Access Network (RAN), and in close proximity to mobile subscribers. End devices can access the applications, services, and data on these edge servers with ultralow latency provided by the application vendors \cite{ouyang2018follow}.

As a new distributed computing paradigm, MEC has brought many research topics to academic and industrial areas such as computation offloading, data caching, and service placement, etc \cite{huda2022survey, zhou2022energy, ouyang2018follow}. The security and privacy issues of the MEC environment have gradually attracted the attention of researchers due to the complexity of the MEC service model, multi-source heterogeneous data, and resource-constrained end devices \cite{ranaweera2021survey}. For example, a malware named "Mirai" manages as many as 400,000 compromised smart devices into a controlled "zombies" network to launch a DDoS attack against edge servers, shutting down over 178,000 domains.
% according to the report of FORESCOUT \cite{forescout.com}, smart healthcare has close to 20\% of devices that have default Server Message Block Protocol (SMB) port 445 open, and close to 12\% have default Remote Desktop Protocol (RDP) port 3389 open.
% These ports potentially become the entrance for malicious manipulate or misuse of attackers. As a "black swan" event, the COVID-19 pandemic makes medical devices are currently targets of interest for cybercriminals.
The number of cyberattacks on such facilities doubled again between November 2020 and January 2021 \cite{threatpost.com}. These fragile IoT devices directly or indirectly lead to this despondent result.

Compared with security and privacy issues in traditional cloud computing, MEC possesses several unique characteristics. First, in order to compete, a growing number of IoT devices deployed in the MEC environment are produced with economically manufactured circuitry that employs weak, guessable, or hardcoded passwords and other brittle security measures. Most such devices are easily preempted, contaminated, and destroyed by malicious users. Therefore, the MEC system is vulnerable to security and privacy threats directly brought by IoT devices. In this paper, \textit{we purposefully incorporate the IoT system's security and privacy issues into the MEC in order to thoroughly investigate the MEC's security and privacy issues}. Second, all end devices requesting edge servers' services must through RAN, and this critical juncture is one of the weakest points in the entire network, resulting in serious communication link security issues, such as eavesdropping, hijacking and Distributed Denial-of-Service (DDoS) attacks\footnote{It is a malicious attempt to interrupt normal traffic to a targeted server, service, or network by flooding the target or its surrounding infrastructure with Internet traffic.} \cite{chen2019deep}. Third, in order to achieve a serviceable platform with dynamic resource allocation capability and ease the burden of the network management, Software Defined Network (SDN)\footnote{It is a networking method that employs software-based controllers or APIs to interface with underlying hardware infrastructure and direct network traffic.} \cite{benton2013openflow}, Network Function Virtualization (NFV)\footnote{It is the replacement of network appliance hardware with virtual machines.} \cite{pattaranantakul2018nfv}, and virtualization technologies are vital for realizing the MEC paradigm. These technologies are viewed as the potential solution to MEC's cost and efficiency problems. However, each of these technologies has its own security and privacy challenges which are easy targets for attackers. Finally, in the MEC environment, user data such as user identity information, location information, and sensitive data is typically stored and processed by an honest-but-curious authorized entity, and the user has no way of knowing whether these semi-trusted authorized entities will secretly obtain the user's private information for the purpose of illegal profit.

To decrease the security and privacy burden associated with MEC, numerous research directions have been explored, such as context-aware security, microservices, and blockchain, etc \cite{ranaweera2021survey}. However, the unique characteristics of AI-based approaches have attracted much more attention than other approaches since it has the advantage of handling a large amount of unpredicted and complex data automatically. \cite{joseph2013machine}. AI technologies have been developed rapidly in the past few decades, from initial laboratory research to various commercial applications. As an important subset of AI, Machine Learning (ML) refers to the concept that computer programs can automatically learn from and adapt to new data without being assisted by humans. With the prosperity of Graph Processing Units (GPU) and big data, ML has completed remarkable achievements in many fields, such as Computer Version (CV), robotic, social media marketing, and gaming \cite{jordan2015machine}, and the simplicity and functionality of deploying learning algorithms in these fields significantly surpass almost all traditional rule-based algorithms. These leading advantages are also affecting the development of the security and privacy field.

According to the AV-TEST report, more than 450,000 new malware are registered every day \cite{av-test}. However, most of the instances are just minor variants of the existing malware.
Nonetheless, the correct identification of these specific malware needs to be based on many complex classification methods, such as hashes, simple rules, or heuristic fingerprints \cite{joseph2013machine}. Fortunately, AI algorithms can manage vast amounts of unpredictable and complex data, which offers a distinct advantage in dealing with sophisticated and developing adversaries in the security industry. In order to meet the challenges of security and privacy, many AI algorithms have been used to protect data privacy and address security issues such as spoofing attacks \cite{wang2017physical}, DoS attacks \cite{manimurugan2021iot}, DDoS attacks \cite{chen2019deep}, intrusions \cite{zhao2017intrusion}, jamming \cite{roopak2019deep}, eavesdropping \cite{khalid2020macler}, and malware \cite{libri2020paella}. A suitable learning algorithm can particularly use the trained model generated from the labeled data to recognize new security and privacy threats. Due to the advantages of learning algorithms, an increasing number of researchers are focusing on how to use them to address security and privacy concerns associated with MEC. The MEC environment integrates a variety of devices, such as IoT devices, mobile devices, and third-party servers, so that MEC naturally has a complicated network architecture, communication links, and various network protocols. In addition, the subscription nodes and the service nodes in the MEC have the defect of resource constraints. Therefore, in such a resource-scarce and heterogeneous distributed environment, traditional AI-based methods can no longer cope with the security and privacy challenges brought by MEC. However, the prosperity of various distributed, lightweight and green learning algorithms has paved the way for deploying such algorithms in the MEC environment.
% Some papers \cite{zhou2019edge,deng2020edge,xu2020edge} push the AI-based approach to the edge and name this new intelligence paradigm as Edge Intelligence (EI). The definition of the "edge" in EI is a more broad concept, which can fit different edge paradigms such as MEC, fog computing, cloudlets, CDN, etc. In this paper, we concentrate on the security and privacy of  "MEC intelligence." 

% As we mentioned above, various AI algorithms, especially the wild use of ML algorithms can be applied to every corner of the MEC system to protect its security and privacy. In spite of ML algorithms have so many advantages, the research on vulnerabilities of ML algorithms is still in its infancy \cite{papernot2016towards}. The deployment of ML algorithms in MEC environment will induce new security and privacy issues.
% The training and prediction of ML models rely on tremendous data. The acquisition of these data is often time-consuming and laborious, and the data may contain sensitive or private information about user groups. At the same time, ML models that consume a lot of time and computing power are increasingly becoming the core assets of various companies.
% How to ensure the security and privacy of ML models and data storage in the MEC environment has sparked widespread interest in scientific and industrial circles \cite{xue2020machine,chio2018machine}. Hence, 
In this paper, \textit{we comprehensively consider security and privacy issues in the ETSI reference architecture from the viewpoints of AI.}

The key contributions of this survey are listed as follows:
\begin{itemize}
    \item We purposefully incorporate the security and privacy of IoT systems as well as the SDN/NFV into the MEC environment to thoroughly investigate the MEC's security and privacy issues.
    \item 
    % Since this paper is from AI's perspective, we not only
    We provide an in-depth review of recent security and privacy issues in the MEC environment and from the layer viewpoints of ETSI reference architecture to thoroughly discuss the solutions for implementing AI technologies.
    \item
    % \item We systematically discuss AI approaches for layer-based MEC security and privacy, and finally,
    At the end of this paper, we meticulously highlight the possible future directions about AI approaches for MEC security and privacy issues.
\end{itemize}

\begin{figure}[!tb]
\centering
\includegraphics[width=3.4in]{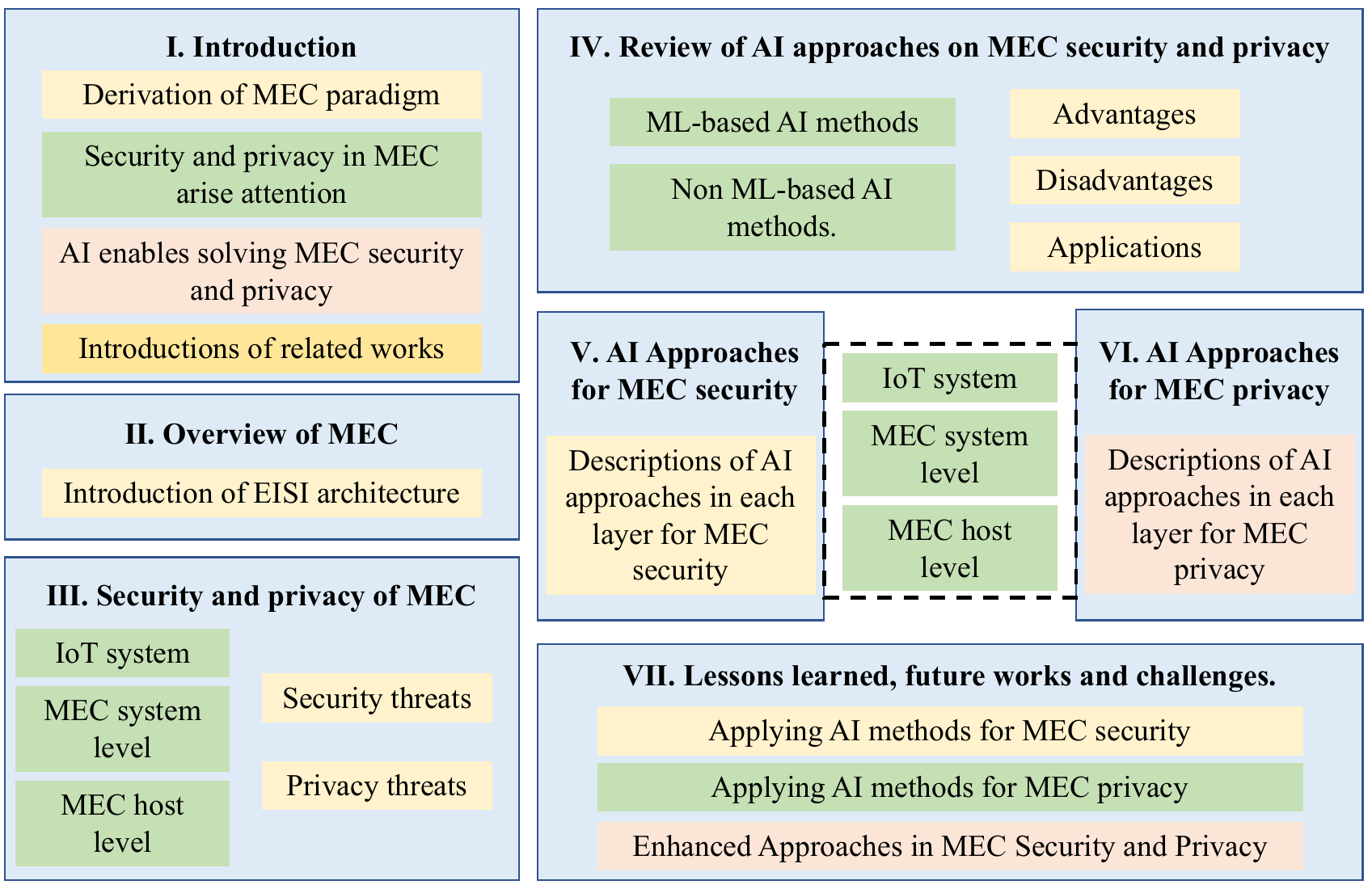}
\caption{Overall architecture of this paper.}
\label{fig:overall architecture}
\end{figure}

Fig. \ref{fig:overall architecture} depicts the overall architecture of this paper. In the following content of this section, we investigate the current survey on discussing MEC's security and privacy and compare the papers that have been published so far and summarize the uniqueness and importance of our work.
% Then, we introduce some prevalent edge computing paradigms. To gain insight into security and privacy issues, we first discuss the detailed layer characteristics of IoT system architecture and the European Telecommunications Standards Institute (ETSI) MEC reference architecture in section \ref{sect:overview}. Additionally, we discuss the symbiotic relationship between the IoT system and MEC. After we catch the global picture of MEC architecture, 
Then, we systematically overview the reference architecture of ETSI MEC in Section \ref{sect:overview}.
In Section \ref{sect:challenge}, we start to comprehensively study the MEC-specific security and privacy issues in the MEC environment. Based on the presented security and privacy issues, in Section \ref{sect:AI perspective}, we thoroughly discuss the most promising AI algorithms, their advantages, disadvantages, and applications in the MEC security and privacy domains. Then, in Section \ref{sect:Layers based security} and \ref{sect:layers based privacy}, we systematically introduce the AI approaches for layer-based MEC's security and privacy issues. In Section \ref{sect:future works}, we summarize the overall work of this paper and propose research problems and future directions. Finally, we draw conclusions in Section \ref{sect:conclusions}.

% Relevant acronyms presented in this survey are tabulated in Table \ref{table:notations}.

\subsection{Related Work}
In order to meet the increasing demand for sensitive applications and the proliferation of IoT devices, some papers have proposed to add an edge side fall between the cloud and users to reduce the pressure on the bandwidth and network traffic of cloud servers and increase the Quality of Services (QoS) of users. The papers \cite{shi2016edge,singh2022ai,satyanarayanan2017emergence,varghese2016challenges,khan2019edge} proposed an initial three-tier framework about edge computing paradigms. The authors of \cite{shi2016edge} defined "edge" as any computing and network resources located between data sources and cloud data centers, and they provided several case studies involving computing offloading, data caching, data processing, and service delivery. 
% Following this, various researchers focused on various edge computing paradigms, including early edge paradigms such as Content Delivery Network (CDN) \cite{satyanarayanan2009case} and cloudlets \cite{jia2015optimal}, as well as emerging edge paradigms such as fog computing \cite{bonomi2011connected}, MEC \cite{abbas2017mobile}, and mobile cloud computing (MCC) \cite{fernando2013mobile}. The goal of these edge paradigms is to overcome the high and unpredictable latency introduced by the traditional cloud computing.

The development of fifth-generation (5G) wireless networks is gaining momentum, with the goal of connecting almost every aspect of life via the network at a significantly faster speed, with extremely low latency and ubiquitous connectivity. Additional to this, 5G networks, based on the 3GPP (The 3rd Generation Partnership Project) 5G specifications \cite{giust2018mec} will be a key future target environment for MEC deployments in the coming years. Mao et al. \cite{mao2017survey} provided an in-depth overview and future research directions for MEC from a 5G communication perspective. The authors of \cite{peng2018survey} presented a comprehensive survey of MEC from the standpoint of service adoption and provision. To meet the stringent latency requirements of applications (e.g., real-time applications) and to reduce energy consumption at User Equipment (UE), the authors \cite{mach2017mobile} proposed a survey about computing offloading issues from different aspects. 
% In recent years, the new intelligence paradigms EI has been successively proposed by Zhou et al. \cite{zhou2019edge}, Zhang et al. \cite{zhang2019openei}, and Wang et al. \cite{wang2020convergence} to fulfill the urgency of extending the AI frontiers to network edge so as to fully realize the potential of the edge big data. In \cite{zhou2019edge}, the concept of EI has been first introduced, and they review the background and motivation and the overarching architectures for AI running at the network edge.
% The concept of the edge introduced in \cite{zhou2019edge,zhang2019openei,wang2020convergence} is in a wide range, which is not discussed MEC from a more detailed perspective. However,

The MEC's ability to ensure security and privacy has been widely questioned due to the use of multiple communication technologies, complex network structures, and multi-source heterogeneous data types \cite{ranaweera2021survey,ali2021multi,chen2019deep}. The papers\cite{ahmed2017mobile,ai2018edge,mach2017mobile,peng2018survey} made no mention of security or privacy as critical aspects of MEC. Although certain surveys, such as \cite{varghese2016challenges,bonomi2011connected,fernando2013mobile,chen2019deep} and \cite{ahmed2017mobile}, mainly focused on security and privacy, the contexts are not concurring to the ETSI standardized MEC architecture and its components. When it comes to security and privacy, the MEC paradigm is closely aligned with ETSI standards, according to a survey published in \cite{ranaweera2021survey}. This survey aimed to guide the research communities on their way toward a feasible MEC deployment. Ali et al. \cite{ali2021multi} proposed a survey about the data security and privacy based on the ETSI standardized MEC architecture. At the same time, some surveys about MEC security and privacy around ML began to appear \cite{subramaniam2019review,chen2019deep,chen2017deep}. In \cite{al2020survey}, they provided a comprehensive survey of ML methods and recent advances in deep learning (DL) methods that can be used to develop enhanced security methods for IoT systems. Authors in \cite{shambour2022progress} specifically summarize the application of some AI technologies in IoT-based hajj and umrah scenarios. Singh et al. \cite{singh2021machine} presented a review of ML for assisting security and privacy issues of Edge Computing (EC), however, they only discussed the naive three-tier architecture and did not consider the security and privacy issues that two critical auxiliary technologies (i.e., SDN/NFV) brought to the EC environment.

To the best of our knowledge, we are the first survey that synthetically investigates ETSI standardized MEC, IoT system as well as their assistive technologies SDN/NFV's security and privacy issues from the perspective of AI. On the one hand, this survey identifies and compares the opportunities, benefits, and drawbacks of various AI approaches for MEC security and privacy. On the other hand, we comprehensively consider various possible AI solutions from the layer viewpoints of ETSI MEC reference architecture. Based on reviewing potential AI applications in the MEC security and privacy context, we discuss and present the identified challenges and future directions.

\section{Overview of Mobile Edge Computing}\label{sect:overview}

European 5G PPP (5G Infrastructure Public Private Partnership) research body recognizes the MEC as one of the key technologies in the development of 5G and beyond 5G technologies \cite{hu2015mobile}. MEC opens up services to mobile users and enterprise entities as well as to adjacent manufactories, these entities now can deliver their resource-intensive and latency-sensitive applications over the mobile network. ETSI and Information Services Group (ISG) proposed the primitive standards and architecture of MEC which illustrated in Fig. \ref{fig:mec}. As an extension of cloud computing, the purpose of MEC is to push the powerful cloud computing capacities onto the MEC servers that are in close proximity with end-users. MEC has the following unique characteristics which differ from the traditional computing paradigms:

\begin{itemize}
    \item \textit{On-Premises:} MEC can run independently of other networks, this is an essential attribution in machine-to-machine (M2M) scenarios.
    \item \textit{Proximity:} The MEC servers are often attached to the base stations or access points close to the end-users, improving real-time response ability.
    \item \textit{Lower Latency:} End-users data can be directly processed on the nearby MEC servers without delivering to the remote cloud. Hence, the end-users QoS and QoE (Quality of Experience) will get a considerable improvement. 
    \item \textit{Location Awareness:} The MEC servers can only serve the specific geographic location’s users covered by the base station to which it is connected.
\end{itemize}

In this subsection, we divide the edge server system into two layers according to the ETSI-published MEC framework and the reference architecture \cite{giust2017multi,ranaweera2021survey}. Each layer has several submodules, and there are three groups of reference points to link different submodules. Where Mx, Mm, and Mp denote reference points connecting to external entities, management reference points, and reference points regarding the MEC platform functionality, respectively.
\subsubsection{MEC System Level}
User Equipment (UE) and Customer Facing Service (CFS) Portal are all external devices subscribed to the edge server services. Users interact with the mobile edge system through the UE application that is instantiated in the UE. Specifically, the CFS Portal enables third-party customers to select and order a variety of edge applications to meet their unique requirements.
Both UE application and CFS Portal connect the edge server through Access Networks (AN). User App Life-Cycle Management Proxy (UALCMP) determines the UE applications can initialize, terminate or relocate mobile edge applications and decides on the granting of these requests to be forwarded to Operations Support System (OSS) and Mobile Edge Orchestrator (MEO).
Then it can send the status information of mobile edge applications to UE applications. In particular, UALCMP can only receive requests which are within the geographic area that MEC servers covered, which means that it needs to fulfill the proximity constraint. OSS receives requests for initiation or termination that forwarded from UE applications and CFS Portal, then decides whether to authorize these requests for further processing on MEO. MEO, as the backbone functional block of mobile system layer, manages both MEC system level and MEC host level. Specifically, it can grant requests forwarded from MEC system level to initiate, terminate or relocate mobile edge applications, and dominate the resources of mobile host layer, such as selecting the appropriate mobile edge host to tackle low-latency or resource-incentive requests.

\subsubsection{MEC Host Level}
MEC host level can be divided into MEC server management level and MEC server level. Mobile Edge Platform Manager (MEPM) as the backbone of MEC host level, is responsible for the management of various functional blocks in Mobile Edge Host (MEH), including MEC platform element management, Mobile Edge (ME) application rules and requirements management (i.e., service authorizations, traffic rules formulation, Domain Name System (DNS) configuration and resource conflict resolution). MEPM is also responsible for managing the life cycle of applications and forwarding information to related applications through MEO. Virtualization Infrastructure Manager (VIM) is mainly responsible for allocating and releasing virtual resources (storage, compute and networking resources) of Virtualization Infrastructure (VI), and using VI resources to create software images for serving CFS Portal and UE applications.
VIM also collects the current status information of VI resources to MEO and MEPM. ME applications are software entities built on the top of VI. In a MEH, the connectivity among different ME applications is established through Local Area Data Network (LADN) \cite{kekki2018mec}. After receiving traffic rules and DNS records from MEPM, MEC Platform uses traffic rules controller and DNS handling to configure the traffic rules of MEC applications and DNS proxy/server. MEC Service contains different services to facilitate MEC applications and MEC platform.

% \subsection{Lessons Learned}

% MEC, as one of the key technologies, provides a promising solution to ease the burden of low latency and computationally intensive requirements of IoT systems by pushing the powerful computation capacity to the edge of the cloud \cite{lyu2017optimal}. In this paper, we include the security of the IoT system into the overall security of the MEC for two main reasons. On the one hand, thousands of IoT devices are one of the main reasons for the prosperous development of MEC. As depicted in Fig. \ref{fig:mec}, most of the requests for CFS and UE application functional blocks will be generated by IoT devices, and many researchers currently investigate interests are combining IoT and MEC \cite{sabella2016mobile, zhao2019novel, sarrigiannis2019online, du2018energy}. On the other hand, the extensive usage of MEC services of IoT systems can easily become a breakthrough to compromise MEC security. Therefore, the security and privacy in the IoT system can greatly affect the security and privacy of the MEC. This is why we have to consider the security of the IoT system as the content of the MEC security.

\begin{figure}[!tb]
\includegraphics[width=3.5in]{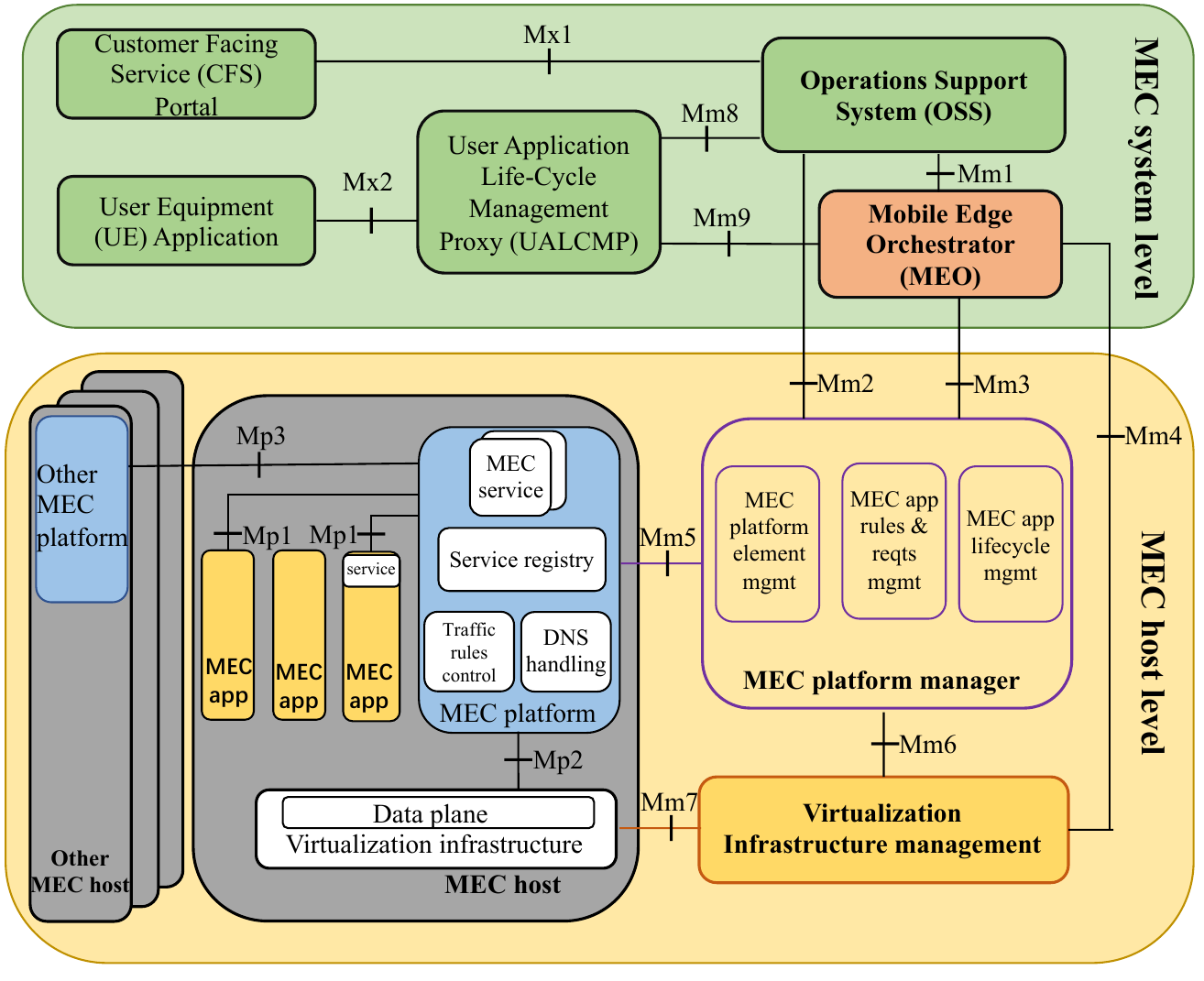}
\caption{MEC reference architecture.}
\label{fig:mec}
\end{figure}

\section{Security and Privacy Challenges of MEC}\label{sect:challenge}
\begin{figure*}[!tb]
\centering\includegraphics[width=7.2in]{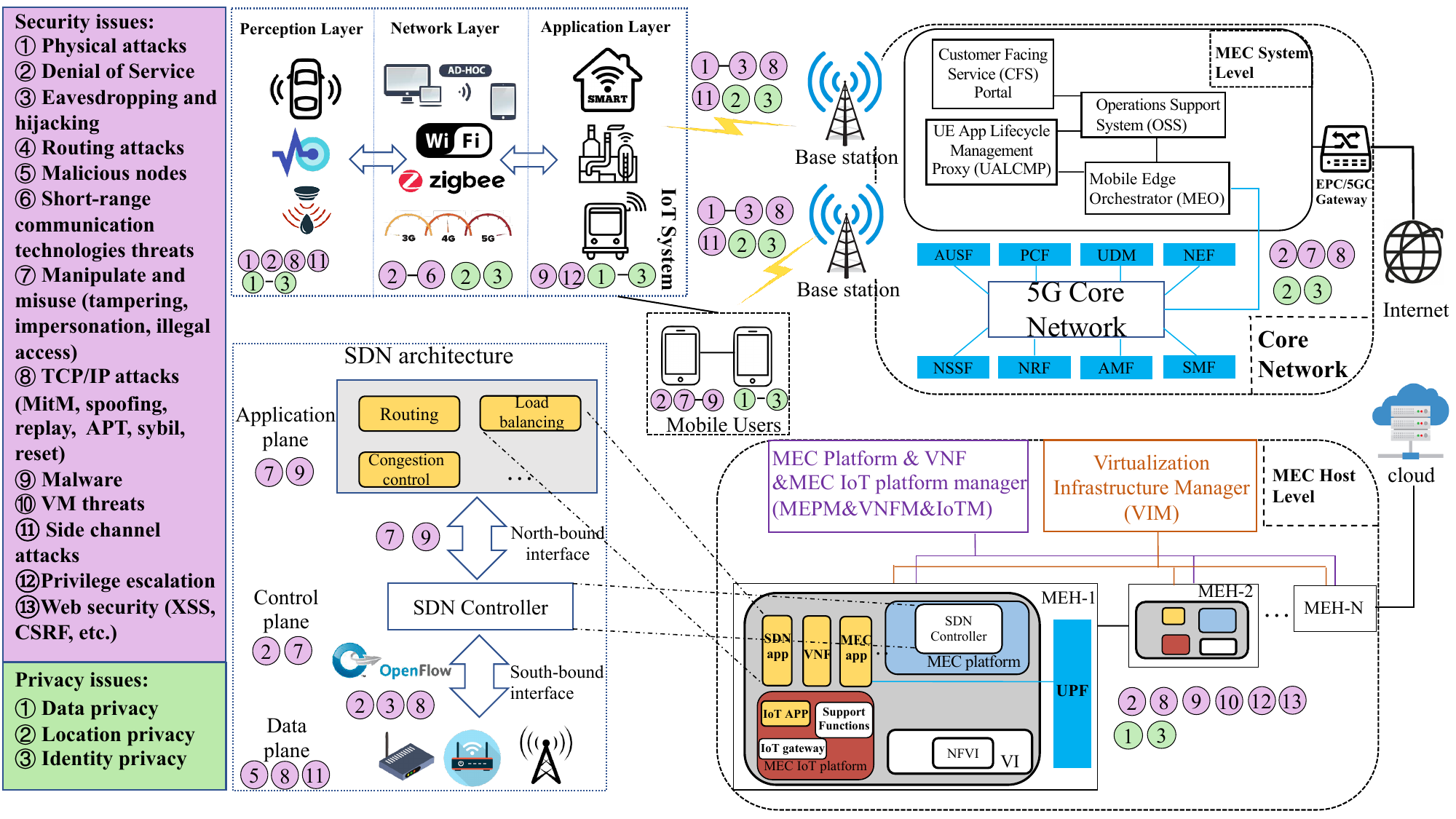}
\caption{Security and privacy threats of a typical MEC deployment.}
\label{fig:security and privacy}
\end{figure*}

Edge computing scenarios are centred on time-sensitive services, such as industrial IoT, autonomous driving, smart cities, etc. As a result, when designing network protocols and topologies, security and privacy are frequently sacrificed in favor of real-time and effective communication. Edge servers with on-demand and close proximity attributions are exposed to the network edge, shortening the distance between the attacker and the MEC physical devices. At the same time, the widely open Application Programming Interfaces (APIs) make it easy for attackers to initiate security threats such as data theft, information tampering, and node intrusion \cite{lounis2020attacks}. Finally, end-users' mobile nature allows them to dynamically join or exit the edge servers that cover them, and frequent topology changes between mobile devices and edge servers will have an impact on network resource management, allowing attackers to launch adaptive attacks by exploiting the calculation transforming between different edge servers. Fig.\ref{fig:security and privacy} \cite{zanzi2019evolving,campolo2018slicing,schiller2018cds} shows the overall architecture of a typical MEC deployment with its security and privacy issues.

% \subsection{MEC-specific Security Threats}
\subsection{Security Threats}
In this subsection, we comprehensively consider the security issues in MEC and IoT systems with their serviceable platforms SDN and NFV.

\subsubsection{IoT Systems Security}
From the perspective of the ETSI MEC architecture, MEC enabled IoT systems consists of two parts, one is IoT devices deployed in the actual environment, and the other is MEC IoT platform hosts as software instance which is migrated to the MEC facilities. IoT devices easily suffer various security risks due to current manufacturing and service vendors' lack of security awareness, lagging security standards, limited software and hardware resources, and the weakness of security protection capabilities. Large-scale cyber attacks targeting or originating from IoT have sparked widespread concern. For example, a malware named Mirai can exploit as many as 400,000 compromised smart devices into a controlled "zombies" network to launch a DDoS attack \cite{cimpanu2016you}. The MEC IoT platform appears as a service provider for MEC applications running on MEC hosts and enabled via the Mp1 interface as depicted in Fig.\ref{fig:mec} \cite{zanzi2019evolving}. In other words, MEC applications can find the MEC IoT platform by querying its service registry, and interact with it using a defined IoT API exposed by the MEC IoT platform.
\paragraph{Perception Layer}
The main goals of the perception layer are to interact with the complicated real world and cooperatively connect heterogeneous sensors to provide diversely intelligent services. The perception layer can also be called the “sensors” layer. Considering the economic pressure brought by the massive deployment of IoT devices, they are generally designed as resource-constrained devices with low power and limited computing resources \cite{zeng2011web}. Therefore, it is a challenge to identify malicious data and authenticate benign equipment through traditional intrusion detection and authentication methods \cite{wang2017physical}.

\textit{DoS/DDoS Attacks:} The notorious DDoS attack, which has been intensively studied in the traditional cloud computing environment, is a new challenge and an open research topic in the MEC environment. Unlike traditional DDoS attacks in the cloud environment that mainly utilize computers as bots, a DDoS attack in the EC environment is often coordinated via control of mobile and IoT devices \cite{he2021game}. There are many applications with different functions installed on mobile user devices, but the security consciousness of different application developers is uneven, and some malicious applications often try to require permissions that are out of scope. Especially for some android devices, although the open-source android framework brings convenience for developers to require APIs to implement various functions, the fragile supervision capabilities of android communities make users easily threatened by malware. Simultaneously, some applications may hijack the user device's microphone in order to collect private information from daily conversations.

\textit{TCP/IP Attacks:} If a fake base station can easily transmit a spoofing synchronizing signal with sufficient high power during the cell selection stage, wireless mobile UE may be drawn to and attempt to camp on the fake base station rather than any legitimate base station \cite{huang2018identifying}.

% \textit{\romannumeral2): DoS/DDoS Attacks}

% Mobile user phone is also vulnerable to DDoS attacks. For instance, an adversary can launch a robocall attack that continuously calls up the victim to make the victim’s mobile phone unable to receive normal phone calls.

\paragraph{Network Layer}
The network layer is the backbone of the IoT system and has three main functions. First, it is responsible for establishing the network topology among IoT devices, then receiving and processing the raw information of the perception layer, and finally, the refined information is transmitted to the application layer to provide services for real-world customers \cite{al2020survey}.

\textit{Ad-hoc Network Security:} Some short-range communication technologies can establish connections between UE equipments such as Bluetooth, WiFi, IrDA, ZigBee, etc. This type of connection is device-to-device which establishes a direct communication link among UE equipments in IoT systems without requiring any edge servers for connection, and we call this type of network as Mobile Ad-hoc Network (MANET) \cite{lounis2020attacks}. These communication technologies allow billions of heterogeneous devices to connect to the backbone and communicate with each other to complete complex and diverse intelligent functions. However, most end devices that adopt these technologies are resource-constrained so that the adversary can violate the route structure, congest wireless channels, and inject malicious nodes with ease.

\textit{Routing Attacks:} While the MANET's nodes roam in a hostile environment, some statical security solutions are unable to adapt to the dynamic change in topology. The routing protocol of MANET such as Dynamic Source Routing (DSR) establishes route paths from a source node to a destination node by exchanging the network topology information between these two nodes. Since all messages are transmitted through wireless channels, so that any intruder can forge a valid routing updates and maliciously furnish incorrect routing states information. Intruders can also modify the route request (RREQ) or route replay (RREP) packets to delete a node, switch the order of the nodes, and append a node. 

\textit{Attack on Short-range Communication Technologies:} In an ad-hoc network, all communication signals must pass through a bandwidth-limited wireless channel, making the network vulnerable to physical-layer threats. The attacker can also eavesdrop and modify the information within the wireless communication channels or impersonate a legitimate participant to intentionally inconsistent wireless channels. Bandwidth-limited wireless channels are also vulnerable to DoS attacks via network-layer packet blasting. These packets exhaust a significant proportion of the computing resources, which introduce the wireless channel contention and network congestion.

\textit{Malicious Nodes:} In a hostile environment, it is impractical for an ad-hoc network to rely on the cooperation of all nodes to complete decentralized control, while malicious nodes can suspend the cooperation algorithm by refusing to respond. This situation also makes some centralized intrusion detection mechanisms incapable of detecting a node failure or a malicious intrusion.

\textit{Side Channel Attacks:} Some IoT devices, such as smart medical devices, wearable devices, and smart home devices, collect raw data about the users' behavior and status {\cite{farooqi2019smart}. These data are much more detailed and accurate, which contain much more sensitive information. As a result, an adversary can exploit these raw data to obtain additional privacy information. For example, Liu et al. \cite{liu2015good} analyze the data collected by the accelerometer sensors in the smartwatch via the new and practical side-channel to successfully infer the user's keystroke behavior \cite{liu2015good}.

\paragraph{Application Layer}
The application layer serves application subscribers by providing intelligent services. For instance, the application layer can provide surveillance video data and humidity measurements to subscribers. The importance of the application layer is to provide users with high-quality intelligent services.

 \textit{Malware:} IoT system applications are most closely connected with users, and they are required to defend against security and privacy threats while providing services for users. However, malware will deliberately obtain extra information beyond its requirements, resulting in the leakage of user's sensitive data and the preempt of IoT devices. For example, Fernandes et al. \cite{fernandes2016security} analyzed the source code of SmartThings apps and found that more than 50\% of the apps on the Samsung smart home platform have significant overprivileged.
% \textit{\romannumeral1):Malware}

% IoT system applications are most closely connected with users, and they are required to defend against security and privacy threats while providing services for users. However, malware will deliberately obtain extra information beyond its requirements, resulting in the leakage of user's sensitive data and the preempt of IoT devices. For example, Fernandes et al. \cite{fernandes2016security} analyzed the source code of SmartThings apps and found that more than 50\% of the apps on the Samsung smart home platform have significant overprivileged.
% \paragraph{Vulnerabilities in Gateway Devices}
% Gateway devices allow traffic to flow in and out of the network and translate the data received from external networks into a format or protocol recognized by devices within the internal network. Since all data must flow through them, it is difficult to avoid common TCP/IP attacks such as eavesdropping, MitM, spoofing, DoS, replay and reset attacks, etc. The widely used gateway devices make it impractical for network administrators to ensure the physical security of each device.
% \paragraph{Physical Attacks}
% Since sensors are widely distributed in some environments such as agriculture and industry, they will be subjected to more stringent physical security. They may be unattended for a long time and maybe directly captured by adversaries \cite{zhao2013survey}. At the same time, some natural disasters such as earthquakes, tsunamis, and tornadoes will also affect the safety of sensors.

% \subsubsection{Mobile Core Network}
\subsubsection{MEC system level with 5G Core Network}
Mobile core network needs to ensure that end devices are securely connected to edge servers through an access interface and run a series of authentication protocols to prevent unauthorized devices. It also needs to provide some security measures such as integrity protection and encryption to protect the communication information from manipulating and eavesdropping on the wireless communication channel. At the same time, the mobile core network is eager for the necessary differentiated security mechanisms to serve various personal businesses and vertical services.

\paragraph{Access Network security}
As part of a mobile telecommunication system, AN resides between end devices and provides wide-area wireless connectivity to edge servers. It traverses the emanated service requests located at the different geographic locations to the edge servers due to the resource-constrained UE. In order to enable high network throughput, ubiquitous connections, and low latency, some novel technologies such as massive multiple-input-multiple-output (MIMO), interference-aware receivers, and advanced coding/modulations are proposed to improve the spectral efficiency \cite{chen2017machine}. The connectivity between the UE  and edge servers through these heterogeneous technologies raises several security concerns that could be exploited by the attacker \cite{ranaweera2021survey}.

\textit{Denial of Service (DoS):} Jamming a wireless channel or compromising a service via DoS is to destroy the availability of AN connected to edge servers. The attacker raises a DoS attack in the MEC environment can only affect the edge servers where the botnets can access \cite{he2021game}. Therefore, it is difficult for the attacker to use their global botnets to launch a tremendous DDoS attack on a specific target edge server. However, the MEC has deployed a large number of latency-tolerant applications, disrupting legitimate users' access to services will greatly affect the QoS of MEC systems. It is difficult to detect such malicious network activities because of the direct connection between MEC systems and end devices \cite{mach2017mobile}. Novel botnet type DDoS attacks will also affect the availability of UALCMP and CFS Portal in the MEC system level.

\textit{Eavesdropping and Hijacking:} A wireless channel between UE equipment and edge servers is prone to cyber risks such as eavesdropping and hijacking due to the broadcast nature of the wireless medium. As a result, wireless communication channels are vulnerable to being hijacked to retrieve information by cyberattacks such as man-in-the-middle (MitM), replay, Advanced Persistent Threat (APT), Sybil, and spoofing attempts \cite{lu2018managing}. Traditional wireless channels mainly use some encryption protocols to secure the transmission process \cite{lu2018managing}. However, the process of information encryption and private key exchanges between channels largely hinders its applicability for massive latency-sensitive applications in the MEC environment.

\paragraph{MEC System Level}
MEC system level as part of the core network of MEC system connecting to end devices, MEH, and 5G core network. It has the following three aspects of functions. First, it determines to grant services for further process and handle services life cycle forwarded from end devices. Second, it guards the resource utilization status, and the configuration of VMs and underlying hardware in the MEH. Finally, it switches control signals with a 5G core network via wireless, wired, or optical. Thus, the MEC system level is critical to attackers for gaining unauthorized access, manipulating or misusing ME applications, etc.

\textit{DoS/DDoS Attacks:} As the UALCMP is the entity handling the life cycle of requests, while the OSS grants the approval for subscribers to use a particular MEC service. The attacks over them can be targeted at congesting the access interfaces between them and UE so that both of them have to be protected from DoS/DDoS attacks.

\textit{Manipulate and Misuse:} Since all the ME application subscribers should be registered in OSS or UALCMP, the attackers could attempt to inject craftily constructed information to manipulate or misuse the functionality of these two. At the same time, malicious UE can inject fake information to impersonate legitimate entities to disrupt the everyday activities of MEO and MEPM. The malicious intrusions are improbable at MEO since MEO is deployed at the in-depth 
MEC system level that is difficult for attackers to reach. However, resource allocation and service manipulation attacks are highly possible such as DNS amplification and VM escape.

\textit{TCP/IP Attacks:} Separate physical hosts for the 5G core network and MEO are more prevalent \cite{costa20175g}, which introduce typical TCP/IP attacks such as eavesdropping, spoofing, DoS, replay, and reset attacks.

\paragraph{SDN Threats}
In the MEC environment, in order to reduce network management costs and improve the network scalability and flexibility, infrastructure providers have introduced SDN. SDN separates the network architecture into a three-tier architecture of application, control and data plane \cite{ahmad2019security}, controlling the network with software to achieve centralized management of the network and programmability of network applications. This convenient network management approach is not only convenient for operators but also for attackers. A capable attacker can preempt SDN applications or tamper with the flow tables in SDN controllers to chaos packets forwarding.

\textit{Applications Plane Threats:} Network functions are implemented as applications in this plane, and these applications are created by using the VI's computing resources in the MEH. These applications aim to provide network services such as QoE, monitoring, load balancing and security, etc. These services host as one type of special MEC application in the MEH and exchange data with the SDN controllers through the north-bound interface. Thus, an authorized malicious application can invade other applications to control the network. This malicious threat can be caused by the open APIs of network equipment, the lack of mutual authentication mechanism between the application plane and the control plane, or the implementation of the wrong access control method for third-party applications\cite{kreutz2013towards}.

\textit{Control Plane Threats:}
The SDN control plane communicates with the application plane and the data plane through the north-bound interface and the south-bound interface respectively, and there can be multiple controllers in SDN. In ETSI MEC architecture, these controllers are as one of the support functions that deployed in the MEC platform \cite{schiller2018cds}.
% Each controller can handle multiple data plane devices by sending flow forwarding rules, and each data plane device can also be controlled by multiple controllers at the same time.
The control plane adopts a centralized management architecture, which relies on programming to achieve overall control of SDN data plane devices. Thus, an attacker can directly send specifically crafted DDoS flows to overwhelm the resources of control plane \cite{shin2013attacking} or IP packets with random header fields to disturb legitimate flow setup \cite{fonseca2012replication}. Also, it is incompetent to secure application authorization, resource usage and tracking while malicious applications in control plane \cite{nadeau2011software}.

\textit{Data Plane Threats:} The main elements of the data plane are switches and routers. They are simple packets forwarding elements without embedded intelligence to take autonomous decisions. These data plane devices communicate with the controller through a standard OpenFlow interface, ensuring the compatibility and interoperability of configuration and communication among different devices. Since the packet forwarding of the data plane depends on the flow tables issued by the control plane, an attacker can control the forwarding of data packets by intercepting or tampering with the flow tables sent to the SDN switches \cite{benton2013openflow}. In addition, three types of attacks may be used to compromise the data plane, including device attack, protocol attack, and side-channel attack \cite{shaghaghi2020software}. Device attacks aim to exploit SDN software or hardware vulnerabilities to compromise the SDN data plane. Protocol attacks exploit network protocol loopholes in the forwarding equipment (e.g., as Border Gateway Protocol (BGP) attacks) to attack the data plane. Side-channel attacks infer the network's forwarding strategy by analyzing the performance metrics of the forwarding equipment. For example, by analyzing the processing time of data packets, attackers can identify forwarding strategies \cite{scott2015survey}.

% \textit{Interface Threats:}
% The configuration complexity of Transport Layer Security (TLS) and Datagram Transport Layer Security (DTLS) in a south-bound interface introduces multitude of attacks such as eavesdropping, DDoS, MitM, etc. The north-bound interface connects the application plane and control plane and is responsible for the information exchange between applications and controllers. Due to the wide variety of applications and the lack of a standard API, the authentication mechanism of the north-bound interface is not comprehensive enough. The main attacks faced by the north-bound interface include data leakage, message tampering, illegal access, identity impersonation, application vulnerabilities, and new vulnerabilities introduced by different applications during cooperation. 

\subsubsection{MEC Host Level Security}
In the MEC paradigm, the MEH is the primary host-level functional entity that performs computational, storage, and networking operations. Furthermore, the User Plane Function (UPF) is a 5G access network entity included within the MEH for integrating the 5G core network into the LADN. Since the SDN security threats and the MEC IoT platform have been illustrated above, we will focus on the virtualization threats and NFV threats in this subsection.

\paragraph{VM Threats} Virtualization integrates physical resources into a resource pool providing various on-demand services, which decreases the high management complexity and operational expenses, and enhances the efficiency for usage and fine-grained control.  However, novel security threats and vulnerabilities introduced in the following contents become one of the major concerns in the MEC environment.

The overall VM entities are actually composed of SDN applications, VNFs, IoT applications, and the MEC applications, which provide services for users or MEC architecture. The threats and vulnerabilities target to the VMs are as follows:

 \textit{Infected VM images:} VM image, as a pre-packaged software file, contains the configuration templates used to initiate the VM instances on demand. The resource renters can either create their own VM images from an image software or downloading VM images stored in the third-party's repository \cite{ahmad2019security}. Thus, this gives the attacker the opportunity to upload VM images injected with malicious code such as a Trojan horse to the third-party repository, and the victim will be infected with the hidden malware after uploading such malicious VM images.
 
\textit{Compromising VM migration:} VM migration allows network operators to optionally initiate, terminate a VM, or move VMs from one physical machine to another. The migration of VM benefits the workload balancing and system management. Due to the dynamic nature of VM and the plaintext presented in migration data, an attacker can easily launch MitM attacks to sniff or tamper the traffic.

\textit{VM hopping:} The attacker gains access to a host with multiple VMs by renting or hacking a guest VM. From the persecuted guest VM, the attacker then compromises other guests' VMs through privileged access to the host. The reason for this could be that the Memory Management Module (MMU) of the hypervisor allows attackers to perform illegal manipulation on the memory pages of other guest VMs based on the access rights they have obtained \cite{pattaranantakul2018nfv}.

\textit{VM escape:} Any interaction between VM and hypervisor may become a potential attack vector \cite{szefer2011eliminating}. VM escape refers to the program running in the virtual machine using the vulnerability of the VM to break through the Virtual Machine Monitor (VMM) or hypervisor. After that, the adversaries may obtain the host OSS management authority, control other virtual machines running on the host machine\cite{dubrulle2015blind}, and completely destroy the original security architecture.
    % \item \textit{VM DoS:} In order to improve the utilization of physical resources, multiple VMs are often created in one physical host to share physical resources (e.g., CPU, GPU, memory, and network bandwidth) with each other. Thus, an adversary can launch a DoS attack in the compromised VM to exhaust all other VMs’ available resources.

\paragraph{NFV Threats}
NFV, as one of the key emerging technologies for 5G networks, consolidates multiple network functions onto the software, running on a range of industry-standard hardware \cite{pattaranantakul2018nfv}. The infrastructure that hosts MEC and NFV is quite similar \cite{hu2015mobile}. Thus, it will be beneficial to reuse the infrastructure and infrastructure management of MEC by hosting both Virtual Network Functions (VNFs) and MEC applications on the same platform. VNFs can be recognized as software that encapsulate network functions in a VM. Therefore, most NFV threats also are inherited from VM threats. NFV-specific security threats may from the NFV management and orchestration (NFV MANO). It is mainly responsible for the orchestration and management of virtual resources, the creation of virtual network functions, and NFV lifecycle management. ETSI has published an official document about NFV MANO to describe how the NFV and its interfaces work without specific details about interface design and implementation \cite{pattaranantakul2018nfv}. Thus, an adversary may exploit the insecure interface to obtain sensitive data or launch a Cross Site Scripting (XSS) or Cross Site Request Forgery (CSRF) attack by injecting a well-constructed script into the web surfaces provided by the management interface.

\subsection{Privacy Threats}
User data in the MEC environment, such as user identity information, location information, and sensitive data, is typically stored and processed by an honest-but-curious authorized entity (e.g., edge data center and infrastructure provider), and the user has no way of knowing whether these semi-trusted authorized entities will secretly obtain the user's private information to achieve the purpose of illegal profit. At the same time, in the open ecosystem of MEC, multiple trust domains are dominated by different infrastructure providers, and users cannot identify which service provider is trustworthy. Thus, MEC, with the complexity and real-time nature of the service mode, multi-source heterogeneous data, and resource-constrained end devices, has more delicacy privacy threats.

In this subsection, we will focus on three aspects of privacy concerns: data, location, and identity privacy.
\subsubsection{Data Privacy}
Users outsource data to edge nodes (i.e., IoT devices or edge servers) with computing resources, which give edge nodes the opportunity to control over the data, introducing the same security risks as cloud computing. It is challenging to ensure the confidentiality and integrity of the data since the complicated communication link may cause data to be lost or maliciously modified. Additionally, through privilege escalation, unauthorized entities may exploit the uploaded data for their own gain. Compared to cloud servers, edge servers have partially circumvented the data security and privacy issues caused by the long-distance transmission of multi-hop routing. However, the applications that are dominated by different application vendors and the access networks that belong to different telecom operators have compelled MEC to introduce more severe data privacy issues, such as coexistence of multiple security domains and data in multiple formats.

\subsubsection{Location Privacy}
LBS refers to using a certain positioning technology (e.g., global positioning system, mobile phone positioning, and positioning through WiFi access points) to provide mobile users with personalization related to their current location service. However, these services frequently collect location data in the background without the user's knowledge or consent. As a result, figuring out how to protect a user's location privacy has become a pressing issue \cite{lee2012efficient}. The leakage of location information can be divided into three major kinds of threats:

\begin{itemize}
    \item \textit{Tracking Threat:} The adversary may obtain continuous location updates, allowing him to pinpoint the user in real time. For example, in Vehicle Ad-hoc networks (VANET), an attacker can eavesdrop on the communication between vehicles to lock and track the target vehicle; the attacker can tamper with the traffic information in the network and publish false information on the network, which creates the illusion of road congestion for other vehicles and affects the route selection of other vehicles.
    \item \textit{Identification Threat:} Even if the adversary only accesses the user's location on a sporadic basis, he may be able to isolate the user's frequently visited locations, such as home and work. The adversary can use these locations as pseudo-identifiers to deduce the user's identity from anonymous location traces \cite{fawaz2014location}.
    \item \textit{Profiling Threat:} The mobility track of the user may not include places that reveal his identity but can be used to profile him by the adversary. Some Location Service Providers (LSPs) collect and analyze the location attributes in the user's request to infer the user's personal information, behavioral preferences, and physical condition \cite{zhang2018data}. For instance, if a user sends a location request that frequently includes a specific hospital location, the LSP can infer the user's physical condition, and the user's home address can be inferred based on the user's resident location at night.
\end{itemize}
\begin{table*}[htbp]
\scriptsize
 \centering
    \caption{AI methods for MEC security and privacy.}
    \label{tab:AI_methods}
    \begin{tabular}{|c|c|c|c|c|}
        \hline
        \rowcolor[gray]{0.8} \bf Classification & \bf Method & \bf Advantages & \bf Disadvantages & \bf Example Application Scenarios \\
        \hline
        \multirow{7}{*}{\makecell[c]{Supervised\\learning}} & kNN & \makecell[c]{Simple, cheap and efficient in \\ detection tasks} & Weak transferability & \makecell[c]{Intrusion detection \cite{liu2022enhanced, syarif2017intrusion, shapoorifard2017intrusion} \\ and privacy-preserving \cite{tan2020lightweight,chen2020privacy}} \\
        \cline{2-5}
        ~ & SVMs & \makecell[c]{High performance with small\\ samples, strong generalization \\ability} & \makecell[c]{Hard to acquire the optimal\\kernel function, sensitive \\ to missing data} & \makecell[c]{Authentication \cite{chen2020automated}, intrusion \cite{kumari2017semi}\\ and malware \cite{xu2016hadm} detection} \\
        \cline{2-5}
        ~ & LR & \makecell[c]{Cheap, capacity of description\\ the relationship between\\ input and output } & \makecell[c]{Easy to be underfitting,\\weak performance with \\missing data and large feature} & \makecell[c]{Intrusion detection \cite{majumder2020smart},\\PUF \cite{laguduva2019machine}} \\
        \cline{2-5}
        ~ & DTs & \makecell[c]{Interpretable inference,\\fast execution} & \makecell[c]{Sensitive to noise,\\ignoring relationship among\\ different attributions in dataset} & \makecell[c]{Intrusion detection \cite{sinclair1999application},\\privacy-preserving \cite{wu2020ensemble,sangaiah2019enforcing} }\\
        \cline{2-5}
        ~ & RFs & \makecell[c]{Parallel efficient framework,\\strong generalization ability,\\and robust to missing data} & \makecell[c]{Performing overfitting on\\ datasets with large noise} & ~ \makecell[c]{Intrusion \cite{liu2020deep,singh2019optimization,guowei2021research} and\\ anomaly \cite{hasan2019attack,wu2019detecting,zhang2020dataset} detection} \\
        %\cline{2-5}
        %~ & NB & \makecell[c]{Cheap, no need of overfull\\parameters, and framework \\without iteration } & \makecell[c]{Requiring proper prior model\\estimation, classification error\\causing by the posterior probability} & Traffic analysis \cite{cao2020packet} \\
        \cline{2-5}
        ~ & CNN & \makecell[c]{Low network complexity,\\reducing weight parameters,\\improving performance by BP\\algorithm and better scalability} & \makecell[c]{Large training cost,\\requiring excessive computation \\ resource and memory } & \makecell[c]{Intrusion detection \cite{tian2019distributed},\\attack recognition \cite{ran2020cloud}\\and privacy-preserving \cite{tian2019lep}} \\
        \cline{2-5}
        ~ & RNN & \makecell[c]{High performance through \\correlation extraction for \\sequential data} & \makecell[c]{Gradient explosion in training \\process with long-time \\sequence dataset} & Malware detection \cite{radhakrishnan2021deep}\\
        \hline
        \multirow{5}{*}{\makecell[c]{Unsupervised\\learning}} & K-Means & \makecell[c]{Fast execution, interpretability\\and performing well in clustering\\with unlabeled data} & \makecell[c]{Sensitive to abnormal data,\\dependence on the value of $K$} & \makecell[c]{Secure links decision \cite{li2019complex},\\privacy-preserving \cite{han2018kclp,liu2020privacy}} \\
        \cline{2-5}
        ~ & PCA & \makecell[c]{Reducing the data dimension,\\eliminating the effect between\\different components}  & \makecell[c]{Ignoring the influence of\\ non-principal components, requiring\\ to be combined with other methods} & \makecell[c]{Attack recognition \cite{huong2021efficient},\\privacy-preserving \cite{osia2018private}}\\
        \cline{2-5}
        ~ & RBM & \makecell[c]{Strong representation capacity,\\high inference performance with\\ unlabeled data} & \makecell[c]{Overfull computation cost,\\ challenges in deployment on-board} & Anomaly detection \cite{benchea2014combining} \\
        \cline{2-5}
        ~ & DBNs & \makecell[c]{Flexible and efficient parallel\\ framework, better scalability} & \makecell[c]{Overfull computation cost,\\slow convergence rate} & \makecell[c]{Intrusion detection \cite{zhao2017intrusion,tian2020intrusion},\\anomaly detection \cite{chen2017deep}} \\
        \cline{2-5}
        ~ & AE & \makecell[c]{Strong generalization capacity\\providing privacy-preserving through\\ the encoding-decoding process} & \makecell[c]{Requiring benign data for\\training in malicious detection} & \makecell[c]{Anomaly detection \cite{schneible2017anomaly},\\privacy-preserving\cite{sadaf2020intrusion}} \\
        \hline
        \multirow{2}{*}{\makecell[c]{Semi-supervised\\learning}} & $S^{3}VM$ & \makecell[c]{Improving the performance with \\scarce labeled data} & \makecell[c]{High computational complexity,\\requiring to resolve the program-\\ming when new data is added} & Anomaly detection \cite{wang2019sample} \\
        \cline{2-5}
        ~ & GANs & \makecell[c]{Generative model based on BP \\algorithm instead of Markov chain, \\generating clearer and faster \\samples} & \makecell[c]{Unstable training process,\\inefficient for discrete data} & \makecell[c]{Anomaly detection \cite{hiromoto2017secure},\\PUF \cite{yoon2020pufgan}} \\
        \hline
        \multirow{2}{*}{RL} & Non-deep RL & \makecell[c]{Framework for sequential decision-\\making problem} & \makecell[c]{Excessive training iterations, risk \\of falling into  suboptimal solution} & \makecell[c]{Malware detection\cite{xiao2017cloud},\\ anti-jamming \cite{aref2017multi}} \\
        \cline{2-5}
        ~ & DRL & \makecell[c]{Efficient inference, high perfor-\\mance in MEC offloading game} & \makecell[c]{Complex model setting and high\\computation cost} & \makecell[c]{Anti-jamming \cite{xiao2018security}\\
        threat edge game \cite{li2021explainable}}\\
        \hline
        \multirow{2}{*}{\makecell[c]{Non-ML based\\methods}} & %\makecell[c]{Symbolic\\Learning} & \makecell[c]{Cheap, easy to implement} & \makecell[c]{Lack of interpretability, few \\available models}  & Intrusion detection \cite{sinclair1999application} \\
        %\cline{2-5}
        \makecell[c]{Bayesian\\Networks} & \makecell[c]{Capacity of dealing with the\\unsure information}  & \makecell[c]{High computational complexity} & \makecell[c]{Intrusion detection \cite{thames2006hybrid},\\privacy-preserving \cite{wright2004privacy}} \\
        \cline{2-5}
        ~ & \makecell[c]{Evolutionary\\Algorithms} & Efficient, potential scalability & \makecell[c]{Complex manual programming,\\ unsuitable to large datasets} & Malware detection \cite{kim2010malware} \\
        \hline
    \end{tabular}
\end{table*}

\subsubsection{Identity Privacy}
\vspace{-1.7pt}
Personal Identifiable Information (PII), also known as user identity, is the information about a person that has been collected, assessed, or used on demand by edge or cloud services \cite{zhang2018data}. Compared with the cloud computing data center located in the core network, edge nodes are located at the edge of the network, enabling the collection of more high-value sensitive information of users, such as location information, lifestyle habits, social relationships, and even health status, etc. A considerable privacy crisis will occur while an attacker can map each private data to the corresponding user. For instance, in early 2018, the Facebook data scandal resulted in the disclosure of 50 million users' PII to a third-party company, Cambridge Analytica, via service providers for "analysis" purposes.

\section{Review of Artificial Intelligence Approaches on MEC Security and Privacy}\label{sect:AI perspective}
In this section, we make a comprehensive review of the MEC security and privacy from the perspective of AI. Firstly, we introduce the classification of ML approaches and summarize advantages, disadvantages as well as applications of the typical approaches. Then, considering the entire AI categories, we present some other applications of non-ML approaches. The summary of the AI-based methods and their corresponding advantages and disadvantages are introduced in Table \ref{tab:AI_methods}.
\subsection{Classification of Machine learning Methods in MEC Security and Privacy}
Discussions and researches around AI have focused on the field of ML in recent years \cite{roy2022analysis}. We divide ML approaches into four categories: supervised learning, unsupervised learning, semi-supervised learning and reinforcement learning. In this subsection, the characteristics of these four categories, classic algorithms and their applications in MEC are discussed in detail.

\subsubsection{Supervised Learning}
% The data collected by plentiful IoT sensors contains a variety of sensitive information that may be exploited by attackers, so precise detection of malicious intrusion in the MEC environment is crucial. In order to effectively protect the security and privacy of MEC, a high-precision and robust classification approach is needed to distinguish benign and malicious activities in the MEC environment.
Supervised learning has attracted more and more attention in the field of the security and privacy protection of MEC due to its outstanding performance in prediction and classification. Generally, utilizing the labeled data as the input of supervised learning, the model can predict the label of test data through learning from the distribution of labels. 
% The main functions of supervised learning are regression and classification, where the regression determines the relativity between the independent and dependent variables, and the classification assigns attribution classes to new instances. In intrusion detection system (IDS), the well-trained supervised learning algorithm can distinguish between normal and abnormal behaviors by learning from the extracted features of the raw data, and its detection accuracy of intrusions is continuously improved as the training progresses \cite{mebawondu2020network}.
We discuss the classic algorithms of supervised learning and their applications in the protection of security and privacy of MEC in the following.
\paragraph{k-Nearest Neighbor (kNN)}
% kNN is an instance-based learning approach without parameters \cite{muhammad2015supervised}. Under the premise of little or even no prior knowledge, kNN can perform classification tasks without probability density \cite{parvin2010modification}. 
% The core of kNN is to obtain the labels of the $k$-nearest neighbor samples of the input instance in the Euclidean distance, and then vote on these labels to determine the classification of the input instance. 
Based on the unsurpassed performance on classification tasks, kNN has been widely used in network malicious intrusion detection \cite{liu2022enhanced, syarif2017intrusion, shapoorifard2017intrusion}. 
% In the MEC environment, a study \cite{soleymani2020trust} proposed a model to detect the non-line of sight (NLoS) condition in VANET by the kNN algorithm based on the feature similarity and symmetry. 
%To protect the security of encrypted images stored in the cloud-assisted edge servers, a secure and verifiable multi-key image search approach was proposed by using the kNN algorithm to secure features in \cite{li2020secure}. 
As to the MEC privacy, computational-hungry algorithms are not suitable for the practical applications with massive data and computation load. Therefore, a lightweight edge-based kNN (EBkNN) model was proposed to protect users' security and privacy in \cite{tan2020lightweight}. 
%To provide guarantees on the efficient and security of the S-HashMap index structure in the intelligent edge network, the kNN algorithm was used in \cite{chen2020privacy} to calculate the similarity between the query vector and index nodes. 
However, the transferability of kNN is poor. A trained model often needs to be retrained to determine the optimal $k$ value when the dataset changes, which will increase the resource consumption of the light-weighted edge devices.
\paragraph{Support Vector Machines (SVMs)}
% SVMs are a classic classification algorithm based on supervised learning, which use hyperplanes to split sample nodes of different classes in the instance space. The support-vector networks were firstly proposed to solve the two-group classification problem in \cite{cortes1995support}. The training process of SVMs is essentially a quadratic optimization problem with bounded constraints and linear equality constraints \cite{joachims1998making}. As a goal of training, it is vital to estimate optimal hyperplanes so that the sample nodes of different classes in the feature space are separated by as large intervals as possible. In addition, the kernel function is a pivotal approach to linearize the nonlinear hyperplane.

Due to the advantages of SVMs in solving linear and non-linear data classification problems, they have been widely used in abnormal intrusion detection in the MEC environment. 
%In \cite{hou2019use}, a study utilized the SVMs method based on Radial Basis Function (RBF) to find the boundary between regular and mutation codes in the Alibaba edge computing system. 
The research \cite{chen2020automated} chose the SVM algorithm to detect cloning attacks and Sybil attacks in industrial edge networks. For the actual data without labels, this work adopted the threshold detection method to generate labels for training, so that the model obtained improved detection accuracy. Nevertheless, for the scenarios with complex data distribution, the optimal kernel function is hard to acquire in SVM, which may decrease the performance of Intrusion Detection System (IDS)\footnote{It is a system that analyzes network traffic for suspicious or unusual activity and generates notifications when it detects it.} \cite{sadaf2020intrusion}, and this disadvantage will be further amplified in the data-heavy MEC environment.

\paragraph{Logistic Regression (LR)}
% LR is a typical ML algorithm, which is mostly used in the binary classification problem. Under the assumption of independent and irrelevant alternatives (IIA), LR can obtain the odds ratio of different categories. LR utilizes the logit of each category (i.e., the natural logarithm of the odds ratio of each category) to fit the decision boundary between each category, then captures the relevance between the decision boundary and the classification probability \cite{sperandei2014understanding}. In LR algorithms, the knowledge of the data distribution is not required, which is usually difficult and costly to acquire in the MEC environment. Besides, LR algorithms have performed well in the description of the relation of the predicted outcome and the input \cite{peng2002introduction}.

% Physical unclonable function (PUF) plays an indispensable role in protecting the IoT devices security. It provides physical digital keys for devices to protect the network from malicious attacks. However, many existing IoT networks choose to sufficiently trust the protection capabilities of PUF, and assume that PUF is completely tamper-proof in the design. 
LR has been widely used in protecting the security and privacy of MEC.
In response to the vulnerability in Physical unclonable function (PUF), %an attack based on LR model and evolution strategies was proposed in \cite{ruhrmair2010modeling}, which simulated PUF well and had basically no difference in performance. The follow-up research
\cite{laguduva2019machine} proved that PUF can be cloned without prior knowledge of its structure. Based on ML, this work proposed a non-invasive attack method against the PUF of edge nodes and the corresponding defence strategy, which adopted LR, RF, artificial neural network (ANN) and merged algorithms, respectively. The attack effectively cloned the PUF in MEC and the countermeasure greatly improved the accuracy of recognizing authentic and cloned PUF. 
%The attack based on the LR model against the XOR arbiter PUF was proposed in \cite{ruhrmair2013puf}, which was an inspiration to designing PUF and protecting MEC security from a hardware perspective. In order to protect the wireless sensor network (WSN) security, authors \cite{majumder2020smart} proposed a LR-based intrusion detection model, which utilized the device power consumption ratio to predict DDoS attacks and MitM attacks. 
However, LR is with the drawback of falling into the dilemma of overfitting that limits the accuracy, and has poor performance with missing data and large feature in MEC environments.

\paragraph{Decision Trees (DTs)}
% DTs are a kind of predict model based on the tree structure, which establish the mapping relationship between object features and values. In DTs, the root node corresponds to the set of samples, the internal node denotes the value of the corresponding classification, and the leaf node represents the decision results. The advantage of DTs is to decompose the complex decision-making process into a combination of simple decisions, thereby providing an interpretable inference scheme \cite{safavian1991survey}. The DTs induction contain three steps: feature selecting, growing and pruning. The feature selecting step splits training samples and determines the feature for inferring. And information gain \cite{quinlan1986induction} is often used to select valuable feature. In the growing step, the root node is assigned to a class according to the split of training samples. The information gain of all features is calculated for a node, then the feature with the largest information gain is selected as the node feature, and child nodes are established according to the different values of the feature. This process continues until the information gain is less than the specified threshold or there are no features to choose. In the pruning step, some sub-trees of the tree structure generated in the growing step are deleted to avoid overfitting the training data \cite{loh2011classification}.

For the data with high-dimensional features in MEC, DTs can reduce the number of features in the internal nodes while ensuring a certain performance in inferring. Based on this, the DTs-based algorithm has been widely used in the network intrusion detection \cite{santhadevi2022eidima}. 
%In order to establish an automatic and efficient network traffic anomaly detection system, authors combined genetic algorithm (GA) and DTs to detect network traffic abnormalities based on the information such as specific IP port numbers in \cite{sinclair1999application}. 
In the edge-cloud computing (ECC) scenario, a private random DT framework based on differential privacy (DP) was used to analyze the impact of different applications on privacy, so as to better coordinate the overall system to ensure the data privacy \cite{wu2020ensemble}. 
%Besides, considering the LBS in the MEC environment, a study \cite{sangaiah2019enforcing} proposed a model of user location confidentiality protection by merging DTs and kNN. 
Whereas, DTs are sensitive to the noise data and ignore the relationship among different attributions in the dataset, and these disadvantages affect the performance of preserving security and privacy seriously. 
\paragraph{Random Forests (RFs)}
% RFs are a kind of classifier that contains multiple DTs. RFs are mainly based on the Bootstrap aggregating (Bagging) algorithm, which is a major ensemble learning (EL) algorithm \cite{breiman1996bagging}. In RFs, there is no correlation between different DTs. Each tree predicts according to its input samples, and finally the class with the most votes is selected as the overall output through voting \cite{ho1995random}. Different from DTs, RFs randomly sample the sample set (bootstrap) to construct the training set of each DT. And by voting to determine the classification results, the robustness of the RFs-based algorithm is improved availably \cite{ho1995random,breiman2001random}. Moreover, based on the structural characteristics of RFs, parallel algorithms can be designed to improve the efficiency of training.

RFs have been widely used in the IoT system for intrusion detection \cite{liu2020deep,singh2019optimization,guowei2021research} and anomaly detection \cite{hasan2019attack,wu2019detecting,zhang2020dataset}. Especially in the detection of DDoS attacks, a mass of previous work has analyzed the performance of RFs \cite{hasan2019attack,farukee2020ddos}. 
%Compared with other ML algorithms (such as LR, SVM, DT, etc.), RFs have been proven to have advantages in terms of accuracy, recall and other performance when features are limited \cite{hasan2019attack}. In intrusion detection, the main factors affecting the accuracy of RFs algorithm detection are the number and depth of decision trees \cite{cheng2018flow}, and the cascade mode combined with other ML approaches can achieve higher detection rate and training efficiency \cite{aung2017analysis}. The centralization mode of SDN leads to its vulnerability. Therefore, RFs have been extended to solve traffic analysis and intrusion detection in software-defined IoT networks \cite{li2018ai,kirutika2019controller}. In addition, 
A research \cite{alrashdi2019ad} proposed an automatic anomaly detection system based on RFs; the proposed system effectively detected anomalies in distributed edge devices. The disadvantage of RFs is that it is prone to overfitting on datasets with large noise.

\paragraph{Convolutional Neural Network (CNN)}
As one of the most classic DL models, CNN has been fully employed in protecting the MEC security and privacy \cite{wang2022lightlog}.
Using the CNN model, Tian et al. \cite{tian2019distributed} proposed a web server attack detection system based on a distributed framework. The Uniform Resource Locator (URL) in the IoT-cloud environment was analyzed and a high detection rate was obtained. 
%CNN was also adopted as a classifier to distinguish benign and malicious packets in IDS \cite{ho2021novel}, and the performance (accuracy) and cost (training overhead) were evaluated and balanced in this work. In order to detect DDoS attacks in IoT networks, a research \cite{roopak2019deep} proposed a DL-based intrusion detection model, and through experimental comparisons proved that the model combined with CNN and long short-term memory (LSTM) performs with the highest accuracy. In addition, Ran et al. \cite{ran2020cloud} proposed a CNN-based attack recognition model in the cloud-edge collaborative environment. The local training and dataset update process were performed on edge devices, and the centralized model was updated at a fixed period by the resources of cloud. In terms of privacy protection, researches \cite{tian2019lep,huang2019lightweight} have built the lightweight privacy-preserving model in MEC based on the feature extraction capabilities of CNN.
However, the problem of CNN is that the amount of calculation under the application of high-dimensional data is huge, and it is difficult to guarantee sufficient computing resources in some resource-constrained MEC environments. Some current neural network scale compression technologies and partition training technologies \cite{li2019edge,zhao2018deepthings} provide support for the deployment of lightweight CNN on mobile terminals.

%Another classic DL model is called Recurrent Neural Network (RNN), which is aimed at sequence type data. When there is a certain correlation between sequences in the data (such as natural language data \cite{lecun2015deep}), different sequences are with different lengths, so it is difficult to process and split the data \cite{cho2014learning}. To address this, RNN is usually adopted for correlation extraction. The key function of RNN is the memory module, which is responsible for storing the value of the intermediate state and using it as the subsequent input \cite{sze2017efficient}. In the MEC environment, the network status and security issues are related over time. For example, the continuous abnormality of network traffic within a period of time may mean that the network has suffered a DoS attack. Therefore, RNN with significant performance in sequence data has been introduced into the MEC environment.
\paragraph{Recurrent Neural Network (RNN)}
Another classic DL model called RNN, which is aimed at sequence type data, has also been fully used in the malware detection in the IoT network \cite{radhakrishnan2021deep,guizani2020network,arivukarasi2020performance,chen2020rnn}. Conventional RNN models use the back-propagation training time (BPTT) to extract sequence data, but it will cause gradient explosion in the training process with long-time sequence data. 
%Based on this, Kim et al. used LSTM to replace the conventional RNN model to implement an IoT intrusion detection model in \cite{kim2016long!}. 
Previous work \cite{diro2018leveraging} applied LSTM to the cloud-edge scenario to achieve distributed network attack detection. In the experiment, the performance of the proposed model was compared with LR, which proved that the utilization of LSTM can fully extract the correlation information from the long-term continuous network state data.
However, the RNN model generally has the disadvantage of gradient explosion, which will limit its performance in long-term sequence training.
%In the sophisticated machine tool field, anomaly detection is the key to machine health management. 
%Lin et al. \cite{lin2019edge} proposed an RNN-based anomaly detection model using edge computing, which employed sequence data such as temperature and current to train the RNN model in the cloud, and executed inference on the edge server. 
\subsubsection{Unsupervised Learning}
%Although supervised learning algorithms can provide high-quality classifiers for MEC security and privacy, they often need effective data (i.e., sufficient labeled datasets) to adequately train the model. Unfortunately, a massive data generated by heterogeneous IoT sensors are gathered in edge severs, which makes the preprocessing and labeling of raw data consume a lot of computing resources. Therefore, the method of directly using unlabeled data for predicting attacks has attracted attention. 
Unsupervised learning is a ML method for unlabeled datasets, which clusters different classes by learning their associations \cite{ghahramani2003unsupervised}. In addition, unsupervised learning can mine the structure information in depth, which may be hidden by labels in supervised learning \cite{karoly2018unsupervised}. Thus, the unsupervised learning algorithm has obvious advantages in the classification task in MEC security and privacy with a large number of unlabeled data. For example, the next-gen cloud security company, Bitglass, exploits unsupervised learning approaches to provide Advanced Threat Protection (ATP) services for detection both known and zero-day threats on cloud applications \cite{wason2020integrated}. The emblematic unsupervised learning methods in MEC security and privacy are listed as follows: 

\paragraph{K-Means}
% K-Means is a simple unsupervised learning approach, which clusters the sample dataset according to the feature similarity iteratively. In the K-Means algorithm, the first step is determine the number of clusters (i.e., the value of $K$). The algorithm estimates centers of the $K$ clusters, and each cluster represents a set of samples with the similar feature. It is worth noting that in order to obtain a better clustering effect, the distance between centroids needs to be established as large as possible \cite{wong1979algorithm}. Then each sample is classified into the cluster corresponding to the closest centroid based on the Euclidean distance. Until all samples are all divided into a specific cluster, the algorithm re-determines the new centroid by calculating the mean of the distances of all samples in the current cluster to improve the clustering. As the distance between other samples in the cluster and the cluster center changes during this process, some samples will be re-divided into other clusters accordingly. This iterative process will continue until no more new samples can replace the centroid of the current cluster \cite{kanungo2002efficient}.

Based on the excellent performance of K-Means in feature clustering on unlabeled data, it has been applied to the field attack detection and privacy protection in MEC. In order to solve the complex link attack problem caused by the open and multi-source characteristic in the MEC environment, Li et al. \cite{li2019complex} proposed a link decision scheme based on attribute attack graphs, which used K-Means to de-redundate complex network alarm information to construct the attribute attack graphs, and utilized the greedy algorithm to make decisions on the link defense by the solution of the minimum dominating set of the attribute attack graph. 
%A previous study \cite{han2018kclp} proposed a K-Means cluster-based location privacy protection model in WSN, using fake source nodes to disguise the real source nodes, and adopting cluster algorithms to generate fake data packets. Moreover, in order to make full use of network resources, proactive caching is usually utilized at the edge sever, and the K-Means algorithm can be used to estimate the popularity of content for caching. But the related information of users may be leaked in the caching process, and the traditional K-Means algorithm cannot cope with the loss of user information. To solve this, a privacy-preserving federated K-Means (PFK) model was proposed \cite{liu2020privacy}. And based on the idea of federated learning (FL) and secret sharing, the model was able to effectively protect the privacy and performed well in the case of user information loss.
However, K-Means is sensitive to the abnormal data, and the progress of K-Means clustering mainly depends on the value of $K$, which is usually set manually. And these disadvantages cause the weak robustness of K-Means-based methods.

\paragraph{Principal Component Analysis (PCA)}
% PCA is an approach of data dimension reduction based on statistic analysis. Different from other traditional ML classifiers, PCA is mainly adopted for the data dimension reduction while keeping original information as much as possible to enhance the interpretability with large datasets \cite{jolliffe2016principal}. What's more, PCA uses the orthogonal transformation to map mutually related features to linearly uncorrelated variables, which are called principal components. Therefore, PCA has been widely adopted in the field of dimension reduction.

Due to the effective feature extraction and dimension reduction capacity for large datasets, PCA fits the MEC environment well. The previous work \cite{huong2021efficient} proposed a multi-attack detection model with low complexity in the cloud-edge environment. The computationally efficient and low-cost PCA was selected as the feature extractor, and the deep neural network (DNN) was used as the classifier. In the experiment, the model was used to detect 10 types of attacks with higher accuracy rate than other ML methods. 
%In order to provide reliable analysis services for private user data, Osia et al. \cite{osia2018private} proposed a DL-based cloud-edge hybrid framework. The framework used the edge server for data preprocessing and provided reliable analysis for the cloud, and utilized the Siamese architecture to fine-tune the server analysis in the middle layer to ensure the data privacy. In the Siamese architecture, PCA was used to extract the dimension of features, which reduced the communication cost and improved the confidentiality through the process of dimension reduction-reconstruction. 
Nevertheless, PCA ignores the influence of non-principal components, which may play a crucial role in inferring. Discarding them directly will decrease the classification accuracy. And PCA is effective for feature extraction, while requires to be combined with other AI-based methods to fully finish the inference task.

\paragraph{Restricted Boltzmann Machine (RBM)}
RBM is a deep generative neural network based on unsupervised learning, which is a variant of the Boltzmann machine, and has been adopted in MEC security protections.
%An RBM model consists of a visible layer and a hidden layer. There is no connection between neural units in a single layer, while the visible layer and the hidden layer are completely connected in the form of wise-pair \cite{marlin2010inductive}. In addition, RBM can fit the distribution of the input, and is usually combined with other ML algorithms to solve classification tasks. However, the combined model aggravates the parameter tuning burden in the training process \cite{larochelle2012learning}.
%Containers are a new type of uninterrupted transportation method in cloud native technology. 
%In order to identify threats and zero-day vulnerabilities in containers, a previous study \cite{kamthania2019novel} proposed a DL model based on RBM to protect the security of applications and workloads in the container environment. 
In terms of malicious detection classifiers, the adopted features are the key factor affecting its accuracy. Benchea et al. \cite{benchea2014combining} used RBM to improve the feature generation, and generated new features in a non-linear manner to train the classifier and improve its performance.
%In addition, for the network anomaly detection system, it needs to be supervised in an adaptive manner due to the complex network status and the changeable attack mode. According to this, a previous research \cite{fiore2013network} proposed a intelligent anomaly detection model based RBM, which inferred the abnormal data by training the model with unlabeled raw data. 
However, the drawback of RBM is also obvious. The training of RBM takes a mass of computation cost, and it is still a challenge to deploy the model on-board in the resource-limited edge sever.

\paragraph{Deep Belief Networks (DBNs)}
% DBNs are a kind of deep model stacked by multiple RBMs. The first two layers of DBNs (i.e., the last RBM in the stack) are undirected, while the remaining layers in the stack are connected in a top-to-down direction \cite{hinton2012deep}. Based on the greedy strategy, DBNs adopt a layer-by-layer training method to train the RBM of each layer from bottom to top, and use the hidden layer of the current layer as the input of the next layer \cite{hinton2006fast}. The layer-by-layer training process can be repeated many times to fully improve the performance of the model. Moreover, the training efficiency can be further improved by combining the training mode of unsupervised learning and the supervised fine-tuning strategy \cite{salakhutdinov2015learning}. 

DBNs are widely employed in the intrusion detection in IoT networks due to the efficient performance with a mass of unlabeled data \cite{hou2016droiddelver,xu2016hadm,zhao2017intrusion,tian2020intrusion}. 
%Chen et al. \cite{chen2017deep} proposed a MEC anomaly detection system based on unsupervised DL, which adopted DBNs to extract features such as location information to detect malicious applications in the network. 
In order to realize the attack detection in the edge transmission network, an unsupervised-based model based on DBNs was proposed in \cite{chen2019deep}. In this work, the Android data package files in edge devices were utilized to extract the permissions information, sensitive program APIs and dynamic information as features, which were adopted by DBNs for attack detection. 
%In addition, Singh et al. \cite{singh2020daas} proposed a Dew Computing as a Service (DaaS) intrusion detection framework for MEC, which used the DBNs model with revamped RBM layers to improve detection accuracy.
Unfortunately, the disadvantages of DBNs are manifested in their high computational cost and slow convergence rate, which limit the performance of DBNs in MEC security and privacy-preserving.
\paragraph{Deep Autoencoder (AE)}
% Deep AE is a neural network model for unlabeled data, which generates the output by encoding the input. AE consists of two major modules: encoding and decoding. In the encoding process, the input sample $x$ is mapped to the feature space $Z$ through the encoder $f$. And in the decoding process, the feature space $Z$ is mapped back to the original sample space through the decoder $g$ to obtain the reconstructed sample $\tilde{x}$ \cite{kingma2019introduction}. It is easy for the Deep AE model to occur overfitting in training for the reason that it tries to learn the unique representation of each sample. One solution is to add random noise to the input layer to enhance the robustness of the model \cite{vincent2010stacked}. Furthermore, the Jacobian matrix paradigm of the encoder can be added to the objective function to enhance its anti-interference performance \cite{rifai2011higher}.

Deep AE is widely used in IDS for the MEC environment due to its feature extraction capabilities \cite{sadaf2020intrusion}, and the process of feature extraction-reconstruction also provides privacy protection for user information in the original data. The previous work \cite{schneible2017anomaly} utilized Deep AE to realize anomaly detection based on the characteristics of MEC. The model was trained in a distributed manner on multiple edge severs. The abnormal data from each edge sever was aggregated to update the model on the central server, and then the updated model was sent to each edge sever to reduce the load on the central server.
While in terms of malicious detection, it requires a large amount of benign data for training an AE model, which is unrealistic in practice.
%In the edge environment of the industrial IoT, Kim et al. \cite{kim2018squeezed} proposed a Squeezed Convolutional Variational AutoEncoder (SCVAE) model to detect the abnormal sequence data. The model was a variant of CNN based on variational AE. The squeezed model reduced the scale and training cost of the model while ensuring the prediction accuracy. In the actual MEC environment, there may be a lack of prior knowledge of certain attacks. A study \cite{tzagkarakis2019botnet} proposed a Deep AE model, which employed a small amount of benign data for training, and performed botnet attack detection based on the error threshold in the reconstruction. In addition, anomaly detection systems \cite{lee2020impact,ghosh2019deep} that use Deep AE to perform feature extraction on edge devices and deploy other ML methods on the central server to directly utilize the extracted features or decoded original features to make predictions have also become a trend. 

\subsubsection{Semi-supervised ML Learning}
% In the MEC environment, a massive amount of raw data is generated at any time and any place. All raw data will become useless if there are no efficient tools to extract and analyze valuable information. In order to make good use of these raw data, a common and standard approach is to use human input or experimentation to point a certain label, tag, or numerical value based on a specific criterion. However, the acquisition of high-quality labeled data is often costly and time-consuming \cite{van2020survey}.

Semi-supervised learning is a technology that falls between supervised learning and unsupervised learning \cite{zhu2005semi}. It can solve the inability to construct a reliable supervised classifier caused by scarce labeled data.
% With massive unlabeled data, semi-supervised learning can allow scarce labeled data further to improve the accuracy and generalization ability of the model. 
Thus, the semi-supervised learning method is well-suited to the MEC scenario, in which a large amount of raw data is generated at any time and in any location but is incapable of being used efficiently. This inherent advantage makes it easier for Semi-supervised learning algorithms to address security and privacy concerns in MEC.

\paragraph{Semi-supervised Support Vector Machine ($S^{3}VM$)}
% $S^{3}VM$ is a semi-supervised learning algorithm that uses sufficient unlabeled data to solve the degradation of performance and generalization of SVM caused by scarce labeled data \cite{bennett1999semi}.

% $S^{3}VM$ is also a kind of large margin classifiers. However, global-optimal and convex supervised SVM turns to non-smooth and non-convex semi-supervised $S^{3}VM$, the computational complexity is also extremely improved. At the same time, the quadratic programming attribute of $S^{3}VM$ makes it necessary to resolve the programming problem as long as new data is added \cite{ding2017overview}.

The discovery of the distributed $S^{3}VM$ architecture \cite{wang2019sample} enables it to solve the security problems in the MEC environment which owns tremendous unlabeled raw data and a small amount of labeled data. Wang et al. \cite{wang2019sample} proposed a hybrid approach including $S^{3}VM$-based appliance pattern matching classifier and the hidden Markov model (HMM)-based energy consumption habit classifier to defend against anomaly intrusion.

\paragraph{Other Non-DL Semi-supervised Approaches}
MEC has a number of characteristics, including real-time processing, low computation costs, and a distributed architecture. Some of the above characteristics are already present in security and privacy-preserving methods based on semi-supervised learning. In \cite{rathore2018semi}, an ELM-based Semi-supervised Fuzzy C-Means (ESFCM) attack detection framework was proposed to secure the IoT system. At the same time, an optimized, low-cost semi-supervised IDS model which combines an Active learning Support Vector Machine (ASVM) and Fuzzy C-Means (FCM) clustering was proposed in \cite{kumari2017semi}. To defend Sybil attacks that appeared in distributed environments, recently, Gong et al. \cite{gong2014sybilbelief} designed a Markov random fields-based, noisy-tolerated and scalable semi-supervised learning approach. The above security defense mechanisms are more or less satisfying the conditions for deployment in the MEC environment. Despite the fact that researchers dislike the current semi-supervised learning method, it cannot be ruled out that it has the potential to become a new mainstream solution to the security and privacy issues in the MEC environment.

\paragraph{Generative Adversarial Networks (GANs)}

GANs have been recently implemented in MEC security and privacy. In \cite{hiromoto2017secure}, a supply chain risk management architecture that mixes machine learning, cryptographic hardware monitoring, and distributed system coordination techniques was proposed to detect normal and abnormal system behaviors in IoT systems. GANs can learn the distribution of attack samples from existing attacks in MEC, thereby generating zero-attack samples that do not exist in the set of known attack samples. Even for degenerate distributions, GANs can generate sharp samples without Markov chains requirement. As a DL-based semi-supervised approach, GANs are suitable for training classifiers with limited ground-truth datasets. Hence, GAN is a promising method for developing security and privacy applications in the MEC environment.
% Dirgantoro et al. \cite{dirgantoro2020generative} utilized the new samples generated by the GANs to overcome the loss of model accuracy caused by a small amount of labeled data.
The latest paper about compressed CGANs models \cite{li2020gan} made it feasible to deploy GANs in the MEC environment.

However, the disadvantages of GANs are also obvious, such as "the Helvetica scenario" and the instability and difficulty of the training process. We must avoid these drawbacks in order to maximize the benefits of GANs and use them to protect the security and privacy of MEC.

\paragraph{Other DL-based Semi-supervised Approaches}

DNN can fit various nonlinear functions appropriately, so it has outstanding contributions in the fields of pattern recognition, data analysis, and control. simultaneously, semi-supervised learning can be well adapted to a small number of labeled data together with a massive number of unlabeled data. Semi-supervised learning combined with the DNN method began to be widely concerned by researchers in order to make good use of the advantage of these two approaches. Although semi-supervised DL sounds like an effective solution to deal with a large amount of unlabeled data in the MEC environment, the detection accuracy is hard to exceed supervised DL. Therefore, only a small amount of work now uses semi-supervised DL in terms of MEC security and privacy. For example, authors in \cite{abdel2021semi} proposed a semi-supervised DL approach for intrusion detection in IoT networks, which combines multi-scale residual temporal convolutional (MS-Res) module to finetune the network capability and traffic attention mechanism to help the model to concentrate on important information in the learning process.

\subsubsection{Reinforcement Learning (RL)}

Researchers have extensively studied popular topics in communication fields such as data caching and offloading in the MEC environment. In recent years, a large number of papers are using DRL algorithms to solve these problems \cite{qian2020reinforcement, wang2020multi,li2021explainable, chen2022physical}. Previous studies (e.g., Integer programming and game-theory methods) only consider one-shot optimization \cite{qiu2020distributed}. However, the process of caching or offloading tasks in the MEC environment is a continuous process. Fortunately, RL provides a promising approach to maximize long-term rewards. At the same time, security and privacy issues have become the bottleneck of mobile caching/offloading, as edge server is always deployed close to users, making attackers also easier to reach the vulnerable position of MEC. Therefore, Xiao et al. \cite{xiao2018security} formally defined a repeated game between MEC systems and attackers and build the RL-based security solutions to defend against jamming and smart attacks in mobile offloading/caching.

However, the RL agent sampling data from the environment is a very inefficient process, and designing an appropriate reward function for unknown environments is challenging. Even if we successfully overcome the aforementioned difficulties, the local optimal is still hard to escape \cite{rlblogpost}.

\subsection{The Other Artificial Intelligence Approaches}
In addition to the previously introduced ML methods, some other non-ML methods (in AI categories, but not in ML categories) have also been applied in MEC security and privacy-preserving.
% Although researches on ML, especially on DL, have been deepened in recent years, AI is a larger conceptual category relatived to ML that intelligentizes a machine so that it can think and perform tasks in a human way. In the MEC environment, a large amount of generated data drives AI to be deployed to the edge to unleash the potential of edge servers \cite{zhou2019edge}. 

%including symbolic learning, Bayesian networks and evolutionary algorithms.
%\subsubsection{Symbolic Learning in MEC Security and Privacy}
%Symbolic learning is a basic class of AI. It generally refers to the intelligent representation of all advanced symbols \cite{haugeland1989artificial}. 
%The output is determined according to the input represented by the symbol, which is generated by the production rules and conceptual hierarchies \cite{chen1995machine}. 
% One of the classic models of symbolic learning is ID3 \cite{quinlan2014c4}, which is based on the example learning mode. The ID3 algorithm gives a general conceptual description of each sample in the structured dataset, and classifies the samples according to their attributes. 
%The previous work \cite{sinclair1999application} utilized one classic model of symbolic learning named ID3 to construct an efficient DT model for the structured data, which partitioned the data into normal and intrusion. However, symbolic learning is lack of interpretability and available models, which limits its applications in the MEC environment.
\subsubsection{Bayesian Networks in MEC Security and Privacy}
Bayesian networks, also named belief networks, are a kind of directed acyclic AI graph model, which are widely used in classification tasks \cite{pourret2008bayesian}. 
%In Bayesian networks, nodes are divided into parent nodes (causes) and child nodes (effects) to represent random variables. The conditional probability value from the parent node to the child node represents the causal relationship between them \cite{sun2006bayesian}.
%The Bayesian network is suitable for scenarios that predict the relative probability between network status information and security conditions in MEC. 
%Utilizing the Bayesian network, a study \cite{thames2006hybrid} proposed a hybrid intelligent IDS combined with the self-organizing mapping, which used the data collected in the network to classify the security status. The hybrid system was with improved accuracy over Bayesian methods. 
To intuitively reflect the relationship between various vulnerabilities, the attack graph in the network can be modelled in IoT networks. Based on the attack graph, some previous works \cite{frigault2008measuring,frigault2008measuring2} proposed to adopt Bayesian network model to find the correlation between vulnerabilities and network status, thereby providing security measurement for the network. 
%In addition, aiming at the Bayesian network for learning distributed heterogeneous data, the authors \cite{wright2004privacy} proposed an effective privacy protection K2 algorithm to construct the network structure for the data of multiple parties. 
However, the structure characteristic brings Bayesian networks with high computational complexity, which increases the burden of resource-limited edge severs.
\subsubsection{Evolutionary Algorithms in MEC Security and Privacy}
Evolutionary algorithms are also an important subset of AI, including genetic algorithm (GA), evolution strategy, neuro-evolution and so on. 
% Inspired by the process of biological evolution, evolutionary algorithms can fit optimization problems well. As a key category of evolutionary algorithms, GA is usually utilized to search for optimal solutions of problems. The current solution of GA is continuously improved after generations of mutations and crossover transformations, and the optimal solution is finally reached when the program converges \cite{chen1995machine}.
%In order to protect the security of the information system, Gupta et al. \cite{gupta2006matching} proposed a security file configuration scheme based on GA, which selected the lowest cost security file to provide the network with the greatest degree of vulnerability coverage. 
In the MEC network, the dependency graph that is used to characterize the dependencies between objects can be employed to represent the relationship between different malware. Based on this, Kim et al. \cite{kim2010malware} proposed a malware detection model by GA, which turned the problem into searching the largest sub-graph isomorphism problem in the dependency graph. Meanwhile, the complexity of the proposed model was greatly decreased by reducing the size of the dependency graph. Nevertheless, the disadvantages of evolutionary algorithms are also obvious. The manual programming is complex which makes the evolutionary algorithms are not suitable in the large dataset.

\begin{table*}[htbp]
    \setlength\tabcolsep{4pt}
    \scriptsize
    \centering
    \caption{Summary of AI-based works at each layer of MEC for security.}
    \label{tab:layer_security}
    \begin{tabular}{|c|c|c|c|c|c|c|c|c|c|}
        \hline
         \multirow{2}{*}{\bf Layers} & \multirow{2}{*}{\bf Works} & \multirow{2}{*}{\bf Attack methods} & \multicolumn{5}{c|}{\bf AI-based methods} & \multirow{2}{*}{\bf {Results}} &  \multirow{2}{*}{\bf \makecell[c]{Applications\\ or \\ scenarios}} \\
        \cline{4-8}
         ~ & ~ & ~ & \cellcolor[gray]{0.8} \makecell[c]{Supervised\\learning} & \cellcolor[gray]{0.8} \makecell[c]{Unsupervised\\learning} & \cellcolor[gray]{0.8} \makecell[c]{Semi-supervised\\learning} &\cellcolor[gray]{0.8}  RL &\cellcolor[gray]{0.8}  \makecell[c]{Non-ML\\methods} & ~ & ~\\
        \hline
        \multirow{11}{*}{\makecell[c]{IoT\\Systems}} & %\cite{wang2017physical} & Spoofing attacks & ELM & ~ & ~ & ~ & ~ & Authentication \\
        %\cline{2-9}
         \cite{chen2019clustering} & \makecell[c]{Spoofing attacks,\\replay attacks} & ~ & Clustering & ~ & ~ & ~ & \makecell[c]{Authentication rate: \\ 100\%}& Authentication \\
        \cline{2-10}
        %~ & \cite{chen2020automated} & \makecell[c]{Hijacking attacks,\\ Sybil attacks} & SVM & ~ & ~ & ~ & ~ & Authentication \\
        %\cline{2-9} 
        ~ & \cite{liao2019multiuser} & Spoofing attacks & DNN & ~ & ~ & ~ & ~ & \makecell[c]{Authentication rate: \\ 100\%} & \makecell[c]{Authentication,\\ PUF} \\
        %\cline{2-9}
        %~ & \cite{chen2019radio} & Spoofing attacks & DTs & ~ & ~ & ~ & ~ & Authentication \\ 
        %\cline{2-9}
        %~ & \cite{wang2017current} & Physical attacks & ELM & ~ & ~ & ~ & ~ & PUF \\ 
        %\cline{2-9}
        %~ & \cite{gwon2013competing} & DoS attacks & ~ & ~ & ~ & Q-learning & ~ & Anti-jamming\\
        \cline{2-10}
        ~ & \cite{aref2017multi} & DoS attacks & ~ & ~ & ~ & RL & ~ & \makecell[c]{Normalized accumulated\\ reward: 2.25 msec} & Anti-jamming \\
        %\cline{2-9}
        %~ & \cite{han2017two} & DoS attacks & DNN & ~ & ~ & DRL & ~ & Anti-jamming\\
        %\cline{2-9}
         %~ & \cite{wang2019sample} & Physical attacks & ~ & ~ & $S^{3}VM$ & ~ & ~ & Anomaly detection\\
        \cline{2-10}
        ~ & 
        %\cite{middlemiss2003feature} & \makecell[c]{DoS attacks,\\ manipulate and misuse}& kNN & ~ & ~ & ~ & GA & IDS \\
        %\cline{2-9}
        %~ &
        \cite{manimurugan2021iot} & \makecell[c]{DoS attacks,\\TCP/IP attacks} & NB & PCA & ~ & ~ & ~ & \makecell[c]{Accuracy: 92.48\%\\ Detection rate: 95.35\%} & IDS\\
        \cline{2-10}
        %~ & \cite{farnaaz2016random} & \makecell[c]{DoS attacks,\\manipulate and misuse} & RF & ~ & ~ & ~ & ~ & IDS \\
        %\cline{2-9}
        %~ & \cite{tian2020intrusion} & \makecell[c]{DoS attacks,\\manipulate and misuse} & ~ & DBN & ~ & ~ & ~ & IDS \\
        %\cline{2-9}
        %~ & \cite{ho2021novel} & \makecell[c]{DoS/DDoS attacks,\\web attacks,\\manipulate and misuse} & CNN & ~ & ~ & ~ & ~ & IDS \\
        %\cline{2-9}
        ~ & \cite{majumder2020smart} & \makecell[c]{DDoS attacks,\\MitM attacks} & LR & ~ & ~ & ~ & ~ & Average accuracy: 74\% & IDS \\
        \cline{2-10}
        %~ & \cite{roopak2019deep} & DDoS attacks & \makecell[c]{CNN,\\LSTM} & ~ & ~ & ~ & ~ & IDS \\
        %\cline{2-9}
        ~ & \cite{sinclair1999application} & DDoS attacks & DT & ~ & ~ & ~ & GA & None quantitative result & IDS \\
        \cline{2-10}
        %~ & \cite{kumari2017semi} &\makecell[c]{DoS attacks,\\manipulate and misuse} & SVM & ~ & FCM & ~ & ~ & IDS \\
        %\cline{2-9}
        %~ & \cite{fiore2013network} & \makecell[c]{DoS attacks,\\manipulate and misuse} & ~ & RBM & ~ & ~ & ~ &  Traffic analysis\\
        %\cline{2-9}
        %~ & \cite{cao2020packet} & \makecell[c]{DoS attacks,\\TCP/IP attacks,\\manipulate and misuse} & NB & ~ & ~ & ~ & ~ & Traffic analysis\\
        %\cline{2-9}
        %~ & \cite{tzagkarakis2019botnet} & DDoS attacks & ~ & AE & ~ & ~ & ~ & Traffic analysis\\
        %\cline{2-9}
        ~ & \cite{kim2010malware} & Malware & ~ & ~ & ~ & ~ & GA & Detection rate: 88.89\% & Malware detection\\
        \cline{2-10}
        %~ & \cite{benchea2014combining} & Malware & ~ & RBM & ~ & ~ & ~ & Malware detection \\
        %\cline{2-9}
        ~ & \cite{xiao2017cloud} & Malware & ~ & ~ & ~ & Q-learning  & ~ & \makecell[c]{Improved detection\\accuracy: 40\%}  & Malware detection \\
        \cline{2-10}
        %~ & \cite{xiao2016mobile} & \makecell[c]{DoS/DDoS attacks,\\spoofing attacks} & ~ & ~ & ~ & Q-learning & ~ & Edge offloading \\
        %\cline{2-9}
        ~ & \cite{xiao2018security} & \makecell[c]{DoS/DDoS attacks,\\spoofing attacks} & ~ & ~ & ~ & RL & ~ & None quantitative result & Edge caching\\
        \cline{2-10}
        ~ & \cite{hou2016droiddelver} & Malware & ~ & DBN & ~ & ~ & ~ & Accuracy: 96.66\% & \makecell[c]{Android malware\\detection} \\
        \cline{2-10}
        ~ & \cite{xu2016hadm} & Malware & SVM & DBN & ~ & ~ & ~ & Accuracy: 94.7\% & \makecell[c]{Android malware\\detection} \\
        %\cline{2-9}
        %~ & \cite{chen2017deep} & Malware & ~ & DBN & ~ & ~ & ~ & \makecell[c]{Android malware\\detection} \\
        \hline
        \multirow{8}{*}{\makecell[c]{MEC \\System\\Level}} & \cite{kim2018squeezed} & Physical attacks & CNN & AE & ~ & ~ & ~ & PR-AUC: 99.20\% & Anomaly detection \\
        \cline{2-10}
        ~ & \cite{sadaf2020intrusion} &\makecell[c]{DoS attacks,\\manipulate and misuse}  & ~ & AE & ~ & ~ & ~ & Accuracy: 95.4\% & IDS \\
        \cline{2-10}
        ~ & \cite{chen2019deep} & \makecell[c]{DoS/DDoS attacks,\\ eavesdropping and hijacking} & ~ & DBN & ~ & ~ & ~ & \makecell[c]{Improved detection\\accuracy: 6\%} & IDS \\
        \cline{2-10}
        ~ & \cite{li2019complex} & \makecell[c]{DoS/DDoS attacks,\\privilege escalation,\\eavesdropping and hijacking} & ~ & K-Means  & ~ & ~ & ~ & \makecell[c]{Overall redundant alarm\\compression rate: 97.2\%} & IDS \\
        \cline{2-10}
        ~ & \cite{rathore2018semi} & \makecell[c]{DoS attacks,\\TCP/IP attacks,\\manipulate and misuse} & ~ & ~ & ESFCM & ~ & ~ & Accuracy: 86.53\% & IDS \\
        \cline{2-10}
        ~ & \cite{li2018ai} & \makecell[c]{DoS attacks,\\manipulate and misuse} & RF & K-Means & ~ & ~ & ~ & \makecell[c]{Accuracy: 96.03\%\\FPR: 1.18\%} & IDS for SDN \\
        \cline{2-10}
        ~ &\cite{tang2016deep} & \makecell[c]{DoS attacks,\\manipulate and misuse} & DNN & ~ & ~ & ~ & ~ & Accuracy: 75.75\% & IDS for SDN \\
        \cline{2-10}
        ~ & \cite{kirutika2019controller} & \makecell[c]{DoS attacks,\\manipulate and misuse,\\TCP/IP attacks} & RF & ~ & ~ & ~ & ~ & \makecell[c]{Precision: 96\%\\Recall: 53.2\%\\Accuracy: 97\%} & IDS for SDN \\
        \hline
        \multirow{8}{*}{\makecell[c]{MEC \\Host\\Level}} & \cite{khalid2020macler} & \makecell[c]{ Eavesdropping and
        hijacking} & MLP & ~ & ~ & ~ & ~ & Accuracy: 96.256\% & \makecell[c]{Hardware Trojan\\ Detection} \\
        \cline{2-10}
        ~ & \cite{yoon2020pufgan} & Manipulate and misuse & ~ & ~ & GAN & ~ & ~ & \makecell[c]{Precision: 95\%\\Recall:20\%} & PUF\\
        \cline{2-10}
        ~ & \cite{diro2018leveraging} & \makecell[c]{DoS attacks,malware,\\physical attacks,\\eavesdropping and hijacking} & LSTM & ~ & ~ & ~ & ~ & Accuracy: 99.91\% & IDS \\
        \cline{2-10}
        ~ & \cite{pan2020justinian} & Manipulate and misuse & ~ & ~ & ~ & RL & ~ & \makecell[c]{Robustness accuracy:\\80\%} & Secure FL \\
        \cline{2-10}
        ~ & \cite{kozik2018scalable} & \makecell[c]{DoS/DDoS attacks,\\TCP/IP attacks} & ELM & ~ & ~ & ~ & ~ & \makecell[c]{Precision: 98\%\\Error rate: 1\%} & IDS \\
        \cline{2-10}
        ~ & \cite{ran2020cloud} & \makecell[c]{DoS attacks,\\manipulate and misuse} & CNN & ~ & ~ & ~ & ~ & Accuracy: 95.8\% & Attacks recognition \\
        \cline{2-10}
        ~ & \cite{huong2021efficient} & \makecell[c]{DoS/DDoS attacks,\\eavesdropping and hijacking} & DNN & PCA & ~ & ~ & ~ & \makecell[c]{Accuracy: 99.9\%\\Detection rate: 100\%} & Attacks recognition \\
        \cline{2-10}
        ~ & \cite{guizani2020network} & Malware & RNN & ~ & ~ & ~ & ~ & Accuracy: 93\% & \makecell[c]{NFV malware\\detection}\\
        \hline
    \end{tabular}
\end{table*}

\section{AI Approaches for Layer Based MEC Security}\label{sect:Layers based security}
In this section, we provide a hierarchical introduction to technologies involving AI approaches for protecting the MEC security. 
Further, we provide a layer-based security solutions in Table \ref{tab:layer_security}. 
%according to the divided five-layer MEC architecture depicted in Section \ref{sect:overview},

\subsection{IoT Systems Security}
\subsubsection{Perception Layer Security}
%The perception layer provides a direct connection between MEC and the real world, which is composed of a massive of heterogeneous sensor devices, responsible for interacting with the environment and collecting raw data. 
%Since the perception layer is located at the border of the overall system, the sensors are directly exposed to aggressive attackers. Consequently, the efficient authentication and protection at the IoT device level is an important issue. In addition, the perception layer is the first threshold for raw data to enter the MEC system. 
Some attackers try to manipulate the device authentication or jam the radio frequency environment in the data collection progress to launch DoS attacks. Lightweight cryptography is widely used in IoT-enabled devices for authentication \cite{alassaf2019enhancing, alassaf2019simulating, kheshaifaty2021engineering}. However, time-varying features in the network limit the performance of such traditional authentication algorithms in large-scale IoT systems. To alleviate these burdens, some researches based on AI methods have been proposed to address these security threats, which are introduced in the following.
\paragraph{IoT Devices Level}
Physical layer authentication (PLA) ensures that the connected device is not malicious. In PLA, physical layer features such as received signal strength (RSS), channel impulse response (CIR), and channel state information (CSI) are adopted to detect spoof attacks \cite{wang2017physical,chen2020automated}. 
%Nevertheless, features of the physical layer are time-varying in the actual dynamic network environment, so the performance of traditional methods is usually limited. Therefore, AI-based methods with extraordinary adaptive capacities are promoted to the field of PLA.
%More specially, Wang et al. \cite{wang2017physical} proposed a PLA scheme based on extreme learning machine (ELM) using the feedforward neural network, whose model parameters can be calculated based on the least squares method instead of complex iterative training. In this work, the dataset from spoofing attacks was used for training to improve the accuracy of the spoofing attacks detection.
%and the experiment proved that the proposed model enhanced the detection accuracy compared with other PLA approaches.
Aiming at the MEC environment, Chen et al. \cite{chen2019clustering} proposed a PLA framework that combined clustering and traditional lightweight symmetric cryptography.
%which striked a trade-off between complexity and authentication rate. 
Based on the advantages of unsupervised clustering approaches in the data without prior knowledge, this framework extracted and clustered CSI features, and then used the symmetric cryptography to achieve authentication of IoT devices. The proposed model significantly improved the authentication rate of the system with a low-complexity structure, and effectively combated spoofing attacks, replay attacks and small integer attacks. 
%Moreover, a study \cite{chen2019radio} proposed a two-layer lightweight radio frequency fingerprint identification (RFFID) model based on ML. The edge layer collected CSI and other features in the network to generate radio frequency fingerprints and authentication decisions, while the cloud layer was responsible for feature extraction and generating DTs model. The proposed model improved efficiency and certification rate compared with other RFFID methods through the edge-cloud collaborative framework.
%Another previous research \cite{chen2020automated} constructed a PLA model against cloning attacks and Sybil attacks in industrial edge networks. In order to save the labor cost, authors utilized the threshold detection method based on channel differences to generate labels for training samples automatically, and used SVM to learn from the labeled data. By setting a reasonable threshold, the model ensured a higher detection accuracy, which provided ideas for the application of supervised ML methods in PLA. 

However, the traditional ML-based methods require to be trained with a mass of samples, which occupies plentiful computationally resources and increases the time consumption. In response to this, Liao et al. \cite{liao2019multiuser} employed the data augmentation algorithm to accelerate the authentication process, and combined with DNN to enhance the accuracy of PLA and the model training efficiency. In the experiment, the author illustrated the proposed model attained a lightweight authentication with higher accuracy compared with the traditional threshold-based PLA method.

%PUF in IoT devices is a security primitive for authentication. Wang et al. \cite{wang2017current} used ELM to implement the PUF with the current mirror array (CMA) circuit. The ELM that based on the random neural network made the generated PUF more powerful and had better adaptability to different challenge scenarios. By designing the actual chip, it was proved that the PUF generated by the proposed model can adapt to different authentication protocols, and reduce the natural bit error rate (BER) while reducing power consumption.

\paragraph{Raw Data Collection Level}
The raw data collected by sensors in MEC is usually transmitted through the cognitive radio network (CRN). Some attackers transmit false signals in the CRN to expend network bandwidth, which causes the received IoT devices to consume a mass of public resources.
%(e.g., computing resources, power consumption, memory, etc.) 
%that reduces the performance of normal tasks. 
%To combat the jamming attack in CRN, traditional methods usually use frequency hopping (FH) technology to expand the spectrum. However, existing smart radio equipment can block the protection of FH \cite{han2017two}. Therefore, intelligent approaches are required against smart attackers.
%Competing mobile network game (CMNG) is a decision-making problem for the party that dominates the frequency spectrum in CRN.
Taking the advantage of RL in game-based problems, 
%Gwon et al. \cite{gwon2013competing} proposed an anti-jamming model based on Q-learning. Jamming games and anti-jamming were integrated to jointly generate the optimal channel access strategy. 
%In this work, the state, action and reward were elaborately designed, and the trade-off between exploration and exploitation of the model was sought. 
%In the experiment, three Q-learning algorithms were compared, and the results proved that the Friend-or-foe Q-learning algorithm can generate better anti-jamming strategies in the distributed ad-hoc network.
\cite{aref2017multi} proposed an anti-jamming model based on multi-agent RL. Multiple CRNs were allowed to transmit signals on the same frequency band, so each CRN was required to discriminate jamming signals and other normal CRN signals. Q-learning was used to generate the jamming-free sub-band frequency strategy, which was with lower complexity and achieved better performance in anti-jamming. 
%Combined with the traditional FH technology, Han et al. \cite{han2017two} proposed an anti-jamming intelligent model in CRN. The DRL algorithm was utilized to generate the optimal strategy for the mobile device to determine whether to stay in the aggressive area, and whether to use FH technology to counter smart jammers. In order to deal with high-dimensional signals, the author combined DNN to improve the training efficiency of Q-learning.

\subsubsection{Network Layer Security}
%The network layer is the main body of the entire MEC framework, and all data collected by the perception layer is preprocessed and transmitted through this layer. 
Numerous communication protocols in the network layer can provide and manage the connectivity between sensors and edge servers. However, the open communication protocol makes it effortless for attackers to seek out the vulnerability during communication. 
%Therefore, the security of the data transmission level is a pivotal issue in the MEC environment with heavy traffic. As another key component of the network layer, middleware provides adaptability for different IoT devices and provides extended functions at the software level. However, malware attacks may cause IoT devices performance degradation or even strike \cite{xiao2018iot}. In addition, in order to release the pressure of centralized servers and realize real-time processing, delay-sensitive tasks can be offloaded to edge servers for big data processing in the network layer, and the large scale contents can be cached in edge severs for downloading. Therefore, ensuring the safety of task offloading and caching is also an issue with vital significance.
% In this section, we introduce the security protection of the network layer from the following aspects:
\paragraph{Data Transmission Level}
An effective solution to anomalies in data transmission is to build an IDS to automatically detect various attacks during communications. 
%IDS issues an alarm in time and takes preventive measures when an event matches a known attack signal. 
%Middlemiss et al. \cite{middlemiss2003feature} proposed a hybrid IDS model based on GA and kNN classifier. 
%The KDD Cup competition dataset was adopted, and a series of TCP packets were used to represent the network connection. 
%The proposed model was efficient in feature selection to classify $24$ attack types with an improved accuracy. 
%In order to improve the accuracy of AI classifiers  availably, the feature selection method is usually employed for data preprocessing. 
In a previous study \cite{manimurugan2021iot}, PCA was used to reduce the dimension of the original data and represent it as new data with the minimum attribute through the principle component load matrix, and the improved NB classifier was utilized to classify network attacks. 
Another work \cite{farnaaz2016random} applied the replacement missing value filter to the feature extraction and utilized the RF classifier to classify attacks with network traffic data. %The results proved that the model acquires a lower false alarm rate and a higher detection rate. 
Whereas, the index of classification accuracy alone is not enough to objectively evaluate the performance of the IDS classifier. It needs to be combined with indexes such as precision, recall rate, and false detection rate to make a comprehensive measurement \cite{hasan2019attack,manimurugan2021iot,farnaaz2016random}.
%Ho et al. \cite{ho2021novel} applied CNN in IDS to enhance the security of the network layer. The convolution operation in the CNN model reduced the dimension of the feature matrix, which enabled IDS to quickly and fully extract the key information in complex network traffic. %In the simulation, the consideration of the true negative rate was added, and the proposed model was compared with other $9$ classifiers, which illustrated the excellent performance of the proposal.

%A previous work \cite{tian2020intrusion} improved the DBN model for network intrusion detection, which used probabilistic mass function (PMF) coding and Min-Max normalization operations to preprocess data. Kullback-Leibler (KL) divergence was exploited in training to avoid over-fitting, so as to improve the detection accuracy of the model.

DoS attacks are the most common type of intrusion in the network layer. In response to this, a previous research \cite{majumder2020smart} adopted LR-based classifiers to detect DDoS attacks and MitM attacks. This work utilized smartphones as edge servers in IoT networks for training. Using the power consumption ratio information of attacking devices and non-attacking devices to predict intrusion, the proposed model achieved a relatively ideal accuracy rate. %Roopak et al. \cite{roopak2019deep} applied CNN and LSTM to DoS attack detection. In the experiment, comprehensive consideration was conducted from the three indicators of accuracy, precision and recall rate, which proved that the hybrid model is superior to other ML methods. 
%Traditional abnormal traffic analysis in the network layer is also a tedious and difficult manual task. 
Another study \cite{sinclair1999application} proposed an intelligent anomaly detection architecture based on GA and the DT classifier. The architecture detected illegal links based on the IP packet header information. 

\paragraph{Offloading and Caching Level}
Ensuring the safety of the offloading and caching process is an important task for maintaining the network layer security. 
%Aiming at cloud offloading scenarios, Xiao et al. \cite{xiao2016mobile} proposed an assisting mobile offloading strategy based on Q-learning, using the secure agent to decide whether providing protection for the offloading data. Mobile devices, smart attackers, and security agents formed a security game, in which the security agent made decisions based on the offloading rate of mobiles and the type of network attack. The proposed game not only increased the task offloading rate but also enhanced the security of offloading. 
%Along with this direction, the 
A previous research \cite{xiao2018security} extended RL to the MEC environment to protect the security of edge caching. In response to attacks in the mobile edge cache, mobile devices adopted RL to make the optimal decision from a limited set of actions to protect the offloading process from interference or perform device authentication. This work proposed a security framework for MEC offloading with RL-based methods, which has enlightening significance for protecting the security of edge offloading and caching.

\subsubsection{Application Layer Security}
The application layer directly provides users with intelligent services. However, the inherent vulnerabilities in the operating system (OS) of smart device pose security threats to the application. 
%Especially for the most widely used Android system in the current handheld devices, the detection of malware has become a key issue. 
In a previous research \cite{hou2016droiddelver}, an Android malware intelligent detection system named DroidDelver based on the DBN model was introduced. In this work, API calls were extracted from the smali code, and different API csalls were categorized into blocks according to their functions. Based on the generated API call blocks, DBN learned the relationship between benign software and malware to detect malware in the system with improved accuracy. Xu et al. \cite{xu2016hadm} proposed a hybrid analysis malware detection system of Android, which extracted features through DBN and combined them with original features to construct a vector set for classification. SVM was utilized as a classifier, and the kernel matrix constructed by the similarity between features was applied for classifying the benign software and malicious applications. 
%Moreover, Chen et al. \cite{chen2017deep} proposed a DBN-based malicious Android application detection framework in the MEC environment to improve the robustness of mobile devices accessing edge servers. The framework used the APK file in the edge server to extract features from functions like required permissions, sensitive APIs, and dynamic behavior. %The DBN model was utilized for feature learning, and the softmax function was applied in the output layer for predicting. 
%Combining the location information of mobile devices, the accuracy of detection was greatly improved compared with other ML methods.

\subsection{MEC System Level Security}
MEC system level provides UE and CFS Portal with available edge sever services.
%By exploiting the computing resource of the edge server, the resource-constrained mobile end can exchange for high-performance attack inference with little communication cost through an end-edge collaborative mode. In addition, SDN provides scalable and flexible management services for MEC system level and application interfaces for third-party operators. However, attackers can pretend to be an operator to attack SDN, thereby threatening the security of MEC system level. 
In this subsection, we summarize the AI-based approaches in MEC system level in the following two aspects:
\subsubsection{End-edge Level}
The collaboration working mode of mobile ends and edge servers provides an efficient solution for intrusion detection in UEs and access networks. 
In a previous work \cite{kim2018squeezed}, the Deep AE model was used for anomaly detection in the industrial IoT network in an end-edge collaborative mode. This work applied a CNN-variational AE model to extract features from the time series state information of sensors and perform anomaly detection. What's more, the memory and computing consumption deployed on the edge sever were reduced by compressing the neural network size. 
In another research \cite{sadaf2020intrusion}, an IDS architecture based on the Deep AE was applied to detect abnormal traffic from sensors to edge servers. In order to improve the accuracy of classification, the model utilized isolation forest to further classify the output of the Deep AE model to find the points of classification errors. This optimized detection architecture based has an enlightening effect on reducing the false alarm rate.

The access network between the edge sever and the mobile end is also with potential vulnerabilities to several attacks. Based on previous works, Chen et al. \cite{chen2019deep} proposed a network attack detection system based on DBN in MEC. The dynamic features were extracted by exploiting the Android file package in the edge server and used for the model training. The loss function was established according to the difference between inputs and the actual outputs, and the loss was reduced by using the BP algorithm to fine-tune the parameters of the neural network. 

Moreover, a mass of high-dimensional and multi-source alarm information is gathered in the edge server \cite{li2019complex}, which is a huge challenge for the processing and storage capabilities of the server. Li et al. \cite{li2019complex} proposed an attack detection method based on the end-edge framework. The alarm information of the edge equipment was de-redundant through the K-Means method. The authors established attribute attack graphs through the correlation between the alarm information to comprehensively analyzed to the potential vulnerabilities of all edge nodes. A greedy decision-making algorithm was utilized to solve the problem of the minimum dominance set in the attribute attack graph to replace the generation of the complex attack linkage strategy. Given these points, this work provides ideas for the feature extraction of high-dimensional and multi-source alarm information in the end-edge level and the consideration of complex attack relevance. 

\subsubsection{SDN Level}
The separation of data plane and control plane in SDN also brings various security threats. In a previous study \cite{li2018ai}, an IDS in SDN based on the combination of K-Means and an improved RF classifier was proposed. The bat algorithm was used for feature selection, where the position of bat was set as the selected feature subset, the iteration of position movement and target search were summarized as the feature selection process, and the fitness function was adopted to evaluate the feature selection process. After the selected features were clustered by K-Means, the flows were classified by using the RF algorithm based on the weighted voting mechanism. The proposed model achieved satisfactory classification accuracy due to the advanced feature selection method while reducing the training cost. 
In \cite{tang2016deep}, authors applied DNN to build an abnormal flow detection system in SDN, which was deployed in the control plane. It monitored the data of all OpenFlow switches and utilized the global network status to detect intrusions. The flow table was modified to propagate the security policy to the switch when detecting intrusions. The DNN model of this system was trained through information such as protocol, duration, and flow bytes as features and achieved improved classification accuracy.

Kirutika et al. \cite{kirutika2019controller} proposed an external monitor based on the RF algorithm for possible attacks in the SDN control plane. 
%In order to ensure the performance of the external monitor, it is necessary to set it separately in a host, and periodically check the control plane. 
The number of incoming/outgoing data packets and lost packets, duration and other network state information were utilized as features to train the RF classifier. The external monitor detected the malicious data and interrupted the controller timely. Consequently, this work ensured the security of SDN from the control plane to prevent the entire SDN system from being corroded by attackers.

\subsection{MEC Host Level Security}
MEC host level is mainly responsible for managing various functions of the host, as well as the collaboration with other hosts and the cloud. 
%While vulnerabilities in the hardware architecture of the edge host itself may provide opportunities for attackers. Therefore, protecting the safety of any party in a collaborative task is with significant importance. At the same time, virtualizaion resources are also a key task of MEC host level, which can provide flexible services for large-scale network security protection. 
We introduce the AI-related works in the MEC host level from the above aspects in the following:
\subsubsection{Edge-host Level}
The edge host security directly affects the robustness of the overall MEC architecture. Khalid et al. \cite{khalid2020macler} proposed a hardware Trojan detection framework in edge hosts by multi-layer perceptrons (MLP). The single power-port current acquisition block was designed in the time-division multiplexed current sensor to reduce the cost of data acquisition. Through the analysis of hardware Trojan benchmarks at the register transfer level in the system-on-chip (SoC), four LEON3 processors from other infrastructure providers were integrated to provide the solution. The proposed model greatly improved the detection accuracy of hardware Trojan compared with the existing methods, and reduced the power consumption on per unit area of the chip. In \cite{yoon2020pufgan}, a GAN-based self-adversarial agent model was proposed to improve the hardware security of edge hosts. The agent utilized vanilla GAN and conditional GAN respectively to attack edge hosts, and the public PUF was used to evaluate the quality of the attack by generating realistic secret keys. The agent reconstructed its underlying security primitives into the public PUF through feedback to improve the security entropy of the system when the attack quality exceeds the setting threshold. 
%In the experiment, the specific hardware equipment and the actual public PUF dataset were applied to illustrate the importance of the current primitive.
\subsubsection{Edge-edge Level}
The secure collaboration mode between edge hosts is crucial for MEC host level. In a research \cite{diro2018leveraging}, LSTM was utilized to detect cyber attacks in the sensors network through the Fog-to-Things architecture. The entire architecture assigned attack detection tasks to edge hosts, then cyber attacks in the cover area were detected through the coordination of parameter update, storage and control between each edge host. In each edge host, the sequential Stochastic Gradient Descent (SGD) algorithm was adopted to calculate local parameters of the LSTM. And the coordination host was responsible for aggregating local parameters from all edge hosts, then updating and returning the global parameters back for updating the local model. %This edge-edge mode provides a collaborative and efficient attacks detection framework. 

Pan et al. \cite{pan2020justinian} proposed a gradient aggregation agent (GAA) model suitable for the MEC environment against Byzantine attacks, using RL to protect the robustness of the distributed learning framework. 
%For the scenario where multiple workers (edge servers) conduct collaborative learning under the command of the master (coordination edge host), a worker is called the Byzantine worker when it is compromised and disrupts the network maliciously, which may cause a devastating blow to the entire learning process. 
The proposed GAA model learned the experience from the interaction among workers and the auxiliary information in the master, and decided the contribution weight of its generated parameters to the overall parameters according to the credit of the worker. For different aggressive environments, the robustness of the GAA model proved its performance in MEC multi-edge collaborative level.
\subsubsection{Edge-cloud Level}
The edge-cloud collaboration mode solves the problem of insufficient edge device resources. In \cite{kozik2018scalable}, ELM was employed to detect attacks in IoT networks. Features about duration, IP, source and destination port numbers were converted into random multi-dimensional space vectors. However, plenty of features occupied the memory and the model training process generated huge energy consumption in the edge sever. To achieve an efficient detection architecture, edge devices were used to randomly project raw data to generate private data and upload the data to the cloud with sufficient resources for model training. 

%Although the collaborative mode has achieved better classification accuracy, it has not fully released the potential of the edge. 
Ran et al. \cite{ran2020cloud} proposed a CNN-based edge-cloud collaborative attack recognition architecture. The initial datasets of multiple edge hosts were set to the same. Then edge hosts were responsible for the raw data preprocessing and feature selection, and utilized the trained CNN locally for attack detection. The edge host with the highest detection rate uploaded its own data to the cloud in a fixed period of time, and the cloud was in charge of updating the dataset and retraining the CNN and sending the improved model to each edge host. This edge-cloud collaborative learning mode improves the detection accuracy and the robustness of the system. In addition, in order to release the workload of the cloud, AI algorithms (such as PCA \cite{huong2021efficient} and Deep AE \cite{sadaf2020intrusion}) can be deployed in edge hosts to pre-learn data, and perform secondary learning through the cloud to fully improve performance.

\subsubsection{NFV level}
NFV provides resource virtualization services for edge hosts. Guizani et al. \cite{guizani2020network} proposed an anti-malware system based on NFV, which used RNN to predict malicious software so that NFV can deploy relevant resources in a timely manner to resist attacks. The raw data collected by the sensor was preprocessed and evaluated in terms of available functions, and feature reduction and data cleaning were employed to filter valuable features. Then the system applied the RNN-LSTM model to learn the preprocessed data, and utilized NFV to virtualize the patch distribution mechanism. %When a malware was detected, there was no need to search and update the IoT device or change the protocol, but directly distributed the patch to the abnormal area according to the bound patching mechanism.

\begin{table*}[htbp]
\scriptsize
    \centering
    \caption{Summary of AI-based work at each layer of MEC for privacy}
    \label{tab:layer_privacy}
    \begin{tabular}{|c|c|c|c|c|c|c|c|c|c|}
        \hline
        \multirow{2}{*}{\bf Layers} & \multirow{2}{*}{\bf Works} & \multirow{2}{*}{\bf Privacy issues} & \multicolumn{5}{c|}{\bf Related AI methods} & \multirow{2}{*}{\bf Results} &  \multirow{2}{*}{\bf \makecell[c]{Applications\\ or \\ scenarios}} \\
        \cline{4-8}
         ~ & ~ & ~ & \cellcolor[gray]{0.8} \makecell[c]{Supervised\\learning} & \cellcolor[gray]{0.8} \makecell[c]{Unsupervised\\learning} & \cellcolor[gray]{0.8} \makecell[c]{Semi-supervised\\learning} &\cellcolor[gray]{0.8}  RL &\cellcolor[gray]{0.8}  \makecell[c]{Non-ML\\methods}& ~ & ~\\
        \hline
        \multirow{8}{*}{\makecell[c]{IoT \\System}} & \cite{han2018kclp} & Location privacy & ~ & K-Means & ~ & ~ & ~ & None quantitative result & WSN privacy\\
        \cline{2-10}
        ~ & \cite{huang2019lightweight} & Data privacy & CNN & ~ & ~ & ~ & ~ & \makecell[c]{Accuracy: 98.27\%\\Encryption runtime: 0.344 s\\Decryption runtime: 0.056 s} & \makecell[c]{Mobile sensing\\ privacy}\\
        \cline{2-10}
        ~ & \cite{tan2020lightweight} & Data privacy & kNN & ~ & ~ & ~ & ~ & Encryption runtime: 0.13 s & \makecell[c]{Encrypted cloud\\ database} \\
        \cline{2-10}
        ~ & \cite{tian2019lep} & Data privacy & CNN & ~ & ~ & ~ & ~ & \makecell[c]{Error detection rate: 99\%\\Encryption runtime: 0.012 s\\Decryption runtime: 0.025 s} & Offloading privacy\\
        \cline{2-10}
        ~ & \cite{yu2018federated} & \makecell[c]{Data privacy,\\location privacy,\\identity privacy} & ~ & AE & ~ & ~ & ~ & Caching efficiency: 15\% & Caching privacy\\
        \cline{2-10}
        ~ & \cite{liu2020privacy} & \makecell[c]{Data privacy,\\location privacy} & ~ & K-Means & ~ & ~ & ~ & \makecell[c]{Runtime: 1209 ms\\Communication overhead: \\1085KB} & Caching privacy\\
        \cline{2-10}
        ~ & 
        %\cite{hsu2020privacy} & Data privacy & ~ & K-Means & ~ & ~ & ~ & Android software privacy \\
        %\cline{2-9}
        \cite{zhao2020privacy} & Identity privacy & CNN & ~ & ~ & ~ & ~ & Accuracy: 97\% & Blockchain\\
        \cline{2-10}
        ~ & \cite{wu2021pecam} & Data privacy & ~ & ~ & GAN & ~ & ~ & Accuracy: 96\% & \makecell[c]{Video streams\\privacy}\\
        \hline
        % \multirow{4}{*}{\makecell[c]{Mobile Edge\\System Layer}} & \cite{qian2019privacy} & Data privacy & ~ & ~ & ~ & ~ & ~ & FL privacy \\
        % \cline{2-9}
        \multirow{3}{*}{\makecell[c]{MEC\\System \\Level}} & \cite{zhao2020p} & Position privacy & ~ & Skip-gram & ~ & ~ & ~ & Identification rate: 83\% & \makecell[c]{Anti-poisoning\\attacks}\\
         \cline{2-10}
        ~ & \cite{mao2018privacy} & Data privacy & DNN & ~ & ~ & ~ & ~ & \makecell[c]{Forward pass time: 0.6s\\Backward pass time: 0.2s} & \makecell[c]{Facial recognition\\privacy}\\
         \cline{2-10}
        ~ & \cite{sangaiah2019enforcing} & Position privacy & DTs, kNN  & ~ & ~ & ~ & ~ & \makecell[c]{Position confidentiality\\assurance: 90\%} & LBS privacy\\
        \hline
        \multirow{4}{*}{\makecell[c]{MEC\\Host \\Level}} & \cite{lu2020privacy} & Data privacy & DNN & ~ & ~ & ~ & ~ & \makecell[c]{Accuracy: 94.23\%\\Compression ratio: 8.77\%} & FL privacy \\
        \cline{2-10} 
        ~ & \cite{wu2020ensemble} & Data privacy & DTs   & ~ & ~ & ~ & ~ & None quantitative result & Edge-cloud privacy\\
        \cline{2-10} 
        ~ & \cite{osia2018private} & Data privacy & ~ & PCA & ~ & ~ & ~ & None quantitative result & Edge-cloud privacy\\
        \cline{2-10}
        ~ & \cite{chen2020privacy} & Data privacy & kNN & ~ & ~ & ~ & ~ & None quantitative result 
 & \makecell[c]{Querying encrypted\\data}\\
        \hline
    \end{tabular}
\end{table*}

\section{AI Approaches for Layer Based MEC privacy}\label{sect:layers based privacy}
Traditional IoT network privacy protection methods are implemented through cryptography or steganography \cite{alotaibi2019secure, shambour2021personal, samkari2019protecting, almutairi2019image}, while advanced AI technology gives more possibilities for privacy-preserving. In this section, we give a hierarchical introduction to AI-based works related to the privacy in the MEC environment. Related works are summarized in the same hierarchical manner as Section \ref{sect:Layers based security} and listed in Table \ref{tab:layer_privacy}.

\subsection{IoT Systems Privacy}
\subsubsection{Perception Layer Privacy}
In the raw data collected by sensors, the location privacy is of great importance. Han et al. \cite{han2018kclp} proposed a location privacy-preserving framework based on K-Means in the IoT network. Fake source nodes were established to transmit packets from real source node, so as to make the attacker confused about the real location of the source node. And fake sink nodes were utilized to conceal the real route through generating fake packets and transmitting them in different directions. In this work, the public and private keys were applied in each node for authentication, and only the real packet was able to be transmitted to the real sink node through the specific route. %The K-Means algorithm was used to cluster according to the determined route and allocate fake packets to the area to increase the safe time.

As the subject of the perception layer, the privacy protection of the sensor is also of the essence. In a previous research \cite{huang2019lightweight}, a lightweight privacy-preserving framework of the mobile sensor was proposed based on CNN. In order to reduce the computing overhead, storage and communication costs on mobile terminals, this work transferred data processing tasks from the cloud to the edge server. The CNN model was publicly available to each edge server for feature extraction. 
The sensor randomly divided the collected image data into two parts for encrypting, which can be restored by superimposing them together. The two encrypted divided parts were respectively used for feature extraction and learning of two edge servers, and the interaction of the two edge servers was controlled by a trusted third party.% The edge server sent the encrypted features after feature extraction to the client for related tasks. This encryption framework between mobile terminals and edge servers provides heuristic ideas for sensor privacy protection.

\subsubsection{Network Layer Privacy}
%As mentioned above, edge offloading and caching are key methods to alleviate the pressure on the network layer. However, the process of outsourcing tasks from mobile devices to edge servers incurs privacy issues, such as user data leakage and data integrity damage. 
In this subsection, we introduce the AI-enabled methods of privacy preserving from the following aspects:
\paragraph{Offloading Level}
ML-based inference tasks on mobile devices are often offloaded to the nearby edge server instead of the cloud sever, which greatly reduces the communication pressure and computational burden. In \cite{tan2020lightweight}, kNN was used to construct a lightweight encrypted cloud database architecture with edge offloading. In this work, data owners uploaded data to the cloud server, data users sent query requests to the cloud, and the cloud utilized kNN for classifying. Pailier encryption system was employed to encrypt data and labels to provide database security, query privacy, and data access mode hiding. %At the same time, the data preprocessing operation was performed at the edge server, which reduced the calculation cost of the cloud server to perform kNN classification.
In another work, Tian et al. \cite{tian2019lep} proposed a lightweight scheme to protect the privacy of the CNN-based inference task in the offloading process. Multiple pairs of encryption and decryption keys were generated and stored in the IoT device in an offline form, and each pair corresponded to a CNN inference task request. In the inference phase, the IoT device offloaded the CNN training task to the nearby edge server online, and the CNN model provided data integrity checks to ensure the correctness of the data. %However, a number of previous offloading privacy protection studies assumed that IoT devices are completely trustworthy \cite{tian2019lep}, and none of them considered the impact of device security, which leave a huge space for research.

\paragraph{Caching Level}
Edge caching provides additional resources for mobile devices to relieve storage pressure. Yu et al. \cite{yu2018federated} proposed a proactive content caching method based on Federated Learning (FL), which performed training with no require of the centralized collection user data. The method used a centralized server to aggregate parameters of distributed edge servers for updating, and each edge server utilized Deep AE for local training. In the absence of user data, the performance of this model was still better than other methods, and it played a vital role in protecting user privacy. In \cite{liu2020privacy}, a privacy-preserving method based on K-Means was used in the caching process of the next generation cellular network. Through FL system, a distributed architecture was adopted to upload user data instead of encrypted user data to the cloud server for training based on the SGD method, so that avoiding data leakage caused by directly uploading data. 
%And the secret sharing technology divided the gradient into multiple random shares, and only a certain number of shares was able to restore the gradient information, so as to achieve the purpose of protecting the privacy of the gradient. 
%In this work, the K-Means algorithm was adopted to perform clustering based on the proactive caching of user information to meet the peak traffic demand. Combined with the privacy-preserving methods, this work provides a reference for efficient and safe caching in the network layer.

\subsubsection{Application Layer Privacy}
%Malware in the operating system will cause the privacy of user information to be violated. In \cite{hsu2020privacy}, an FL-based privacy-preserving system was proposed to protect the Android system from malicious software. The K-Means algorithm was used to cluster according to the topic number-proportion value pair of the APK file, and the federated SVM was introduced to provide MPC and trained at the edge sever. In addition, the classifier was cooperatively trained by mobile devices to implement a lightweight and efficient privacy-preserving system.

Blockchain is a well-known shared database in application layer. Due to the data in blockchain cannot be tampered, it can be applied for the privacy-preserving of user information.
Zhao et al. \cite{zhao2020privacy} proposed a CNN-based FL system combined with blockchain to provide a user data privacy protection model for home appliance manufacturers. Firstly, the user's smart phone collected data from home appliances, and conducted local-level training of the model in a collaborative manner with the edge server. Then users signed the trained model and uploaded the model to the blockchain to protect it from tampering. In addition, the manufacturer used the DP technology to calculate an average model for the models collected from users and preserve their privacy.

In addition, video streaming and analytics (VSM) is also a popular application in IoT system. In order to protect the user privacy contained in the required data in the VSA, Wu et al. \cite{wu2021pecam} proposed an enhanced privacy protection system. Using the steganography of the GAN model at the front end, the system was able to reverse the privacy-enhanced video without changing the back end without auxiliary data. In addition, by combining with system optimization technology, the system fully reduced the network bandwidth and realized efficient real-time processing on the hardware. 
%And this work enlightens the research on preserving privacy of video streams.
\subsection{MEC System Level Privacy}
%In order to improve the QoS provided by the edge server in the MEC, the user's historical information is required, while it also contains the user's privacy. To solve this, Qian et al. \cite{qian2019privacy} proposed a privacy-aware service placement model between IoT devices and edge servers. The model constructed the problem of whether to place the service on the edge server as a binary problem, and learned the user's preference for the service through ML algorithms. Due to the characteristics of FL, there was no need to transmit the user's private information, but to train the model through the user's preferences. The edge server fed back the updated parameters to the mobile device, and retrained on the device side to perform related operations. For the service placement problem, a greedy algorithm was adopted to maximize the objective function. This model provides privacy-preserving services for users in the MEC environment.

Aiming at protecting edge servers from privacy damage caused by malicious users’ poisoning attacks, Zhao et al. \cite{zhao2020p} proposed a privacy-preserving model based on a feature learning model named Skip-gram. In the edge server, the social relationship between different users was extracted and an inferred social graph was established. The model predicted the location of the poisoning through the best map between the social graph and the social graph.
In other work, Mao et al. \cite{mao2018privacy} proposed a privacy-preserving model of DNN facial recognition based on DP mechanism. In order to reduce training costs, the DNN model was partitioned from the convolutional layer, where the first part was deployed on the user side and the second part was arranged on the edge server side. The output of the user-side model was input to the edge server, and the calculated loss information was fed back to the user-side to update the gradient. %In this model, the user side utilized part of the DNN activation output with artificial noise instead of directly transmitting the original image data to the edge server, which provides a privacy protection idea based on the partition mechanism.

Moreover, in a previous research \cite{sangaiah2019enforcing}, a combined approach based on DTs and kNN was used to identity the user's position, and the hidden Markov model was applied to estimate the user's destination and location tracking sequence. This approach was implemented in the MEC environment to ensure the timeliness and confidentiality of the delivery of LBS and provide privacy-preserving LBS for roaming users. G-Means clustering method was adopted to extract effective location features from the redundant information of the device, and DTs and kNN were merged to solve the position tracking sequence ordering problem. Then the edge sever utilized excessive locations to predict the location through hidden Markov models. 
%Through experiments, the proposed model provided users with high LBS confidentiality.

\subsection{MEC Host Level Privacy}
In MEC host level, related privacy-preserving researches focus on the realization of two collaborative levels, edge-edge and edge-cloud. 
%In particular, the distributed characteristic of FL enables servers of different levels to obtain information according to their own authority, so that the privacy of users with higher authority is preserved in a more secure central server. 
\subsubsection{Edge-edge Level}
Lu et al. \cite{lu2020privacy} proposed a joint learning mechanism based on FL to provide privacy protection support for joint work scenarios of multiple edge servers. Edge servers were used to train the ML-based model, and the parameter server was responsible for collecting parameters of all edge servers and updating global parameters. In this work, the parameter privacy of the entire model was protected because the edge server only communicated with the parameter server and had no authority to obtain the information from other edge servers. Meanwhile, in order to improve the training efficiency of the model, a gradient sparsification method was adopted to compress the interaction between edge servers and the parameter server so as to reduce the communication consumption and obtain effective compression space at a tiny cost of accuracy.

\subsubsection{Edge-cloud Level}
The edge-cloud hybrid architecture is an efficient privacy-preserving framework. In a research \cite{wu2020ensemble}, a privacy-preserving mechanism in the edge-cloud environment based on DTs was proposed. On account of the DP algorithm, the edge server as the collector of the adversary dataset was required to not distinguish the difference from the normal dataset. In this mechanism, the cloud server was used to build a private random DTs model, and the edge server was responsible for collecting data and adding it to the random DTs. %When private random DTs were established, the cloud platform provided label inference functions.

In another work, Osia et al. \cite{osia2018private} proposed an edge-cloud hybrid privacy-preserving framework that used edge server resources to reduce the cloud processing latency and improve the processing performance. The framework consisted of two modules. The privacy edge module was responsible for extracting features from the original image data, and the cloud server module was in charge of inferring and feeding back to users. In the privacy edge module, a Siamese privacy framework was adopted, and the PCA model was employed to reduce the dimension of the data to extract features and ensure the accuracy of feature extraction through fine-tuning. In the cloud server, the CNN model was utilized to perform inference based on the extracted low-dimensional features, and the use of low-dimensional features instead of the raw user data to transfer to the cloud also ensured the secure user data privacy.

\section{Lessons Learned, Future Works and Challenges}\label{sect:future works}
% we summarize the work of this paper. The two aspects of using AI methods to ensure MEC security as well as privacy-preserving and coping with the ML security and privacy challenges in MEC are concluded. The security and privacy issues involved in Section \ref{sect:challenge} is summarized concretely,
In this section, according to the security and privacy concerns and related AI-based solution, we summarize specific research problems and propose promising future directions to give our insights to future researchers.

% \subsection{AI Methods for MEC Security and Privacy}
\subsection{AI Methods for MEC Security}
The integration of various complex heterogeneous technologies in MEC makes it face multitudinous threats. Additionally, although powerful AI methods can provide efficient classifiers and feature extraction tools for MEC security protection, the deployment of these methods in the MEC environment is also extremely challenging. In this subsection, we summarize the existing research issues and future directions in using AI methods to solve various security threats in MEC.

\subsubsection{IoT Systems Security}
The heavily deployed IoT networks provide MEC with a wide range of interconnection and low-latency applications, while also brings the following challenges:
%However, a mass of IoT devices, heterogeneous communication technologies and applications are integrated in IoT networks, resulting in various network vulnerabilities, which may be obtained by attackers to invade MEC or even the entire IoT network. 
(1) attackers can steal sensitive information from IoT devices in collecting raw data through side channel attacks, or directly perform physical attacks to disrupt the normal operation of IoT devices, and the extensively applied gateway devices also increase the risk of TCP/IP attacks; (2) for the reason that the application layer of IoT directly provides users with related services, it is also indispensable to perform effective malware detection in IoT networks. 

\paragraph{A General IDS Architecture for Enhancing Trust Based Approaches in Ad-hoc Networks}

\textit{Research Problems:}
Existing IDS for the security of wired networks could be used in wireless contexts. However, the intrinsic characteristics of MANET may limit their application, such as the absence of centralised infrastructure, limited bandwidth, and mobility of the nodes, etc.

\textit{Future Directions:}
Considering the intrinsic characteristics of MANET, a general IDS architecture should have the following properties in the future: 1) A self-defence mechanism is of immense importance for defending repeated false alarms by sending a flood of irrelevant packets to the IDS host; 2) It should require little system resources to execute and not hinder system performance by adding overhead; 3) It should run continuously and keep up transparent to the system and the users; 4) The IDS should abide by standard to be cooperative and open, such as the standard alert format Intrusion Detection Message Exchange Format (IDMEF) and a protocol for transporting such alerts Intrusion Detection Exchange Protocol (IDXP) \cite{albers2002security}.

\paragraph{Efficient Malware Detection Tools in IoT
 Networks}

\textit{Research Problems:}
The malware detection tools can effectively prevent threatening objects in the IoT network from destroying software-level security. However, various widely popular applications in the practical IoT environment accelerate the evolution of malware, which greatly limits the performance of traditional detection methods, and cannot guarantee the false alarm rate for benign system files.

\textit{Future Directions:}
Future researches on malware detection should mainly focus on combining traditional malware detection tools with powerful AI models. The key to detection lies in heuristic features such as file properties, code fragments, and file hashes that distinguish benign from malware. Therefore, some DL-based Natural Language Processing (NLP) models can be used to extract and process various feature information in OpCode, and construct malware detection as a binary classification problem. In addition, in order to adapt to the continuous emergence of system files and malware in the IoT network, the dependency graph can be used to express the relationship between software, and solve the state-of-the-art Graph Neural Network (GNN) model to extract features and graph reduction operations to reduce the amount of computation.

\paragraph{MEC-IoT Security Protocols with Machine Learning}

\textit{Research Problems:}
The Intelligent Transportation System (ITS) is a typical application of MEC-IoT systems, which consists of  advanced sensors and control systems. ITS relies on the interconnectivity of various devices to process real-time data flow and transmit it to assure secure and efficient digital services. Such data flow is plaintext that is prone to eavesdropping and hijacking. Unmanned Aerial Vehicles (UAV) are another example in which the aim is to conserve battery life while offloading computational or storage information to MEC servers for processing. As a result, using strong cryptographic primitives or lengthy security processes would be impossible \cite{porambage2018survey}. The recent works either consider security protocols or machine learning technologies to secure the MEC-IoT systems. However, an elaborately designed method by aligning security protocols with machine learning can be more effective in protecting the security of MEC-IoT systems.

\textit{Future Directions:}
Although the ML-based IDS can cope well with abnormal data traffic, the required massive data collection in the interconnective IoT systems raises privacy concerns during both the training and prediction stages. One promising solution is secretly sharing the data with light-weighted cryptography protocols and evaluating the data with three-party computation. However, there are several research challenges to be solved in the future. First, the three-party computation is only suitable for computation over a $\mathbb{Z}_{2^k}$ ring. But both the training data and intermediate parameters of machine learning are decimal values that are unable to handle modular arithmetic. Second, secret sharing is costly and quickly becomes a major performance bottleneck in resource-constrained MEC-IoT systems.

\subsubsection{Mobile Core Network Security}
The mobile core network is the provider of the main functions of MEC. 
%At the MEC system level, requests from the UE and CFS Portal need to be handled and processed accordingly, and the OSS and MEO modules determine whether to authorize these service requests. Therefore, ensuring the security of the connection between the edge server and the UE at the MEC system level is the primary prerequisite for the mobile core network to combat DoS/DDoS attacks. 
Although the in-depth modules of the MEC system (e.g., MEO) are difficult to be invaded by attackers, manipulated and misused attacks can inject fake data into the authentication module to disrupt the normal operation of the edge server. Moreover, threats in SDN are also hot research topics in the mobile core network.

\paragraph{Ensuring Reliable End-Edge Connections in Access Network}

\textit{Research Problems:}
As mentioned above, the end-edge reliable connection constitutes the first barrier to the access network security. In addition to the rational use of service request authentication provided by modules such as OSS and MEO at the MEC system level, AI-based IDS should also be combined to protect the connection from being disrupted.

\textit{Future Directions:} 
In the future, researchers can focus on choosing optimal AI models and features for detecting attacks between ends and edges. In order to obtain the sensitive information uploaded by mobile ends to the edge server, attackers usually inject malicious data packets into the end-edge connection. Therefore, IDS at the end-edge level should focus on the analysis of network traffic, analyze network dynamic information from the captured data packets, and take countermeasures when abnormalities are detected. In order to perform efficient feature extraction on large-scale data streams, DL models (such as DNN and DBN) can be used as favourable tools for building IDS in the future.

\paragraph{Detecting and Mitigating DDoS Attack on SDN Control Plane}

\textit{Research Problems:}
SDN offers unparalleled programming which enables network administrators to dynamically customize and control their networks within the MEC environment. One of the security concerns is DDoS attacks which drain the network capacity of SDN control plane by sending heavy traffic.

\textit{Future Directions:}
Although the advantage of the SDN control plane is that it can get the global view of the entire network, the control plane is insufficiently scalable to support high-frequency flow requests. A promising solution is to leverage the scalability and easy customization of virtualized software functions and adopt appropriate ML technologies and rule-based schemes to safeguard the centralized SDN control plane.

\subsubsection{Mobile Host Level Security}
%At the mobile edge host level, resource-constrained edge servers usually complete AI-based reasoning tasks through collaboration (edge-edge mode or edge-cloud mode), although it affords an efficient distributed solution for MEC, it also brings security and privacy issues. 
%In addition, as a key function of the MEC host level, virtualization can allocate resources according to demand, which reduces operating costs while increasing the flexibility of MEC services. 
When considering MEC security, the multi-faceted threats cannot be ignored: (1) VMs composed of various heterogeneous technologies may suffer from infected VM images, compromising VM migration, VM hopping, VM escape, and VM DoS attacks; (2) there are specific threats in NFV MANO when adopting the NFV technology to encapsulate the MEC function into VMs; (3) the hypervisor responsible for managing VMs may have traditional TCP/IP attacks, eavesdropping and hijacking attacks, and incur VM hopping, VM escape, and VM DoS.

% \textit{\romannumeral1) Securing Collaborative Edge-Edge/-Cloud Modes}

% \textit{Research Problems:}
% The distributed architecture of MEC facilitates the realization of FL technology in security tasks. Deploying complex neural networks on multiple edge and cloud severs to collaboratively improve the accuracy of IDS has become a popular paradigm \cite{diro2018leveraging,pan2020justinian}. However, it has become a crucial challenge to coordinate the parameter aggregation and update of different edge severs to maximize the performance of the entire FL system.

% \textit{Future Directions:}
% The adaptive algorithm can be developed to generate the parameter update strategy according to the contribution of different edge severs, which will become the emphasis of future work. 
% Furthermore, the update of the AI model is completed in the central server (may be a powerful edge node \cite{diro2018leveraging} or the cloud \cite{ran2020cloud}), so there is a risk of privacy leakage and malicious attacks in this process. Therefore, protecting the security and data privacy of the central server in FL will be the consideration of future researchers.

\paragraph{Coping with Inherent Threats in VMs}

\textit{Research Problems:}
The various inherent threats in VMs are open challenges, which bring risks to taking advantages of VM to virtualize edge network resources. Consequently, coping with inherent threats in VMs is getting more attention from researchers.

\textit{Future Directions:}
In order to deal with infected VM images, future researchers can enhance the security of VM images by combining traditional encryption algorithms and the firewall technology. For compromising VM migration, future research should also focus on the encryption of the plaintext transmitted during the migration process, and it is also meaningful for intrusion detection in the migration channel. For VM hopping, future countermeasures may be to deploy effective authentication algorithms in MMU to prevent malicious intrusion. And for VM escape, a valuable research direction is to enhance the security of the VMM and maintain the management authority of the VM by integrating additional monitors. In addition, for the above threats as well as VM DoS, an effective potential solution is to monitor through the AI-based IDS. For example, some commonly used lightweight AI classifiers such as kNN, RFs, and NB can be adapted to the IDS for VMs.

\paragraph{MEC Secure Resource Management by NFV}

\textit{Research Problems:}
The key function of NFV is to provide flexible resource management services for the network. In order to safely use NFV to manage MEC resources, the threats in the VM must first be resolved, which has been explained in detail above. And it is also a promising research problem to solve the threats of NFV.

\textit{Future Directions:}
In the future, we can regard for implementing NFV in MEC by adopting DRL that has been proven to perform well in the wireless resource allocation game. The security of the NFV interface should be enhanced to counter the threats from the NFV MANO. In addition, an adaptive and powerful hypervisor can be established, and IDS based on ML approaches (e.g., DNN, K-Means, DBN, etc.) with powerful feature extraction ability can be embedded in it to detect VM threats and possible TCP/IP attacks as well as eavesdropping and hijacking attacks in the future.

\paragraph{Developing Intelligent IDS by NFV}

\textit{Research Problems:}
Generally, AI-based IDS consumes a lot of computing resources and occupies a certain amount of storage space, which makes the deployment in resource-constrained edge devices challenging. Therefore, adopting NFV to provide flexible resource management for intelligent IDS is a meaningful research topic.

\textit{Future Directions:}
Future researchers can consider using the flexible resource management framework brought by NFV to assist in the realization of delay-sensitive and computationally intensive attack detection and defense tasks. In order to improve the emergency response speed of IDS, future research directions can adopt the NFV technology to deploy the required resources in time to counter attacks. For different types of attacks, the corresponding defense methods (e.g., configuring system patches and performing reasonable security domain partition) can be virtualized, and the bound virtualization function can be directly deployed to the abnormal area when a certain type of attack is detected by the intelligent IDS.

\subsubsection{Mobile Users Security}
The security of mobile users' devices is also a challenging subject. 
%While the popularization of smart devices facilitates people, it also collects a large amount of private information closely related to users. The security of UE directly affects the robustness of the front-end of the MEC architecture.
The possible threats on the UE side include malware and hijacking, DoS/DDoS attacks, and TCP/IP attacks, which bring a new research perspective to the security protection of MEC.

\paragraph{Ensuring UE Hardware-Level Security}

\textit{Research Problems:}
Ensuring the mobile users security from the hardware level of devices will also become an enlightenment for the future work. PUF technology embedded in mobile devices is attracting attentions. It can provide identity authentication for devices through electronic circuits. Unfortunately, advanced non-intrusive attacks \cite{laguduva2019machine} have been able to clone the PUF in MEC, which greatly threaten network security.

\textit{Future Directions:}
It is a prospective research direction to utilize ML method with great performance in classification to recognize the cloned PUF. In addition, deploying PUF on the chip also faces the dilemma of a trade-off between performance and energy consumption. Future researchers can devote themselves to applying ML methods to generate the optimal PUF configuration for different application scenarios.

% \textit{\romannumeral2) Solving Threats in UE's OS}

% \textit{Research Problems:}
% Vulnerabilities in the user's smart device OS also create an opportunity for malware to be injected into MEC. In particular, the currently most widely used Android system is vulnerable to security due to its open source characteristics. Therefore, effective detection of vulnerabilities in OS and timely patching of them in the MEC environment is a potential research problem in the future.

% \textit{Future Directions:}
% % Future researchers can apply ML methods to realize threat detection in users' smart devices.
% Since APIs in the UE system contain application information, it can be applied as an effective feature for ML model training. In the future, it can be considered to utilize the DL model to extract features such as sensitive APIs from complex APK files, and combine classification algorithms such as SVM and LR to distinguish normal applications from malware. In addition, in addition to the Android system, protecting the security of other OSs in devices of MEC is also an encouraging future research direction.

\subsection{AI Methods for MEC Privacy}
%According to the review in this survey, there is still a lot of room for further research on the privacy-preserving of MEC. 
%Compared with the cloud computing, MEC has a more flexible processing framework, lower service delay and cheaper cost, but it comes at the cost of more privacy leakage risks. 
As mentioned in Section \ref{sect:challenge}, the open MEC ecological environment has introduced multi-party infrastructure providers, which makes MEC services face the following privacy issues: (1) tasks offloading makes user's sensitive information may be stolen or tampered in the communication link during the offloading process; (2) the LBS brings a great test to MEC's privacy-preserving; (3) users' highly sensitive PII arouses subscribers' concerns about MEC.

\subsubsection{Preserving Data Collection Privacy}

\textit{Research Problems:}
In IoT networks, raw data collected by a large number of sensors urgently require to be uploaded to edge nodes for processing to offer corresponding services. However, the existing communication technologies in sensor networks (e.g., Bluetooth, WiFi, NFC, etc.) cannot provide users with completely trusted communication links. Therefore, protecting the privacy of sensitive information contained in the raw data is a meaningful research problem.

\textit{Future Directions:} 
In the future, an appropriate research direction is to use the DP technology between sensors and edge nodes to add random disturbances to the raw data for encryption without affecting the task effect. The end-to-end encryption technology HE can also provide researchers with a solution to protect the privacy of the raw data, which has a lower computational complexity. In addition, it is also an effective privacy-preserving solution to divide the raw data into multiple parts at the sensor, and upload encrypted parts to multiple edge nodes for further processing. It is worth noting that this solution requires a trusted third party to coordinate the collaboration between different edge nodes, and it requires careful design by future researchers.

\subsubsection{Ensuring LBS Privacy}

\textit{Research Problems:}
LBS is a basic service of IoT, and the location information is also required for multiple services. As mentioned in Section \ref{sect:challenge}, the user's location involves key personal information, and the current smart devices make it easy to leak location privacy. More importantly, the leakage of location privacy in some application scenarios that rely on location information (such as VANET) will cause unpredictable consequences. Therefore, the LBS that provides privacy guarantees in the MEC is imminent.

\textit{Future Directions:}
Future research directions can focus on using encryption technologies such as DP and HE to encrypt user location information in UE. In addition, because the routing information of the network also contains location privacy, attackers can trace the source according to the routing information. Therefore, future researchers can devote themselves to developing methods to encrypt routing information, such as adding fake nodes and injecting fake source data to ad-hoc networks to confuse attackers about the real route.

\subsubsection{Privacy-Preserving Approaches for Edge Offloading and Caching}

\textit{Research Problems:}
Edge offloading and caching are key functions of the MEC architecture, which enables mobile users to obtain nearby resources to handle delay-sensitive and computationally intensive tasks. A large amount of data containing sensitive information is offloaded or cached on the edge server, which brings numerous privacy issues.

\textit{Future Directions:}
Future researchers can devote themselves to the research of edge encryption databases, and store user information that is offloaded or cached to edge servers with guaranteed privacy. And some ML models (such as kNN, CNN, SVM, etc.) can be selected as the classifier of the encrypted database to classify according to the popularity of the content. The encrypted data and labels will provide reliable query and access services for edge offloading and caching. In addition, privacy-preserving research for some classic edge offloading and caching applications (for example, video streaming \cite{zhang2021elf}) is also a promising research direction.

\subsection{Enhanced Approaches in MEC Security and Privacy}
In addition to utilizing AI approaches to solve the security and privacy issues in MEC, how to enhance the performance of AI-based methods are also with great promise. 

\subsubsection{Integrating AI and Third-party Technology}

\textit{Research Problems:}
Lightweight edge servers have less abundant computing power and storage than cloud servers with centralized resources. Therefore, it is also an important challenge to deploy computationally intensive AI models at the edge reasonably, which determines the practicality of the research model. Especially for the DL model, the training of the neural network requires a large amount of computing resources, and this is overloaded for a single edge server. Therefore, the DL model deployment scheme based on partition training \cite{mohammed2020distributed,zeng2020coedge} and neural network hardware acceleration technology \cite{li2019edge,zhao2018deepthings} has become a trend.

\textit{Future Directions:}
The pivotal challenge of adopting partition training is how to adaptively divide the neural network into multiple partitions for training on multiple edge servers. And for the neural network hardware acceleration technology, the main methods include matrix decomposition, pruning, layer reduction, etc., which are mainly based on the idea of scale compression. The main challenge is to select an appropriate scale compression method for the model, and combine it with partition training to reduce model complexity and storage space while assuring the accuracy. The deployment of neural networks at the hardware level is also an attractive topic. Researches \cite{krestinskaya2019neuromemristive,lammie2020memtorch} of using memristors to realize DNN will provide basic support for improving the performance of the neural networks on MEC severs.

\section{Conclusion} \label{sect:conclusions}
MEC is becoming the main computation paradigm with its lightweight and efficient architecture. In order to duly extract valuable information from a mass of raw data and make relevant decisions, AI is deployed in MEC to provide intelligent data-related services. As the security and privacy issues attract more attention, AI-based technologies play a crucial role in protecting the security and privacy in the MEC environment.

This survey introduces the MEC architecture from a holistic perspective, and separates the layers from the MEC enabled IoT System and edge sever system levels. For the functions of each layer, related security, and privacy threats are explained in detail. After that, the methods related to MEC security and privacy are summarized from the perspective of AI, the ML-based methods are classified and mainly explained, and the inherent challenges in ML are also discussed. In addition, the non-ML AI approaches for MEC security and privacy are also summarized. Then, the related MEC security and privacy-preserving AI-based works are systematically discussed in each layer. At last, this survey summarizes the ideas and challenges of existing works from the three aspects of AI-based methods, MEC framework, and security and privacy challenges in AI, and envisages future research directions. This survey aims to provide researchers with an overview of MEC security and privacy-preserving from the perspective of AI, and to encourage the development of related researches.

\section*{Acknowledgment}
This work is supported by National Natural Science Foundation of China (NSFC) under grants No. 61972448, 62172068, and 61802048, and Hong Kong Research Grants Council, General Research Fund (GRF) under Grant 11203523.

% if have a single appendix:
%\appendix[Proof of the Zonklar Equations]
% or
%\appendix  % for no appendix heading
% do not use \section anymore after \appendix, only \section*
% is possibly needed

% use appendices with more than one appendix
% then use \section to start each appendix
% you must declare a \section before using any
% \subsection or using \label (\appendices by itself
% starts a section numbered zero.)
%

% \appendices
% \section{Proof of the First Zonklar Equation}
% Appendix one text goes here.

% % you can choose not to have a title for an appendix
% % if you want by leaving the argument blank
% \section{}
% Appendix two text goes here.

% use section* for acknowledgment

% Can use something like this to put references on a page
% by themselves when using endfloat and the captionsoff option.
\ifCLASSOPTIONcaptionsoff
  \newpage
\fi

% trigger a \newpage just before the given reference
% number - used to balance the columns on the last page
% adjust value as needed - may need to be readjusted if
% the document is modified later
%\IEEEtriggeratref{8}
% The "triggered" command can be changed if desired:
%\IEEEtriggercmd{\enlargethispage{-5in}}

% references section

% can use a bibliography generated by BibTeX as a .bbl file
% BibTeX documentation can be easily obtained at:
% http://mirror.ctan.org/biblio/bibtex/contrib/doc/
% The IEEEtran BibTeX style support page is at:
% http://www.michaelshell.org/tex/ieeetran/bibtex/
%\bibliographystyle{IEEEtran}
% argument is your BibTeX string definitions and bibliography database(s)
%\bibliography{IEEEabrv,../bib/paper}
%
% <OR> manually copy in the resultant .bbl file
% set second argument of \begin to the number of references
% (used to reserve space for the reference number labels box)
% \begin{thebibliography}{1}

% \bibitem{IEEEhowto:kopka}
% H.~Kopka and P.~W. Daly, \emph{A Guide to \LaTeX}, 3rd~ed.\hskip 1em plus
%   0.5em minus 0.4em\relax Harlow, England: Addison-Wesley, 1999.
% \end{thebibliography}
\bibliographystyle{IEEEtran}
\bibliography{ref.bib}

% Generated by IEEEtran.bst, version: 1.14 (2015/08/26)
\begin{thebibliography}{100}
\providecommand{\url}[1]{#1}
\csname url@samestyle\endcsname
\providecommand{\newblock}{\relax}
\providecommand{\bibinfo}[2]{#2}
\providecommand{\BIBentrySTDinterwordspacing}{\spaceskip=0pt\relax}
\providecommand{\BIBentryALTinterwordstretchfactor}{4}
\providecommand{\BIBentryALTinterwordspacing}{\spaceskip=\fontdimen2\font plus
\BIBentryALTinterwordstretchfactor\fontdimen3\font minus \fontdimen4\font\relax}
\providecommand{\BIBforeignlanguage}[2]{{%
\expandafter\ifx\csname l@#1\endcsname\relax
\typeout{** WARNING: IEEEtran.bst: No hyphenation pattern has been}%
\typeout{** loaded for the language `#1'. Using the pattern for}%
\typeout{** the default language instead.}%
\else
\language=\csname l@#1\endcsname
\fi
#2}}
\providecommand{\BIBdecl}{\relax}
\BIBdecl

\bibitem{mao2017survey}
Y.~Mao, C.~You, J.~Zhang, K.~Huang, and K.~B. Letaief, ``A survey on mobile edge computing: The communication perspective,'' \emph{IEEE Communications Surveys \& Tutorials}, vol.~19, no.~4, pp. 2322--2358, 2017.

\bibitem{hu2015mobile}
Y.~C. Hu, M.~Patel, D.~Sabella, N.~Sprecher, and V.~Young, ``Mobile edge computing—a key technology towards 5g,'' \emph{ETSI white paper}, vol.~11, no.~11, pp. 1--16, 2015.

\bibitem{ouyang2018follow}
T.~Ouyang, Z.~Zhou, and X.~Chen, ``Follow me at the edge: Mobility-aware dynamic service placement for mobile edge computing,'' \emph{IEEE Journal on Selected Areas in Communications}, vol.~36, no.~10, pp. 2333--2345, 2018.

\bibitem{huda2022survey}
S.~A. Huda and S.~Moh, ``Survey on computation offloading in uav-enabled mobile edge computing,'' \emph{Journal of Network and Computer Applications}, p. 103341, 2022.

\bibitem{zhou2022energy}
H.~Zhou, Z.~Zhang, Y.~Wu, M.~Dong, and V.~C. Leung, ``Energy efficient joint computation offloading and service caching for mobile edge computing: A deep reinforcement learning approach,'' \emph{IEEE Transactions on Green Communications and Networking}, 2022.

\bibitem{ranaweera2021survey}
P.~Ranaweera, A.~D. Jurcut, and M.~Liyanage, ``Survey on multi-access edge computing security and privacy,'' \emph{IEEE Communications Surveys \& Tutorials}, 2021.

\bibitem{threatpost.com}
B.~Bracken, ``Cyberattacks on healthcare spike 45\% since november,'' \url{https://threatpost.com/cyberattacks-healthcare-spike-ransomware/162770/}, accessed August 3, 2021.

\bibitem{chen2019deep}
Y.~Chen, Y.~Zhang, S.~Maharjan, M.~Alam, and T.~Wu, ``Deep learning for secure mobile edge computing in cyber-physical transportation systems,'' \emph{IEEE Network}, vol.~33, no.~4, pp. 36--41, 2019.

\bibitem{benton2013openflow}
K.~Benton, L.~J. Camp, and C.~Small, ``Openflow vulnerability assessment,'' in \emph{Proceedings of the second ACM SIGCOMM workshop on Hot topics in software defined networking}, 2013, pp. 151--152.

\bibitem{pattaranantakul2018nfv}
M.~Pattaranantakul, R.~He, Q.~Song, Z.~Zhang, and A.~Meddahi, ``Nfv security survey: From use case driven threat analysis to state-of-the-art countermeasures,'' \emph{IEEE Communications Surveys \& Tutorials}, vol.~20, no.~4, pp. 3330--3368, 2018.

\bibitem{joseph2013machine}
A.~D. Joseph, P.~Laskov, F.~Roli, J.~D. Tygar, and B.~Nelson, ``Machine learning methods for computer security (dagstuhl perspectives workshop 12371),'' in \emph{Dagstuhl Manifestos}, vol.~3, no.~1.\hskip 1em plus 0.5em minus 0.4em\relax Schloss Dagstuhl-Leibniz-Zentrum fuer Informatik, 2013.

\bibitem{jordan2015machine}
M.~I. Jordan and T.~M. Mitchell, ``Machine learning: Trends, perspectives, and prospects,'' \emph{Science}, vol. 349, no. 6245, pp. 255--260, 2015.

\bibitem{av-test}
AV-TEST, ``Malware,'' \url{http://www.av-test.org/en/statistics/malware/}.

\bibitem{wang2017physical}
N.~Wang, T.~Jiang, S.~Lv, and L.~Xiao, ``Physical-layer authentication based on extreme learning machine,'' \emph{IEEE Communications Letters}, vol.~21, no.~7, pp. 1557--1560, 2017.

\bibitem{manimurugan2021iot}
S.~Manimurugan, ``Iot-fog-cloud model for anomaly detection using improved na{\"\i}ve bayes and principal component analysis,'' \emph{Journal of Ambient Intelligence and Humanized Computing}, pp. 1--10, 2021.

\bibitem{zhao2017intrusion}
G.~Zhao, C.~Zhang, and L.~Zheng, ``Intrusion detection using deep belief network and probabilistic neural network,'' in \emph{2017 IEEE International Conference on Computational Science and Engineering (CSE) and IEEE International Conference on Embedded and Ubiquitous Computing (EUC)}, vol.~1.\hskip 1em plus 0.5em minus 0.4em\relax IEEE, 2017, pp. 639--642.

\bibitem{roopak2019deep}
M.~Roopak, G.~Y. Tian, and J.~Chambers, ``Deep learning models for cyber security in iot networks,'' in \emph{2019 IEEE 9th annual computing and communication workshop and conference (CCWC)}.\hskip 1em plus 0.5em minus 0.4em\relax IEEE, 2019, pp. 0452--0457.

\bibitem{khalid2020macler}
F.~Khalid, S.~R. Hasan, S.~Zia, O.~Hasan, F.~Awwad, and M.~Shafique, ``Macler: Machine learning-based runtime hardware trojan detection in resource-constrained iot edge devices,'' \emph{IEEE Transactions on Computer-Aided Design of Integrated Circuits and Systems}, vol.~39, no.~11, pp. 3748--3761, 2020.

\bibitem{libri2020paella}
A.~Libri, A.~Bartolini, and L.~Benini, ``paella: Edge ai-based real-time malware detection in data centers,'' \emph{IEEE Internet of Things Journal}, vol.~7, no.~10, pp. 9589--9599, 2020.

\bibitem{shi2016edge}
W.~Shi, J.~Cao, Q.~Zhang, Y.~Li, and L.~Xu, ``Edge computing: Vision and challenges,'' \emph{IEEE internet of things journal}, vol.~3, no.~5, pp. 637--646, 2016.

\bibitem{singh2022ai}
A.~Singh, S.~C. Satapathy, A.~Roy, and A.~Gutub, ``Ai-based mobile edge computing for iot: Applications, challenges, and future scope,'' \emph{Arabian Journal for Science and Engineering}, pp. 1--31, 2022.

\bibitem{satyanarayanan2017emergence}
M.~Satyanarayanan, ``The emergence of edge computing,'' \emph{Computer}, vol.~50, no.~1, pp. 30--39, 2017.

\bibitem{varghese2016challenges}
B.~Varghese, N.~Wang, S.~Barbhuiya, P.~Kilpatrick, and D.~S. Nikolopoulos, ``Challenges and opportunities in edge computing,'' in \emph{2016 IEEE International Conference on Smart Cloud (SmartCloud)}.\hskip 1em plus 0.5em minus 0.4em\relax IEEE, 2016, pp. 20--26.

\bibitem{khan2019edge}
W.~Z. Khan, E.~Ahmed, S.~Hakak, I.~Yaqoob, and A.~Ahmed, ``Edge computing: A survey,'' \emph{Future Generation Computer Systems}, vol.~97, pp. 219--235, 2019.

\bibitem{giust2018mec}
F.~Giust, G.~Verin, K.~Antevski, J.~Chou, Y.~Fang, W.~Featherstone, F.~Fontes, D.~Frydman, A.~Li, A.~Manzalini \emph{et~al.}, ``Mec deployments in 4g and evolution towards 5g,'' \emph{ETSI White paper}, vol.~24, no. 2018, pp. 1--24, 2018.

\bibitem{peng2018survey}
K.~Peng, V.~Leung, X.~Xu, L.~Zheng, J.~Wang, and Q.~Huang, ``A survey on mobile edge computing: Focusing on service adoption and provision,'' \emph{Wireless Communications and Mobile Computing}, vol. 2018, 2018.

\bibitem{mach2017mobile}
P.~Mach and Z.~Becvar, ``Mobile edge computing: A survey on architecture and computation offloading,'' \emph{IEEE Communications Surveys \& Tutorials}, vol.~19, no.~3, pp. 1628--1656, 2017.

\bibitem{ali2021multi}
B.~Ali, M.~A. Gregory, and S.~Li, ``Multi-access edge computing architecture, data security and privacy: A review,'' \emph{IEEE Access}, 2021.

\bibitem{ahmed2017mobile}
E.~Ahmed and M.~H. Rehmani, ``Mobile edge computing: opportunities, solutions, and challenges,'' 2017.

\bibitem{ai2018edge}
Y.~Ai, M.~Peng, and K.~Zhang, ``Edge computing technologies for internet of things: a primer,'' \emph{Digital Communications and Networks}, vol.~4, no.~2, pp. 77--86, 2018.

\bibitem{bonomi2011connected}
F.~Bonomi, ``Connected vehicles, the internet of things, and fog computing,'' in \emph{The eighth ACM international workshop on vehicular inter-networking (VANET), Las Vegas, USA}.\hskip 1em plus 0.5em minus 0.4em\relax sn, 2011, pp. 13--15.

\bibitem{fernando2013mobile}
N.~Fernando, S.~W. Loke, and W.~Rahayu, ``Mobile cloud computing: A survey,'' \emph{Future generation computer systems}, vol.~29, no.~1, pp. 84--106, 2013.

\bibitem{subramaniam2019review}
P.~Subramaniam and M.~J. Kaur, ``Review of security in mobile edge computing with deep learning,'' in \emph{2019 Advances in Science and Engineering Technology International Conferences (ASET)}.\hskip 1em plus 0.5em minus 0.4em\relax IEEE, 2019, pp. 1--5.

\bibitem{chen2017deep}
Y.~Chen, Y.~Zhang, and S.~Maharjan, ``Deep learning for secure mobile edge computing,'' \emph{arXiv preprint arXiv:1709.08025}, 2017.

\bibitem{al2020survey}
M.~A. Al-Garadi, A.~Mohamed, A.~K. Al-Ali, X.~Du, I.~Ali, and M.~Guizani, ``A survey of machine and deep learning methods for internet of things (iot) security,'' \emph{IEEE Communications Surveys \& Tutorials}, vol.~22, no.~3, pp. 1646--1685, 2020.

\bibitem{shambour2022progress}
M.~K. Shambour and A.~Gutub, ``Progress of iot research technologies and applications serving hajj and umrah,'' \emph{Arabian Journal for Science and Engineering}, vol.~47, no.~2, pp. 1253--1273, 2022.

\bibitem{singh2021machine}
S.~Singh, R.~Sulthana, T.~Shewale, V.~Chamola, A.~Benslimane, and B.~Sikdar, ``Machine-learning-assisted security and privacy provisioning for edge computing: A survey,'' \emph{IEEE Internet of Things Journal}, vol.~9, no.~1, pp. 236--260, 2021.

\bibitem{giust2017multi}
F.~Giust, X.~Costa-Perez, and A.~Reznik, ``Multi-access edge computing: An overview of etsi mec isg,'' \emph{IEEE 5G Tech Focus}, vol.~1, no.~4, p.~4, 2017.

\bibitem{kekki2018mec}
S.~Kekki, W.~Featherstone, Y.~Fang, P.~Kuure, A.~Li, A.~Ranjan, D.~Purkayastha, F.~Jiangping, D.~Frydman, G.~Verin \emph{et~al.}, ``Mec in 5g networks,'' \emph{ETSI white paper}, vol.~28, pp. 1--28, 2018.

\bibitem{lounis2020attacks}
K.~Lounis and M.~Zulkernine, ``Attacks and defenses in short-range wireless technologies for iot,'' \emph{IEEE Access}, vol.~8, pp. 88\,892--88\,932, 2020.

\bibitem{zanzi2019evolving}
L.~Zanzi, F.~Cirillo, V.~Sciancalepore, F.~Giust, X.~Costa-Perez, S.~Mangiante, and G.~Klas, ``Evolving multi-access edge computing to support enhanced iot deployments,'' \emph{IEEE Communications Standards Magazine}, vol.~3, no.~2, pp. 26--34, 2019.

\bibitem{campolo2018slicing}
C.~Campolo, R.~dos Reis~Fontes, A.~Molinaro, C.~Esteve~Rothenberg, and A.~Iera, ``Slicing on the road: Enabling the automotive vertical through 5g network softwarization,'' \emph{Sensors}, vol.~18, no.~12, p. 4435, 2018.

\bibitem{schiller2018cds}
E.~Schiller, N.~Nikaein, E.~Kalogeiton, M.~Gasparyan, and T.~Braun, ``Cds-mec: Nfv/sdn-based application management for mec in 5g systems,'' \emph{Computer Networks}, vol. 135, pp. 96--107, 2018.

\bibitem{cimpanu2016you}
C.~Cimpanu, ``You can now rent a mirai botnet of 400,000 bots,'' \emph{BleepingComputer. com}, vol.~24, 2016.

\bibitem{zeng2011web}
D.~Zeng, S.~Guo, and Z.~Cheng, ``The web of things: A survey,'' \emph{JCM}, vol.~6, no.~6, pp. 424--438, 2011.

\bibitem{he2021game}
Q.~He, C.~Wang, G.~Cui, B.~Li, R.~Zhou, Q.~Zhou, Y.~Xiang, H.~Jin, and Y.~Yang, ``A game-theoretical approach for mitigatingedge ddos attack,'' \emph{IEEE Transactions on Dependable and Secure Computing}, 2021.

\bibitem{huang2018identifying}
K.-W. Huang and H.-M. Wang, ``Identifying the fake base station: A location based approach,'' \emph{IEEE Communications Letters}, vol.~22, no.~8, pp. 1604--1607, 2018.

\bibitem{farooqi2019smart}
N.~Farooqi, A.~Gutub, and M.~O. Khozium, ``Smart community challenges: enabling iot/m2m technology case study,'' \emph{Life Sci J}, vol.~16, no.~7, pp. 11--17, 2019.

\bibitem{liu2015good}
X.~Liu, Z.~Zhou, W.~Diao, Z.~Li, and K.~Zhang, ``When good becomes evil: Keystroke inference with smartwatch,'' in \emph{Proceedings of the 22nd ACM SIGSAC Conference on Computer and Communications Security}, 2015, pp. 1273--1285.

\bibitem{fernandes2016security}
E.~Fernandes, J.~Jung, and A.~Prakash, ``Security analysis of emerging smart home applications,'' in \emph{2016 IEEE symposium on security and privacy (SP)}.\hskip 1em plus 0.5em minus 0.4em\relax IEEE, 2016, pp. 636--654.

\bibitem{chen2017machine}
S.~Chen, R.~Ma, H.-H. Chen, H.~Zhang, W.~Meng, and J.~Liu, ``Machine-to-machine communications in ultra-dense networks—a survey,'' \emph{IEEE Communications Surveys \& Tutorials}, vol.~19, no.~3, pp. 1478--1503, 2017.

\bibitem{lu2018managing}
X.~Lu, D.~Niyato, N.~Privault, H.~Jiang, and P.~Wang, ``Managing physical layer security in wireless cellular networks: A cyber insurance approach,'' \emph{IEEE Journal on Selected Areas in Communications}, vol.~36, no.~7, pp. 1648--1661, 2018.

\bibitem{costa20175g}
X.~Costa-Perez, A.~Garcia-Saavedra, X.~Li, T.~Deiss, A.~De~La~Oliva, A.~Di~Giglio, P.~Iovanna, and A.~Moored, ``5g-crosshaul: An sdn/nfv integrated fronthaul/backhaul transport network architecture,'' \emph{IEEE wireless communications}, vol.~24, no.~1, pp. 38--45, 2017.

\bibitem{ahmad2019security}
I.~Ahmad, S.~Shahabuddin, T.~Kumar, J.~Okwuibe, A.~Gurtov, and M.~Ylianttila, ``Security for 5g and beyond,'' \emph{IEEE Communications Surveys \& Tutorials}, vol.~21, no.~4, pp. 3682--3722, 2019.

\bibitem{kreutz2013towards}
D.~Kreutz, F.~M. Ramos, and P.~Verissimo, ``Towards secure and dependable software-defined networks,'' in \emph{Proceedings of the second ACM SIGCOMM workshop on Hot topics in software defined networking}, 2013, pp. 55--60.

\bibitem{shin2013attacking}
S.~Shin and G.~Gu, ``Attacking software-defined networks: A first feasibility study,'' in \emph{Proceedings of the second ACM SIGCOMM workshop on Hot topics in software defined networking}, 2013, pp. 165--166.

\bibitem{fonseca2012replication}
P.~Fonseca, R.~Bennesby, E.~Mota, and A.~Passito, ``A replication component for resilient openflow-based networking,'' in \emph{2012 IEEE Network operations and management symposium}.\hskip 1em plus 0.5em minus 0.4em\relax IEEE, 2012, pp. 933--939.

\bibitem{nadeau2011software}
T.~Nadeau and P.~Pan, ``Software driven networks problem statement,'' \emph{Network Working Group Internet-Draft, Sep}, vol.~30, 2011.

\bibitem{shaghaghi2020software}
A.~Shaghaghi, M.~A. Kaafar, R.~Buyya, and S.~Jha, ``Software-defined network (sdn) data plane security: issues, solutions, and future directions,'' \emph{Handbook of Computer Networks and Cyber Security}, pp. 341--387, 2020.

\bibitem{scott2015survey}
S.~Scott-Hayward, S.~Natarajan, and S.~Sezer, ``A survey of security in software defined networks,'' \emph{IEEE Communications Surveys \& Tutorials}, vol.~18, no.~1, pp. 623--654, 2015.

\bibitem{szefer2011eliminating}
J.~Szefer, E.~Keller, R.~B. Lee, and J.~Rexford, ``Eliminating the hypervisor attack surface for a more secure cloud,'' in \emph{Proceedings of the 18th ACM conference on Computer and communications security}, 2011, pp. 401--412.

\bibitem{dubrulle2015blind}
P.~Dubrulle, R.~Sirdey, P.~Dore, M.~Aichouch, and E.~Ohayon, ``Blind hypervision to protect virtual machine privacy against hypervisor escape vulnerabilities,'' in \emph{2015 IEEE 13th International Conference on Industrial Informatics (INDIN)}.\hskip 1em plus 0.5em minus 0.4em\relax IEEE, 2015, pp. 1394--1399.

\bibitem{lee2012efficient}
K.~C. Lee, B.~Zheng, C.~Chen, and C.-Y. Chow, ``Efficient index-based approaches for skyline queries in location-based applications,'' \emph{IEEE Transactions on Knowledge and Data Engineering}, vol.~25, no.~11, pp. 2507--2520, 2012.

\bibitem{fawaz2014location}
K.~Fawaz and K.~G. Shin, ``Location privacy protection for smartphone users,'' in \emph{Proceedings of the 2014 ACM SIGSAC Conference on Computer and Communications Security}, 2014, pp. 239--250.

\bibitem{zhang2018data}
J.~Zhang, B.~Chen, Y.~Zhao, X.~Cheng, and F.~Hu, ``Data security and privacy-preserving in edge computing paradigm: Survey and open issues,'' \emph{IEEE access}, vol.~6, pp. 18\,209--18\,237, 2018.

\bibitem{liu2022enhanced}
G.~Liu, H.~Zhao, F.~Fan, G.~Liu, Q.~Xu, and S.~Nazir, ``An enhanced intrusion detection model based on improved knn in wsns,'' \emph{Sensors}, vol.~22, no.~4, p. 1407, 2022.

\bibitem{syarif2017intrusion}
A.~R. Syarif and W.~Gata, ``Intrusion detection system using hybrid binary pso and k-nearest neighborhood algorithm,'' in \emph{2017 11th International Conference on Information \& Communication Technology and System (ICTS)}.\hskip 1em plus 0.5em minus 0.4em\relax IEEE, 2017, pp. 181--186.

\bibitem{shapoorifard2017intrusion}
H.~Shapoorifard and P.~Shamsinejad, ``Intrusion detection using a novel hybrid method incorporating an improved knn,'' \emph{Int. J. Comput. Appl}, vol. 173, no.~1, pp. 5--9, 2017.

\bibitem{tan2020lightweight}
Y.~Tan, W.~Wu, J.~Liu, H.~Wang, and M.~Xian, ``Lightweight edge-based knn privacy-preserving classification scheme in cloud computing circumstance,'' \emph{Concurrency and Computation: Practice and Experience}, vol.~32, no.~19, p. e5804, 2020.

\bibitem{chen2020privacy}
Q.~Chen, K.~Fan, K.~Zhang, H.~Wang, H.~Li, and Y.~Yang, ``Privacy-preserving searchable encryption in the intelligent edge computing,'' \emph{Computer Communications}, vol. 164, pp. 31--41, 2020.

\bibitem{chen2020automated}
S.~Chen, Z.~Pang, H.~Wen, K.~Yu, T.~Zhang, and Y.~Lu, ``Automated labeling and learning for physical layer authentication against clone node and sybil attacks in industrial wireless edge networks,'' \emph{IEEE Transactions on Industrial Informatics}, vol.~17, no.~3, pp. 2041--2051, 2020.

\bibitem{kumari2017semi}
V.~V. Kumari and P.~R.~K. Varma, ``A semi-supervised intrusion detection system using active learning svm and fuzzy c-means clustering,'' in \emph{2017 International Conference on I-SMAC (IoT in Social, Mobile, Analytics and Cloud)(I-SMAC)}.\hskip 1em plus 0.5em minus 0.4em\relax IEEE, 2017, pp. 481--485.

\bibitem{xu2016hadm}
L.~Xu, D.~Zhang, N.~Jayasena, and J.~Cavazos, ``Hadm: Hybrid analysis for detection of malware,'' in \emph{Proceedings of SAI Intelligent Systems Conference}.\hskip 1em plus 0.5em minus 0.4em\relax Springer, 2016, pp. 702--724.

\bibitem{majumder2020smart}
A.~J. Majumder, J.~D. Miller, C.~B. Veilleux, and A.~A. Asif, ``Smart-power: A smart cyber-physical system to detect iot security threat through behavioral power profiling,'' in \emph{2020 IEEE 44th Annual Computers, Software, and Applications Conference (COMPSAC)}.\hskip 1em plus 0.5em minus 0.4em\relax IEEE, 2020, pp. 1041--1049.

\bibitem{laguduva2019machine}
V.~Laguduva, S.~A. Islam, S.~Aakur, S.~Katkoori, and R.~Karam, ``Machine learning based iot edge node security attack and countermeasures,'' in \emph{2019 IEEE Computer Society Annual Symposium on VLSI (ISVLSI)}.\hskip 1em plus 0.5em minus 0.4em\relax IEEE, 2019, pp. 670--675.

\bibitem{sinclair1999application}
C.~Sinclair, L.~Pierce, and S.~Matzner, ``An application of machine learning to network intrusion detection,'' in \emph{Proceedings 15th Annual Computer Security Applications Conference (ACSAC'99)}.\hskip 1em plus 0.5em minus 0.4em\relax IEEE, 1999, pp. 371--377.

\bibitem{wu2020ensemble}
X.~Wu, X.~Xu, F.~Dai, J.~Gao, G.~Ji, and L.~Qi, ``An ensemble of random decision trees with personalized privacy preservation in edge-cloud computing,'' in \emph{2020 International Conferences on Internet of Things (iThings) and IEEE Green Computing and Communications (GreenCom) and IEEE Cyber, Physical and Social Computing (CPSCom) and IEEE Smart Data (SmartData) and IEEE Congress on Cybermatics (Cybermatics)}.\hskip 1em plus 0.5em minus 0.4em\relax IEEE, 2020, pp. 779--786.

\bibitem{sangaiah2019enforcing}
A.~K. Sangaiah, D.~V. Medhane, T.~Han, M.~S. Hossain, and G.~Muhammad, ``Enforcing position-based confidentiality with machine learning paradigm through mobile edge computing in real-time industrial informatics,'' \emph{IEEE Transactions on Industrial Informatics}, vol.~15, no.~7, pp. 4189--4196, 2019.

\bibitem{liu2020deep}
Z.~Liu, N.~Su, Y.~Qin, J.~Lu, and X.~Li, ``A deep random forest model on spark for network intrusion detection,'' \emph{Mobile Information Systems}, vol. 2020, 2020.

\bibitem{singh2019optimization}
R.~K. Singh, S.~Dalal, V.~K. Chauhan, and D.~Kumar, ``Optimization of far in intrusion detection system by using random forest algorithm,'' in \emph{Proceedings of 2nd International Conference on Advanced Computing and Software Engineering (ICACSE)}, 2019.

\bibitem{guowei2021research}
Z.~Guowei, Y.~Hui, Y.~ZHUANG, G.~Yue, X.~ZHANG, and Q.~Shuang, ``Research on network intrusion detection method of power system based on random forest algorithm,'' in \emph{2021 13th International Conference on Measuring Technology and Mechatronics Automation (ICMTMA)}.\hskip 1em plus 0.5em minus 0.4em\relax IEEE, 2021, pp. 374--379.

\bibitem{hasan2019attack}
M.~Hasan, M.~M. Islam, M.~I.~I. Zarif, and M.~Hashem, ``Attack and anomaly detection in iot sensors in iot sites using machine learning approaches,'' \emph{Internet of Things}, vol.~7, p. 100059, 2019.

\bibitem{wu2019detecting}
M.~Wu, Z.~Song, and Y.~B. Moon, ``Detecting cyber-physical attacks in cybermanufacturing systems with machine learning methods,'' \emph{Journal of intelligent manufacturing}, vol.~30, no.~3, pp. 1111--1123, 2019.

\bibitem{zhang2020dataset}
W.~Zhang, J.~Wang, G.~Han, S.~Huang, Y.~Feng, and L.~Shu, ``A dataset accuracy weighted random forest algorithm for iot fault detection based on edge computing and blockchain,'' \emph{IEEE Internet of Things Journal}, 2020.

\bibitem{tian2019distributed}
Z.~Tian, C.~Luo, J.~Qiu, X.~Du, and M.~Guizani, ``A distributed deep learning system for web attack detection on edge devices,'' \emph{IEEE Transactions on Industrial Informatics}, vol.~16, no.~3, pp. 1963--1971, 2019.

\bibitem{ran2020cloud}
Q.~Ran, X.~Ju, X.~Zhang, and Y.~Zhang, ``Cloud edge cooperative attack recognition based on cnn,'' in \emph{Journal of Physics: Conference Series}, vol. 1693, no.~1.\hskip 1em plus 0.5em minus 0.4em\relax IOP Publishing, 2020, p. 012143.

\bibitem{tian2019lep}
Y.~Tian, J.~Yuan, S.~Yu, and Y.~Hou, ``Lep-cnn: A lightweight edge device assisted privacy-preserving cnn inference solution for iot,'' \emph{arXiv preprint arXiv:1901.04100}, 2019.

\bibitem{radhakrishnan2021deep}
G.~Radhakrishnan, K.~Srinivasan, S.~Maheswaran, K.~Mohanasundaram, D.~Palanikkumar, and A.~Vidyarthi, ``A deep-rnn and meta-heuristic feature selection approach for iot malware detection,'' \emph{Materials Today: Proceedings}, 2021.

\bibitem{li2019complex}
Q.~Li, S.~Meng, S.~Zhang, J.~Hou, and L.~Qi, ``Complex attack linkage decision-making in edge computing networks,'' \emph{IEEE Access}, vol.~7, pp. 12\,058--12\,072, 2019.

\bibitem{han2018kclp}
G.~Han, H.~Wang, M.~Guizani, S.~Chan, and W.~Zhang, ``Kclp: A k-means cluster-based location privacy protection scheme in wsns for iot,'' \emph{IEEE Wireless Communications}, vol.~25, no.~6, pp. 84--90, 2018.

\bibitem{liu2020privacy}
Y.~Liu, Z.~Ma, Z.~Yan, Z.~Wang, X.~Liu, and J.~Ma, ``Privacy-preserving federated k-means for proactive caching in next generation cellular networks,'' \emph{Information Sciences}, vol. 521, pp. 14--31, 2020.

\bibitem{huong2021efficient}
T.~T. Huong, T.~P. Bac, D.~M. Long, B.~D. Thang, T.~D. Luong, and N.~T. Binh, ``An efficient low complexity edge-cloud framework for security in iot networks,'' in \emph{2020 IEEE Eighth International Conference on Communications and Electronics (ICCE)}.\hskip 1em plus 0.5em minus 0.4em\relax IEEE, 2021, pp. 533--539.

\bibitem{osia2018private}
S.~A. Osia, A.~S. Shamsabadi, A.~Taheri, H.~R. Rabiee, and H.~Haddadi, ``Private and scalable personal data analytics using hybrid edge-to-cloud deep learning,'' \emph{Computer}, vol.~51, no.~5, pp. 42--49, 2018.

\bibitem{benchea2014combining}
R.~Benchea and D.~T. Gavrilu{\c{t}}, ``Combining restricted boltzmann machine and one side perceptron for malware detection,'' in \emph{International conference on conceptual structures}.\hskip 1em plus 0.5em minus 0.4em\relax Springer, 2014, pp. 93--103.

\bibitem{tian2020intrusion}
Q.~Tian, D.~Han, K.-C. Li, X.~Liu, L.~Duan, and A.~Castiglione, ``An intrusion detection approach based on improved deep belief network,'' \emph{Applied Intelligence}, vol.~50, pp. 3162--3178, 2020.

\bibitem{schneible2017anomaly}
J.~Schneible and A.~Lu, ``Anomaly detection on the edge,'' in \emph{MILCOM 2017-2017 IEEE Military Communications Conference (MILCOM)}.\hskip 1em plus 0.5em minus 0.4em\relax IEEE, 2017, pp. 678--682.

\bibitem{sadaf2020intrusion}
K.~Sadaf and J.~Sultana, ``Intrusion detection based on autoencoder and isolation forest in fog computing,'' \emph{IEEE Access}, vol.~8, pp. 167\,059--167\,068, 2020.

\bibitem{wang2019sample}
X.~Wang, I.~Yang, and S.-H. Ahn, ``Sample efficient home power anomaly detection in real time using semi-supervised learning,'' \emph{IEEE Access}, vol.~7, pp. 139\,712--139\,725, 2019.

\bibitem{hiromoto2017secure}
R.~E. Hiromoto, M.~Haney, and A.~Vakanski, ``A secure architecture for iot with supply chain risk management,'' in \emph{2017 9th IEEE International Conference on Intelligent Data Acquisition and Advanced Computing Systems: Technology and Applications (IDAACS)}, vol.~1.\hskip 1em plus 0.5em minus 0.4em\relax IEEE, 2017, pp. 431--435.

\bibitem{yoon2020pufgan}
J.~Yoon and H.~Lee, ``Pufgan: Embracing a self-adversarial agent for building a defensible edge security architecture,'' in \emph{IEEE INFOCOM 2020-IEEE Conference on Computer Communications}.\hskip 1em plus 0.5em minus 0.4em\relax IEEE, 2020, pp. 904--913.

\bibitem{xiao2017cloud}
L.~Xiao, Y.~Li, X.~Huang, and X.~Du, ``Cloud-based malware detection game for mobile devices with offloading,'' \emph{IEEE Transactions on Mobile Computing}, vol.~16, no.~10, pp. 2742--2750, 2017.

\bibitem{aref2017multi}
M.~A. Aref, S.~K. Jayaweera, and S.~Machuzak, ``Multi-agent reinforcement learning based cognitive anti-jamming,'' in \emph{2017 IEEE Wireless Communications and Networking Conference (WCNC)}.\hskip 1em plus 0.5em minus 0.4em\relax IEEE, 2017, pp. 1--6.

\bibitem{xiao2018security}
L.~Xiao, X.~Wan, C.~Dai, X.~Du, X.~Chen, and M.~Guizani, ``Security in mobile edge caching with reinforcement learning,'' \emph{IEEE Wireless Communications}, vol.~25, no.~3, pp. 116--122, 2018.

\bibitem{li2021explainable}
H.~Li, J.~Wu, H.~Xu, G.~Li, and M.~Guizani, ``Explainable intelligence-driven defense mechanism against advanced persistent threats: A joint edge game and ai approach,'' \emph{IEEE Transactions on Dependable and Secure Computing}, vol.~19, no.~2, pp. 757--775, 2021.

\bibitem{thames2006hybrid}
J.~L. Thames, R.~Abler, and A.~Saad, ``Hybrid intelligent systems for network security,'' in \emph{Proceedings of the 44th annual Southeast regional conference}, 2006, pp. 286--289.

\bibitem{wright2004privacy}
R.~Wright and Z.~Yang, ``Privacy-preserving bayesian network structure computation on distributed heterogeneous data,'' in \emph{Proceedings of the tenth ACM SIGKDD international conference on Knowledge discovery and data mining}, 2004, pp. 713--718.

\bibitem{kim2010malware}
K.~Kim and B.-R. Moon, ``Malware detection based on dependency graph using hybrid genetic algorithm,'' in \emph{Proceedings of the 12th annual conference on Genetic and evolutionary computation}, 2010, pp. 1211--1218.

\bibitem{roy2022analysis}
P.~K. Roy, S.~Saumya, J.~P. Singh, S.~Banerjee, and A.~Gutub, ``Analysis of community question-answering issues via machine learning and deep learning: State-of-the-art review,'' \emph{CAAI Transactions on Intelligence Technology}, 2022.

\bibitem{santhadevi2022eidima}
D.~Santhadevi and B.~Janet, ``Eidima: Edge-based intrusion detection of iot malware attacks using decision tree-based boosting algorithms,'' in \emph{High Performance Computing and Networking}.\hskip 1em plus 0.5em minus 0.4em\relax Springer, 2022, pp. 449--459.

\bibitem{farukee2020ddos}
M.~B. Farukee, M.~Z. Shabit, M.~R. Haque, and A.~S. Sattar, ``Ddos attack detection in iot networks using deep learning models combined with random forest as feature selector,'' in \emph{International Conference on Advances in Cyber Security}.\hskip 1em plus 0.5em minus 0.4em\relax Springer, 2020, pp. 118--134.

\bibitem{alrashdi2019ad}
I.~Alrashdi, A.~Alqazzaz, E.~Aloufi, R.~Alharthi, M.~Zohdy, and H.~Ming, ``Ad-iot: Anomaly detection of iot cyberattacks in smart city using machine learning,'' in \emph{2019 IEEE 9th Annual Computing and Communication Workshop and Conference (CCWC)}.\hskip 1em plus 0.5em minus 0.4em\relax IEEE, 2019, pp. 0305--0310.

\bibitem{wang2022lightlog}
Z.~Wang, J.~Tian, H.~Fang, L.~Chen, and J.~Qin, ``Lightlog: A lightweight temporal convolutional network for log anomaly detection on the edge,'' \emph{Computer Networks}, vol. 203, p. 108616, 2022.

\bibitem{li2019edge}
E.~Li, L.~Zeng, Z.~Zhou, and X.~Chen, ``Edge ai: On-demand accelerating deep neural network inference via edge computing,'' \emph{IEEE Transactions on Wireless Communications}, vol.~19, no.~1, pp. 447--457, 2019.

\bibitem{zhao2018deepthings}
Z.~Zhao, K.~M. Barijough, and A.~Gerstlauer, ``Deepthings: Distributed adaptive deep learning inference on resource-constrained iot edge clusters,'' \emph{IEEE Transactions on Computer-Aided Design of Integrated Circuits and Systems}, vol.~37, no.~11, pp. 2348--2359, 2018.

\bibitem{guizani2020network}
N.~Guizani and A.~Ghafoor, ``A network function virtualization system for detecting malware in large iot based networks,'' \emph{IEEE Journal on Selected Areas in Communications}, vol.~38, no.~6, pp. 1218--1228, 2020.

\bibitem{arivukarasi2020performance}
M.~Arivukarasi and A.~Antonidoss, ``Performance analysis of malicious url detection by using rnn and lstm,'' in \emph{2020 Fourth International Conference on Computing Methodologies and Communication (ICCMC)}.\hskip 1em plus 0.5em minus 0.4em\relax IEEE, 2020, pp. 454--458.

\bibitem{chen2020rnn}
C.-Y. Chen, L.-A. Chen, Y.-Z. Cai, and M.-H. Tsai, ``Rnn-based ddos detection in iot scenario,'' in \emph{2020 International Computer Symposium (ICS)}.\hskip 1em plus 0.5em minus 0.4em\relax IEEE, 2020, pp. 448--453.

\bibitem{diro2018leveraging}
A.~Diro and N.~Chilamkurti, ``Leveraging lstm networks for attack detection in fog-to-things communications,'' \emph{IEEE Communications Magazine}, vol.~56, no.~9, pp. 124--130, 2018.

\bibitem{ghahramani2003unsupervised}
Z.~Ghahramani, ``Unsupervised learning,'' in \emph{Summer School on Machine Learning}.\hskip 1em plus 0.5em minus 0.4em\relax Springer, 2003, pp. 72--112.

\bibitem{karoly2018unsupervised}
A.~I. K{\'a}roly, R.~Full{\'e}r, and P.~Galambos, ``Unsupervised clustering for deep learning: A tutorial survey,'' \emph{Acta Polytechnica Hungarica}, vol.~15, no.~8, pp. 29--53, 2018.

\bibitem{wason2020integrated}
R.~Wason, ``An integrated casb implementation model to enhance enterprise cloud security,'' 2020.

\bibitem{hou2016droiddelver}
S.~Hou, A.~Saas, Y.~Ye, and L.~Chen, ``Droiddelver: An android malware detection system using deep belief network based on api call blocks,'' in \emph{International conference on web-age information management}.\hskip 1em plus 0.5em minus 0.4em\relax Springer, 2016, pp. 54--66.

\bibitem{zhu2005semi}
X.~J. Zhu, ``Semi-supervised learning literature survey,'' 2005.

\bibitem{rathore2018semi}
S.~Rathore and J.~H. Park, ``Semi-supervised learning based distributed attack detection framework for iot,'' \emph{Applied Soft Computing}, vol.~72, pp. 79--89, 2018.

\bibitem{gong2014sybilbelief}
N.~Z. Gong, M.~Frank, and P.~Mittal, ``Sybilbelief: A semi-supervised learning approach for structure-based sybil detection,'' \emph{IEEE Transactions on Information Forensics and Security}, vol.~9, no.~6, pp. 976--987, 2014.

\bibitem{li2020gan}
M.~Li, J.~Lin, Y.~Ding, Z.~Liu, J.-Y. Zhu, and S.~Han, ``Gan compression: Efficient architectures for interactive conditional gans,'' in \emph{Proceedings of the IEEE/CVF Conference on Computer Vision and Pattern Recognition}, 2020, pp. 5284--5294.

\bibitem{abdel2021semi}
M.~Abdel-Basset, H.~Hawash, R.~K. Chakrabortty, and M.~J. Ryan, ``Semi-supervised spatio-temporal deep learning for intrusions detection in iot networks,'' \emph{IEEE Internet of Things Journal}, 2021.

\bibitem{qian2020reinforcement}
Y.~Qian, R.~Wang, J.~Wu, B.~Tan, and H.~Ren, ``Reinforcement learning-based optimal computing and caching in mobile edge network,'' \emph{IEEE Journal on Selected Areas in Communications}, vol.~38, no.~10, pp. 2343--2355, 2020.

\bibitem{wang2020multi}
X.~Wang, Z.~Ning, and S.~Guo, ``Multi-agent imitation learning for pervasive edge computing: a decentralized computation offloading algorithm,'' \emph{IEEE Transactions on Parallel and Distributed Systems}, vol.~32, no.~2, pp. 411--425, 2020.

\bibitem{chen2022physical}
L.~Chen, S.~Tang, V.~Balasubramanian, J.~Xia, F.~Zhou, and L.~Fan, ``Physical-layer security based mobile edge computing for emerging cyber physical systems,'' \emph{Computer Communications}, vol. 194, pp. 180--188, 2022.

\bibitem{qiu2020distributed}
X.~Qiu, W.~Zhang, W.~Chen, and Z.~Zheng, ``Distributed and collective deep reinforcement learning for computation offloading: A practical perspective,'' \emph{IEEE Transactions on Parallel and Distributed Systems}, vol.~32, no.~5, pp. 1085--1101, 2020.

\bibitem{rlblogpost}
A.~Irpan, ``Deep reinforcement learning doesn't work yet,'' \url{https://www.alexirpan.com/2018/02/14/rl-hard.html}, 2018.

\bibitem{pourret2008bayesian}
O.~Pourret, P.~Na{\"\i}m, and B.~Marcot, \emph{Bayesian networks: a practical guide to applications}.\hskip 1em plus 0.5em minus 0.4em\relax John Wiley \& Sons, 2008.

\bibitem{frigault2008measuring}
M.~Frigault, L.~Wang, A.~Singhal, and S.~Jajodia, ``Measuring network security using dynamic bayesian network,'' in \emph{Proceedings of the 4th ACM workshop on Quality of protection}, 2008, pp. 23--30.

\bibitem{frigault2008measuring2}
M.~Frigault and L.~Wang, ``Measuring network security using bayesian network-based attack graphs,'' in \emph{2008 32nd Annual IEEE International Computer Software and Applications Conference}.\hskip 1em plus 0.5em minus 0.4em\relax IEEE, 2008, pp. 698--703.

\bibitem{chen2019clustering}
Y.~Chen, H.~Wen, J.~Wu, H.~Song, A.~Xu, Y.~Jiang, T.~Zhang, and Z.~Wang, ``Clustering based physical-layer authentication in edge computing systems with asymmetric resources,'' \emph{Sensors}, vol.~19, no.~8, p. 1926, 2019.

\bibitem{liao2019multiuser}
R.-F. Liao, H.~Wen, S.~Chen, F.~Xie, F.~Pan, J.~Tang, and H.~Song, ``Multiuser physical layer authentication in internet of things with data augmentation,'' \emph{IEEE Internet of Things Journal}, vol.~7, no.~3, pp. 2077--2088, 2019.

\bibitem{kim2018squeezed}
D.~Kim, H.~Yang, M.~Chung, S.~Cho, H.~Kim, M.~Kim, K.~Kim, and E.~Kim, ``Squeezed convolutional variational autoencoder for unsupervised anomaly detection in edge device industrial internet of things,'' in \emph{2018 international conference on information and computer technologies (icict)}.\hskip 1em plus 0.5em minus 0.4em\relax IEEE, 2018, pp. 67--71.

\bibitem{li2018ai}
J.~Li, Z.~Zhao, R.~Li, and H.~Zhang, ``Ai-based two-stage intrusion detection for software defined iot networks,'' \emph{IEEE Internet of Things Journal}, vol.~6, no.~2, pp. 2093--2102, 2018.

\bibitem{tang2016deep}
T.~A. Tang, L.~Mhamdi, D.~McLernon, S.~A.~R. Zaidi, and M.~Ghogho, ``Deep learning approach for network intrusion detection in software defined networking,'' in \emph{2016 international conference on wireless networks and mobile communications (WINCOM)}.\hskip 1em plus 0.5em minus 0.4em\relax IEEE, 2016, pp. 258--263.

\bibitem{kirutika2019controller}
K.~Kirutika, V.~Vetriselvi, R.~Parthasarathi, and G.~S.~V. Rao, ``Controller monitoring system in software defined networks using random forest algorithm,'' in \emph{2019 International Carnahan Conference on Security Technology (ICCST)}.\hskip 1em plus 0.5em minus 0.4em\relax IEEE, 2019, pp. 1--6.

\bibitem{pan2020justinian}
X.~Pan, M.~Zhang, D.~Wu, Q.~Xiao, S.~Ji, and Z.~Yang, ``Justinian's gaavernor: Robust distributed learning with gradient aggregation agent,'' in \emph{29th $\{$USENIX$\}$ Security Symposium ($\{$USENIX$\}$ Security 20)}, 2020, pp. 1641--1658.

\bibitem{kozik2018scalable}
R.~Kozik, M.~Chora{\'s}, M.~Ficco, and F.~Palmieri, ``A scalable distributed machine learning approach for attack detection in edge computing environments,'' \emph{Journal of Parallel and Distributed Computing}, vol. 119, pp. 18--26, 2018.

\bibitem{alassaf2019enhancing}
N.~Alassaf, A.~Gutub, S.~A. Parah, and M.~Al~Ghamdi, ``Enhancing speed of simon: A light-weight-cryptographic algorithm for iot applications,'' \emph{Multimedia Tools and Applications}, vol.~78, no.~23, pp. 32\,633--32\,657, 2019.

\bibitem{alassaf2019simulating}
N.~Alassaf and A.~Gutub, ``Simulating light-weight-cryptography implementation for iot healthcare data security applications,'' \emph{International Journal of E-Health and Medical Communications (IJEHMC)}, vol.~10, no.~4, pp. 1--15, 2019.

\bibitem{kheshaifaty2021engineering}
N.~Kheshaifaty and A.~Gutub, ``Engineering graphical captcha and aes crypto hash functions for secure online authentication,'' \emph{Journal of Engineering Research}, 2021.

\bibitem{farnaaz2016random}
N.~Farnaaz and M.~Jabbar, ``Random forest modeling for network intrusion detection system,'' \emph{Procedia Computer Science}, vol.~89, pp. 213--217, 2016.

\bibitem{huang2019lightweight}
K.~Huang, X.~Liu, S.~Fu, D.~Guo, and M.~Xu, ``A lightweight privacy-preserving cnn feature extraction framework for mobile sensing,'' \emph{IEEE Transactions on Dependable and Secure Computing}, 2019.

\bibitem{yu2018federated}
Z.~Yu, J.~Hu, G.~Min, H.~Lu, Z.~Zhao, H.~Wang, and N.~Georgalas, ``Federated learning based proactive content caching in edge computing,'' in \emph{2018 IEEE Global Communications Conference (GLOBECOM)}.\hskip 1em plus 0.5em minus 0.4em\relax IEEE, 2018, pp. 1--6.

\bibitem{zhao2020privacy}
Y.~Zhao, J.~Zhao, L.~Jiang, R.~Tan, D.~Niyato, Z.~Li, L.~Lyu, and Y.~Liu, ``Privacy-preserving blockchain-based federated learning for iot devices,'' \emph{IEEE Internet of Things Journal}, vol.~8, no.~3, pp. 1817--1829, 2020.

\bibitem{wu2021pecam}
H.~Wu, X.~Tian, M.~Li, Y.~Liu, G.~Ananthanarayanan, F.~Xu, and S.~Zhong, ``Pecam: privacy-enhanced video streaming and analytics via securely-reversible transformation,'' in \emph{Proceedings of the 27th Annual International Conference on Mobile Computing and Networking}, 2021, pp. 229--241.

\bibitem{zhao2020p}
P.~Zhao, H.~Huang, X.~Zhao, and D.~Huang, ``P 3: Privacy-preserving scheme against poisoning attacks in mobile-edge computing,'' \emph{IEEE Transactions on Computational Social Systems}, vol.~7, no.~3, pp. 818--826, 2020.

\bibitem{mao2018privacy}
Y.~Mao, S.~Yi, Q.~Li, J.~Feng, F.~Xu, and S.~Zhong, ``A privacy-preserving deep learning approach for face recognition with edge computing,'' in \emph{Proc. USENIX Workshop Hot Topics Edge Comput.(HotEdge)}, 2018, pp. 1--6.

\bibitem{lu2020privacy}
X.~Lu, Y.~Liao, P.~Lio, and P.~Hui, ``Privacy-preserving asynchronous federated learning mechanism for edge network computing,'' \emph{IEEE Access}, vol.~8, pp. 48\,970--48\,981, 2020.

\bibitem{alotaibi2019secure}
M.~Alotaibi, D.~Al-hendi, B.~Alroithy, M.~AlGhamdi, and A.~Gutub, ``Secure mobile computing authentication utilizing hash, cryptography and steganography combination,'' \emph{Journal of Information Security and Cybercrimes Research}, vol.~2, no.~1, pp. 73--82, 2019.

\bibitem{shambour2021personal}
M.~Shambour and A.~Gutub, ``Personal privacy evaluation of smart devices applications serving hajj and umrah rituals,'' \emph{Journal of Engineering Research}, 2021.

\bibitem{samkari2019protecting}
H.~Samkari and A.~Gutub, ``Protecting medical records against cybercrimes within hajj period by 3-layer security,'' \emph{Recent Trends Inf Technol Appl}, vol.~2, no.~3, pp. 1--21, 2019.

\bibitem{almutairi2019image}
S.~Almutairi, A.~Gutub, and M.~Al-Ghamdi, ``Image steganography to facilitate online students account system,'' \emph{Rev. Bus. Technol. Res}, vol.~16, no.~2, pp. 43--49, 2019.

\bibitem{albers2002security}
P.~Albers, O.~Camp, J.-M. Percher, B.~Jouga, L.~Me, and R.~S. Puttini, ``Security in ad hoc networks: a general intrusion detection architecture enhancing trust based approaches.'' in \emph{Wireless Information Systems}, 2002, pp. 1--12.

\bibitem{porambage2018survey}
P.~Porambage, J.~Okwuibe, M.~Liyanage, M.~Ylianttila, and T.~Taleb, ``Survey on multi-access edge computing for internet of things realization,'' \emph{IEEE Communications Surveys \& Tutorials}, vol.~20, no.~4, pp. 2961--2991, 2018.

\bibitem{zhang2021elf}
W.~Zhang, Z.~He, L.~Liu, Z.~Jia, Y.~Liu, M.~Gruteser, D.~Raychaudhuri, and Y.~Zhang, ``Elf: accelerate high-resolution mobile deep vision with content-aware parallel offloading,'' in \emph{Proceedings of the 27th Annual International Conference on Mobile Computing and Networking}, 2021, pp. 201--214.

\bibitem{mohammed2020distributed}
T.~Mohammed, C.~Joe-Wong, R.~Babbar, and M.~Di~Francesco, ``Distributed inference acceleration with adaptive dnn partitioning and offloading,'' in \emph{IEEE INFOCOM 2020-IEEE Conference on Computer Communications}.\hskip 1em plus 0.5em minus 0.4em\relax IEEE, 2020, pp. 854--863.

\bibitem{zeng2020coedge}
L.~Zeng, X.~Chen, Z.~Zhou, L.~Yang, and J.~Zhang, ``Coedge: Cooperative dnn inference with adaptive workload partitioning over heterogeneous edge devices,'' \emph{IEEE/ACM Transactions on Networking}, vol.~29, no.~2, pp. 595--608, 2020.

\bibitem{krestinskaya2019neuromemristive}
O.~Krestinskaya, A.~P. James, and L.~O. Chua, ``Neuromemristive circuits for edge computing: A review,'' \emph{IEEE transactions on neural networks and learning systems}, vol.~31, no.~1, pp. 4--23, 2019.

\bibitem{lammie2020memtorch}
C.~Lammie, W.~Xiang, B.~Linares-Barranco, and M.~R. Azghadi, ``Memtorch: An open-source simulation framework for memristive deep learning systems,'' \emph{arXiv preprint arXiv:2004.10971}, 2020.

\end{thebibliography}
% \vspace{-1.6cm}
\begin{IEEEbiography}[{\includegraphics[width=1in,height=1.25in,clip,keepaspectratio]{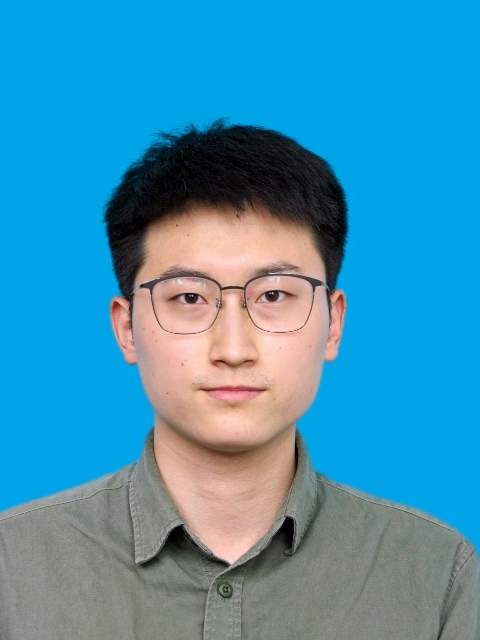}}]{Cheng Wang} received his B.S. and M.S. degrees from Lanzhou University, China, in 2017 and 2020. He is a PhD student at Hubei Engineering Research Center on Big Data Security, School of Cyber Science and Engineering of Huazhong University of Science and Technology, China. His research interests include web security, edge computing and recommender system.
\end{IEEEbiography}

% \vspace{-1.8cm}

\begin{IEEEbiography}[{\includegraphics[width=1in,height=1.25in,clip,keepaspectratio]{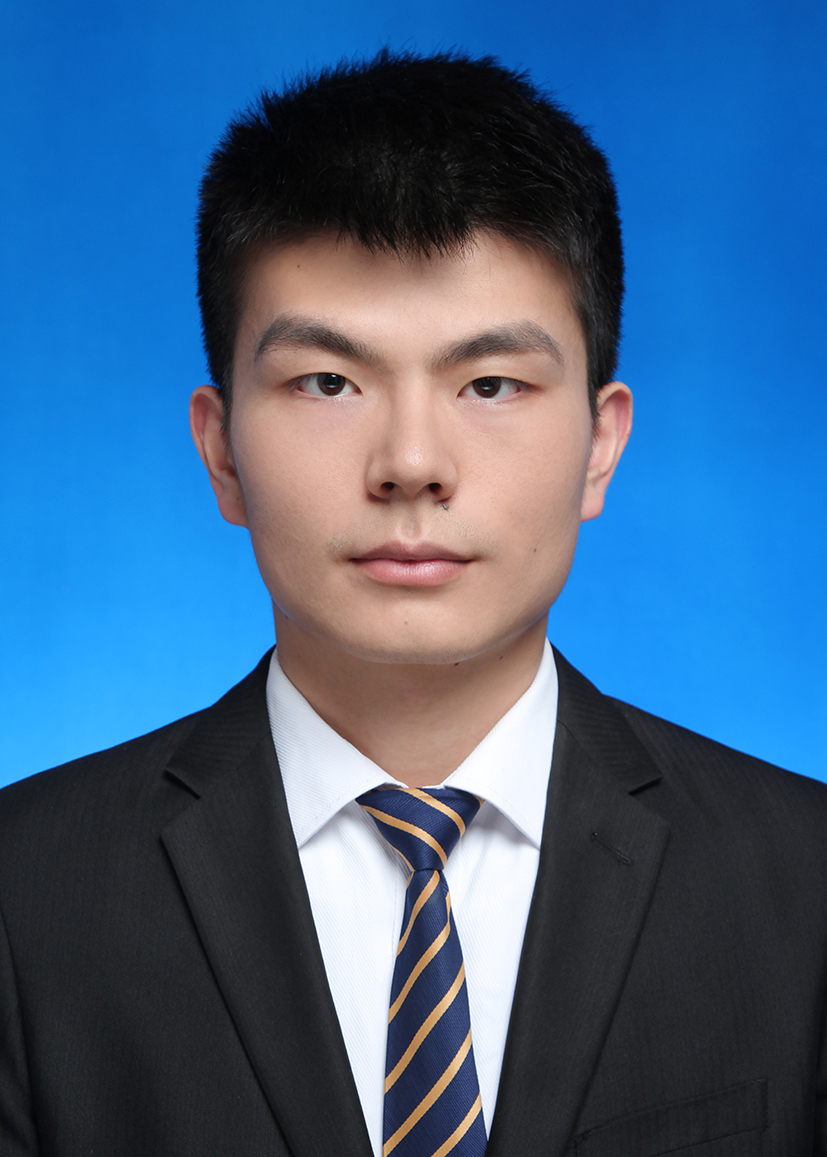}}]{Zenghui Yuan} received his M.S. degrees from Huazhong University of Science and Technology, China, in 2021. He is a PhD student at Hubei Engineering Research Center on Big Data Security, School of Cyber Science and Engineering of Huazhong University of Science and Technology, China. His main research interests include deep learning, reinforcement learning and edge computing.
\end{IEEEbiography}
% \vspace{-1.8cm}

\begin{IEEEbiography}[{\includegraphics[width=1in,height=1.25in,clip,keepaspectratio]{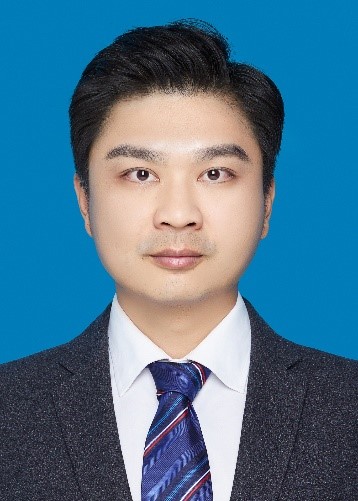}}]{Pan Zhou} (S'07–M'14-SM'20) is currently a full professor and PhD advisor with Hubei Engineering Research Center on Big Data Security, School of Cyber Science and Engineering, Huazhong University of Science and Technology (HUST), Wuhan, P.R. China. He received his Ph.D. in the School of Electrical and Computer Engineering at the Georgia Institute of Technology (Georgia Tech) in 2011, Atlanta, USA. He received his B.S. degree in the Advanced Class of HUST, and a M.S. degree in the Department of Electronics and Information Engineering from HUST, Wuhan, China, in 2006 and 2008, respectively. He held honorary degree in his bachelor and merit research award of HUST in his master study. He was a senior technical member at Oracle Inc., America, during 2011 to 2013, and worked on Hadoop and distributed storage system for big data analytics at Oracle Cloud Platform. He has published more than 170 refereed papers in international leading journals and key conferences in the area of security and privacy, big data analytics, machine learning, mobile computing and networks, including: IEEE/ACM TON, TKDE, TDSC, TIFS, JSAC, TMC, TPDS, TIT, TIP, TSE, TNNLS, TCOMP, TMM, TII, TAI, TCOM, TWC, TNSE, TVT, TETCI, ICDE, INFOCOM, CVPR, ICCV, ICDCS, ICPP, ACM MM, TOS, TKDD, NEURIPS, AAAI, IJCAI, NAACL, COLING, PoPETs/PETS, SECON, CIKM, ECAI, etc. He received the “Rising Star in Science and Technology of HUST” in 2017. He is currently an associate editor of IEEE Transactions on Network Science and Engineering. He is a Senior member of IEEE.
\end{IEEEbiography}

% \vspace{-1.8cm}

\begin{IEEEbiography}[{\includegraphics[width=1in,height=1.25in,clip,keepaspectratio]{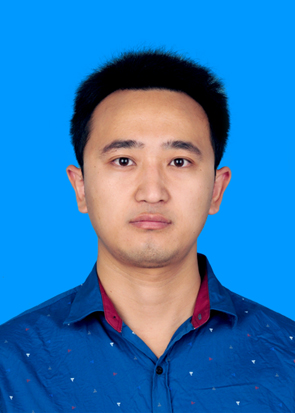}}]{Zichuan Xu} received the BSc and ME degrees
from Dalian University of Technology, China,
in 2008 and 2011, respectively, and the PhD degree
from the Australian National University, in 2016, all
in computer science. He was a research associate
with the University College London. He currently
is an associate professor in the School of Software,
Dalian University of Technology. His research interests include cloud computing, software-defined
networking, wireless sensor networks, algorithmic
game theory, and optimization problems.
\end{IEEEbiography}
% \vspace{-1.8cm}

\begin{IEEEbiography}[{\includegraphics[width=1in,height=1.25in,clip,keepaspectratio]{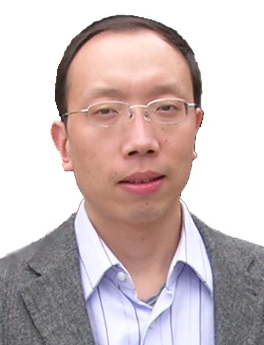}}]{Ruixuan Li} (Member, IEEE) received the B.S.,M.S., and Ph.D. degrees in computer science from the Huazhong University of Science and Technology, China, in 1997, 2000, and 2004, respectively. From 2009 to 2010, he was a Visiting Researcher with the Department of Electrical and Computer Engineering, University of Toronto. He is currently a Professor with the School of Computer Science and Technology, Huazhong University of Science and Technology. His research interests include cloud and edge computing, big data management, and distributed system security. He is a member of ACM.
\end{IEEEbiography}

% \vspace{-1.8cm}

\begin{IEEEbiography}[{\includegraphics[width=1in,height=1.25in,clip,keepaspectratio]{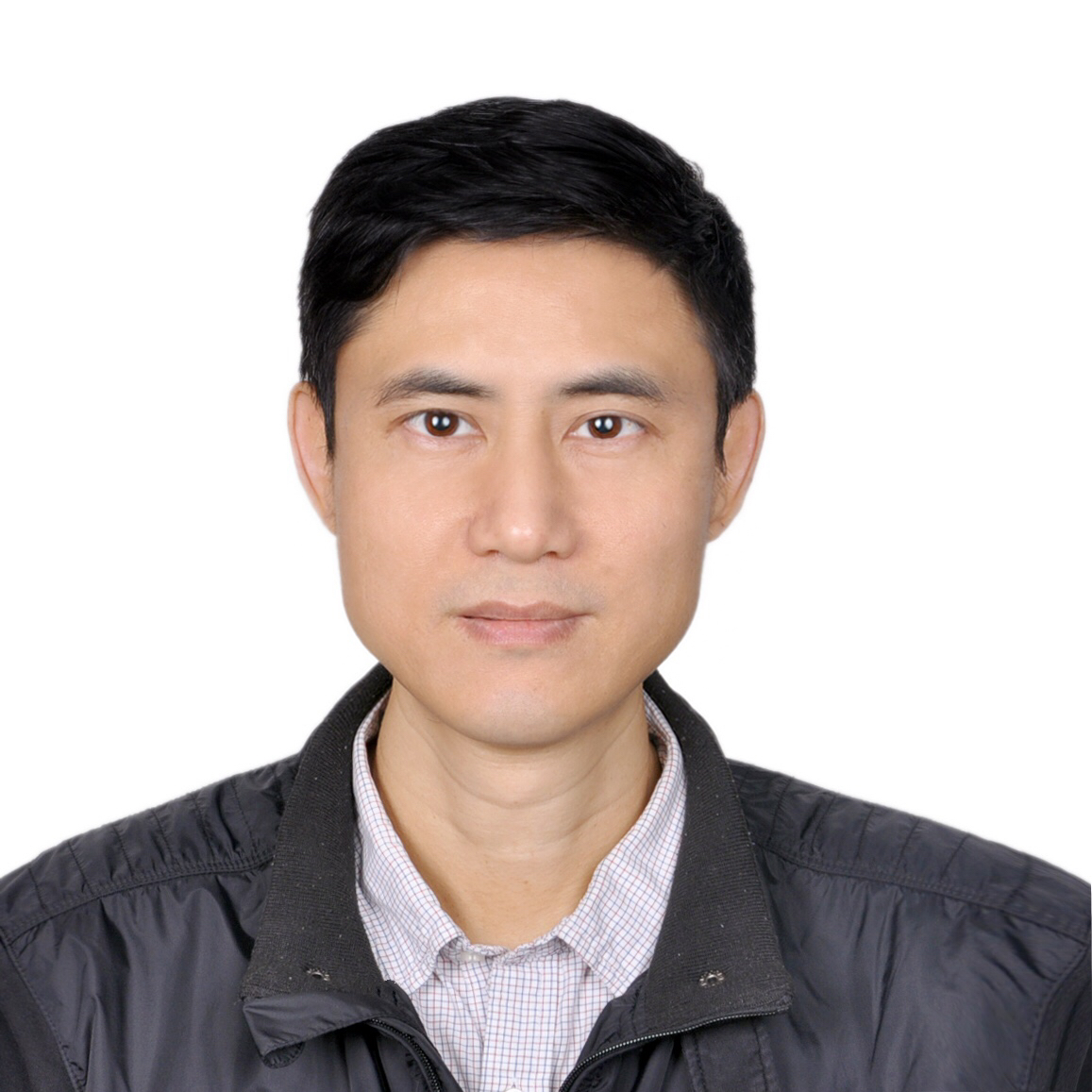}}]{Dapeng Oliver Wu}  (Fellow, IEEE)  Dapeng Oliver Wu (Fellow, IEEE) received the
B.E. degree in electrical engineering from the
Huazhong University of Science and Technology,
Wuhan, China, in 1990, the M.E. degree in electrical
engineering from the Beijing University of Posts and
Telecommunications, Beijing, China, in 1997, and
the Ph.D. degree in electrical and computer engineering from Carnegie Mellon University, Pittsburgh,
PA, USA, in 2003.

He is currently a Professor with the Department
of Electrical and Computer Engineering, University
of Florida, Gainesville, FL, USA; and the Director of the NSF Center
for Big Learning. His research interests are in the areas of networking,
communications, signal processing, computer vision, machine learning, smart
grid, and information and networks security.

Dr. Wu was an Elected Member of the Multimedia Signal Processing
Technical Committee and the IEEE Signal Processing Society from January
2009 to December 2012. He was elected as a Distinguished Lecturer by
the IEEE Vehicular Technology Society in 2016. He received the University
of Florida Term Professorship Award in 2017, the University of Florida
Research Foundation Professorship Award Limits in 2009, the AFOSR Young
Investigator Program (YIP) Award in 2009, the ONR Young Investigator
Program (YIP) Award in 2008, the NSF CAREER Award in 2007, the IEEE
TRANSACTIONS ON CIRCUITS AND SYSTEMS FOR VIDEO TECHNOLOGY
(CSVT) Best Paper Award in 2001, and the Best Paper Awards in IEEE
GLOBECOM 2011 and the International Conference on Quality of Service
in Heterogeneous Wired/Wireless Networks (QShine) 2006. He has served as
the Technical Program Committee (TPC) Chair for IEEE INFOCOM 2012,
the IEEE International Conference on Communications (ICC 2008), and the
Signal Processing for Communications Symposium; and a member of executive committee and/or technical program committee of over 100 conferences.
He has served as the Chair for the Award Committee, the Mobile and Wireless
Multimedia Interest Group (MobIG), the Technical Committee on Multimedia
Communications, and the IEEE Communications Society. He has served
as the Editor-in-Chief for IEEE TRANSACTIONS ON NETWORK SCIENCE
AND ENGINEERING; an Editor-at-Large for IEEE OPEN JOURNAL OF THE
COMMUNICATIONS SOCIETY; the Founding Editor-in-Chief for journal of
Advances in Multimedia; and an Associate Editor for IEEE TRANSACTIONS
ON CLOUD COMPUTING, IEEE TRANSACTIONS ON COMMUNICATIONS,
IEEE TRANSACTIONS ON SIGNAL AND INFORMATION PROCESSING OVER
NETWORKS, IEEE Signal Processing Magazine, IEEE TRANSACTIONS ON
CIRCUITS AND SYSTEMS FOR VIDEO TECHNOLOGY, IEEE TRANSACTIONS ON WIRELESS COMMUNICATIONS, and IEEE TRANSACTIONS ON
VEHICULAR TECHNOLOGY. He is also a Guest-Editor for the Special Issue
on Cross-Layer Optimized Wireless Multimedia Communications and the
Special Issue on Airborne Communication Networks of IEEE JOURNAL ON
SELECTED AREAS IN COMMUNICATIONS (JSAC).
\end{IEEEbiography}
% biography section
% 
% If you have an EPS/PDF photo (graphicx package needed) extra braces are
% needed around the contents of the optional argument to biography to prevent
% the LaTeX parser from getting confused when it sees the complicated
% \includegraphics command within an optional argument. (You could create
% your own custom macro containing the \includegraphics command to make things
% simpler here.)
%\begin{IEEEbiography}[{\includegraphics[width=1in,height=1.25in,clip,keepaspectratio]{mshell}}]{Michael Shell}
% or if you just want to reserve a space for a photo:

% \begin{IEEEbiography}{Michael Shell}
% Biography text here.
% \end{IEEEbiography}

% % if you will not have a photo at all:
% \begin{IEEEbiographynophoto}{John Doe}
% Biography text here.
% \end{IEEEbiographynophoto}

% insert where needed to balance the two columns on the last page with
% biographies
%\newpage

% \begin{IEEEbiographynophoto}{Jane Doe}
% Biography text here.
% \end{IEEEbiographynophoto}

% You can push biographies down or up by placing
% a \vfill before or after them. The appropriate
% use of \vfill depends on what kind of text is
% on the last page and whether or not the columns
% are being equalized.

%\vfill

% Can be used to pull up biographies so that the bottom of the last one
% is flush with the other column.
%\enlargethispage{-5in}

% that's all folks
\end{document}